\documentclass[useAMS]{tMPH2e}

\begin{document}

\markboth{Pablo Echenique and J. L. Alonso}{A review of Hartree-Fock SCF}

\title{A mathematical and computational review of\\ Hartree-Fock SCF methods in
Quantum Chemistry}

\author{Pablo Echenique$^{\ast}$$\dag$${\ddag}$\thanks{$^\ast$Corresponding
        author. Email: echenique.p@gmail.com
\vspace{6pt}} and J. L. Alonso$\dag$${\ddag}$\\\vspace{6pt}
  $\dag$ Departamento de F{\'{\i}}sica Te\'orica, Universidad de Zaragoza,\\
   Pedro Cerbuna 12, 50009, Zaragoza, Spain.\\
  $\ddag$ Instituto de Biocomputaci\'on y F{\'{\i}}sica de los Sistemas
   Complejos (BIFI),\\ Edificio Cervantes, Corona de Arag\'on 42, 50009,
   Zaragoza, Spain.\\\vspace{6pt}\received{May 2007\,} }

\maketitle

\begin{abstract}

We present here a review of the fundamental topics of Hartree-Fock
theory in Quantum Chemistry. From the molecular Hamiltonian, using and
discussing the Born-Oppenheimer approximation, we arrive to the
Hartree and Hartree-Fock equations for the electronic problem. Special
emphasis is placed in the most relevant mathematical aspects of the
theoretical derivation of the final equations, as well as in the
results regarding the existence and uniqueness of their
solutions. All Hartree-Fock versions with different spin
restrictions are systematically extracted from the general case, thus
providing a unifying framework. Then, the discretization of the
one-electron orbitals space is reviewed and the Roothaan-Hall
formalism introduced. This leads to a exposition of the basic
underlying concepts related to the construction and selection of
Gaussian basis sets, focusing in algorithmic efficiency issues.
Finally, we close the review with a section in which the most relevant
modern developments (specially those related to the design of
linear-scaling methods) are commented and linked to the issues
discussed.  The whole work is intentionally introductory and rather
self-contained, so that it may be useful for non experts that aim to
use quantum chemical methods in interdisciplinary
applications. Moreover, much material that is found scattered in the
literature has been put together here to facilitate comprehension and
to serve as a handy reference.

\bigskip

\centerline{\bfseries Table of contents}\medskip

\hbox to \textwidth{\hsize\textwidth\hspace{-16pt}\vbox{\noindent\hsize18pc
\\
{1.}    Introduction\\
{2.}    Molecular Hamiltonian and atomic units\\
{3.}    The Born-Oppenheimer approximation\\
{4.}    The variational method\\
{5.}    The statement of the problem\\
{6.}    The Hartree approximation\\
{7.}    The Hartree-Fock approximation\\
\\} \hspace{2pt}\vbox{\noindent\hsize18pc
\\
{8.}    The Roothaan-Hall equations\\
{9.}    Introduction to Gaussian basis sets\\
{10.}   Modern developments:
 An introduction \\ to linear-scaling methods \\
{Appendix A:}    Functional derivatives\\
{Appendix B:}    Lagrange multipliers\\
\\
\\
      }}
\end{abstract}

\section[Introduction]
        {Introduction}
\label{sec:QC_introduction}

In the hot field of computer simulation of biological macromolecules,
available potential energy functions are often not accurate enough to
properly describe complex processes such as the folding of proteins
\cite{Sko2005PNAS,Sno2005ARBBS,Sch2005SCI,Gin2005NAR,Bon2001ARBBS,Hao1999COSB,Ech2007COP}.
In order to improve the situation, it is convenient to extract ab
initio information from quantum mechanical calculations with the hope
of being able to devise less computationally demanding methods that
can be used to tackle large systems. In this spirit, the effective
potential for the nuclei calculated in the non-relativistic
Born-Oppenheimer approximation is typically considered as a good
reference to assess the accuracy of cheaper potentials
\cite{Mor2006JPCB,Jen2005ARCC,Mac2004JCC,Mor2004PNAS,Bor2003JPCB,Fri1998COSB,Bea1997JACS}.
The study of molecules at this level of theoretical detail and the
design of computationally efficient approximations for solving the
demanding equations that appear constitute the major part of the field
called \emph{quantum chemistry} \cite{Bar2000PAC,Sim1991JPC}. In this
work, we voluntarily circumscribe ourselves to the basic formalism
needed for the ground-state quantum chemical calculations that are
typically performed in this context. For more general expositions, we
refer the reader to any of the thorough accounts in
refs.~\cite{Lev1999BOOK,Jen1998BOOK,Sza1996BOOK}.

In sec.~\ref{sec:QC_molecular_H}, we introduce the molecular
Hamiltonian and a special set of units (the atomic ones) that are
convenient to simplify the equations. In
sec.~\ref{sec:QC_Born-Oppenheimer}, we present in an axiomatic way the
concepts and expressions related to the separation of the electronic
and nuclear problems in the Born-Oppenheimer scheme. In
sec.~\ref{sec:QC_variational}, we introduce the variational method
that underlies the derivation of the basic equations of the Hartree
and Hartree-Fock approximations, discussed in
sec.~\ref{sec:QC_Hartree} and~\ref{sec:QC_Hartree-Fock}
respectively. The computational implementation of the Hartree-Fock
approximation is tackled in sec.~\ref{sec:QC_Roothaan}, where the
celebrated Roothaan-Hall equations are derived. In
sec.~\ref{sec:QC_basis_sets}, the main issues related to the
construction and selection of Gaussian basis sets are discussed, and,
finally, in sec.~\ref{sec:QC_modern_developments}, the hottest areas
of modern research are briefly reviewed and linked to the issues in
the rest of the work, with a special emphasis in the development
of linear-scaling methods.

\section[Molecular Hamiltonian and atomic units]
         {Molecular Hamiltonian and atomic units}
\label{sec:QC_molecular_H}

Since 1960, the international scientific community has agreed on an
`official' set of basic units for measurements: \emph{Le
  \underline{S}yst\`eme \underline{I}nternational d'Unit\'es}, or SI
for short (see {\texttt{http://www.bipm.org/en/si/}} and
ref.~\cite{Tay1995TR}). The meter (m), the kilogram (kg), the
second~(s), the ampere (A), the kelvin (K), the mole (mol), the joule
(J) and the pascal~(Pa) are examples of SI units.

Sticking to the SI scheme, the non-relativistic quantum mechanical
Hamiltonian operator of a molecule consisting of $N_{N}$ nuclei (with
atomic numbers $Z_{\alpha}$ and masses $M_{\alpha}$,
$\alpha=1,\ldots,N_{N}$) and $N$ electrons (i.e., the \emph{molecular
Hamiltonian}) is expressed as\footnote{\label{foot:no_spin} Note that
the non-relativistic molecular Hamiltonian does not depend on
spin-like variables.}:

\begin{eqnarray}
\label{eq:chQC_ham_tot_SI}
\hat{H}&=&\mbox{}-\sum_{\alpha=1}^{N_{N}}
            \frac{{\hbar}^{2}}{2M_{\alpha}}{\nabla}_{\alpha}^{2}-
          \sum_{i=1}^{N}
            \frac{{\hbar}^{2}}{2m_{e}}{\nabla}_{i}^{2}+
          \frac{1}{2}\sum_{\alpha \ne \beta}
            \left ( \frac{e^{2}}{4{\pi}{\epsilon}_{0}} \right )
            \frac{Z_{\alpha}Z_{\beta}}{|{\bm R}_{\beta}-{\bm R}_{\alpha}|}
            \nonumber \\
       &&\mbox{} - \sum_{i=1}^{N}\sum_{\alpha=1}^{N_{N}}
            \left ( \frac{e^{2}}{4{\pi}{\epsilon}_{0}} \right )
            \frac{Z_{\alpha}}{|{\bm R}_{\alpha}-{\bm r}_{i}|}+
          \frac{1}{2}\sum_{i \ne j}
            \left ( \frac{e^{2}}{4{\pi}{\epsilon}_{0}} \right )
            \frac{1}{|{\bm r}_{j}-{\bm r}_{i}|} \ ,
\end{eqnarray}

where $\hbar$ stands for $h/2\pi$, being $h$ Planck's constant,
$m_{e}$ denotes the electron mass, $e$~the proton charge, ${\bm
r}_{i}$ the position of the $i$-th electron, ${\bm R}_{\alpha}$ that
of the $\alpha$-th nucleus, ${\epsilon}_{0}$ the vacuum permittivity
and ${\nabla}_{i}^{2}$ the Laplacian operator with respect to the
coordinates of the $i$-th particle.

Although using a common set of units presents obvious communicative
advantages, when circumscribed to a particular field of science, it is
common to appeal to non-SI units in order to simplify the most
frequently used equations by getting rid of some constant factors that
always appear grouped in the same ways and, thus, make the numerical
values in any calculation of the order of unity. In the field of
quantum chemistry, {\it atomic units} (see
table~\ref{tab:atomic_units}), proposed in ref.~\cite{Har1927PCPS} and
named in ref.~\cite{Shu1959NAT}, are typically used. In these units,
eq.~(\ref{eq:chQC_ham_tot_SI}) is substantially simplified to

\begin{eqnarray}
\label{eq:chQC_ham_tot_au}
\hat{H}&=&\mbox{} -\sum_{\alpha=1}^{N_{N}}
            \frac{1}{2M_{\alpha}}{\nabla}_{\alpha}^{2}-
          \sum_{i=1}^{N}
            \frac{1}{2}{\nabla}_{i}^{2}+
          \frac{1}{2}\sum_{\alpha \ne \beta}
            \frac{Z_{\alpha}Z_{\beta}}{|{\bm R}_{\beta}-{\bm R}_{\alpha}|}
       \nonumber \\
       &&\mbox{} -\sum_{i=1}^{N}\sum_{\alpha=1}^{N_{N}}
            \frac{Z_{\alpha}}{|{\bm R}_{\alpha}-{\bm r}_{i}|}+
          \frac{1}{2}\sum_{i \ne j}
            \frac{1}{|{\bm r}_{j}-{\bm r}_{i}|} \ .
\end{eqnarray}

\begin{table}
\tbl{
Atomic units up to five significant
digits. Taken from the National Institute of Standards and Technology
(NIST) web page at {\texttt{http://physics.nist.gov/cuu/Constants/}}.
Note that only four independent units are required in a 
mechanical-plus-electromagnetic system. The rest of them can be
easily obtained from any such four. For example, using the units
in the table, the relations $\hbar = 1$ and $1/(4\pi\epsilon_0)=1$
result.}
{\begin{tabular}{ll}
\toprule
Unit of mass: &
mass of the electron $= m_{e} = 9.1094 \cdot 10^{-31}$ kg \\
Unit of charge: &
charge on the proton $= e = 1.6022 \cdot 10^{-19}$ C \\
Unit of length: &
1 bohr
$= a_{0} = \frac{4{\pi}{\epsilon}_{0}{\hbar}^{2}}{m_{e}e^{2}} = 0.52918$ \AA
$= 5.2918 \cdot 10^{-11}$ m\\
Unit of energy: &
1 hartree $= \frac{{\hbar}^{2}}{m_{e}a_{0}^{2}} = 627.51$ kcal/mol
$= 4.3597 \cdot 10^{-18}$ J\\
\botrule
\end{tabular}}
\label{tab:atomic_units} 
\end{table}

Since all the relevant expressions in quantum chemistry are derived in
one way or another from the molecular Hamiltonian, the simplification
brought up by the use of atomic units propagates to the whole
formalism. Consequently, they shall be the choice all throughout this
work.

\begin{table}
\tbl{Energy units conversion factors
to five significant digits. Taken from the National Institute of
Standards and Technology (NIST) web page at
{\texttt{http://physics.nist.gov/cuu/Constants/}}. The table must be
read by rows. For example, the value $4.1838$, in the third row,
fourth column, indicates that 1 kcal/mol = \mbox{4.1838 kJ/mol.}}
{\begin{tabular}{lccccc}
\toprule
            &              1 hartree &                   1 eV &    1 kcal/mol          &   1 kJ/mol             &         1 cm$^{-1}$ \\
\colrule
1 hartree   &                      1 & 27.211                 & 627.51                 & 262.54                 & 219470              \\
1 eV        & $3.6750 \cdot 10^{-2}$ &                      1 &        23.061          &     96.483           &              8065.5 \\
1 kcal/mol  & $1.5936 \cdot 10^{-3}$ & $4.3363 \cdot 10^{-2}$ &             1          &     4.1838          &              349.75 \\
1 kJ/mol    & $3.8089 \cdot 10^{-4}$ & $1.0364 \cdot 10^{-2}$ & $2.3902 \cdot 10^{-1}$ &          1             &              83.595 \\
1 cm$^{-1}$ & $4.5560 \cdot 10^{-6}$ & $1.2398 \cdot 10^{-4}$ & $2.8592 \cdot 10^{-3}$ & $1.1962 \cdot 10^{-2}$ &                   1 \\
\botrule
\end{tabular}}
\label{tab:unit_conversion}
\end{table}

Apart from the atomic units and the SI ones, there are some other
miscellaneous units that are often used in the literature: the
\emph{{\aa}ngstr\"om}, which is a unit of length defined as \mbox{1
\AA $\, = 10^{-10}$ m}, and the units of energy \emph{cm}$^{-1}$
(which reminds about the spectroscopic origins of quantum chemistry
and, even, quantum mechanics), \emph{electronvolt} (eV),
\emph{kilocalorie per mole} (kcal/mol) and \emph{kilojoule per mole}
(kJ/mol).  The last two are specially used in the field of
macromolecular simulations and quantify the energy of a mole of
entities; for example, if one asserts that the torsional barrier
height for $\mathrm{H}_{2}\mathrm{O}_{2}$ is $\sim 7$ kcal/mol, one is
really saying that, in order to make a mole of
$\mathrm{H}_{2}\mathrm{O}_{2}$ (i.e., $N_{\mathrm{A}} \simeq 6.0221
\cdot 10^{23}$ molecules) rotate $180^{\mathrm{o}}$ around the O--O
bond, one must spend $\sim 7$ kcal. For the conversion factors between
the different energy units, see table~\ref{tab:unit_conversion}.

Finally, to close this section, we rewrite
eq.~(\ref{eq:chQC_ham_tot_au}) introducing some self-explanatory
notation that will be used in the subsequent discussion:

\begin{subequations}
\label{eq:chQC_ham_tot_op}
\begin{align}
& \hat{H}=\hat{T}_{N}+\hat{T}_{e}+\hat{V}_{NN}+\hat{V}_{eN}+\hat{V}_{ee} \  ,
  \label{eq:chQC_ham_tot_op_a} \\
& \hat{T}_{N} := - \sum_{\alpha=1}^{N_{N}} \frac{1}{2M_{\alpha}}{\nabla}_{\alpha}^{2} \ ,
  \label{eq:chQC_ham_tot_op_b} \\
& \hat{T}_{e} := - \sum_{i=1}^{N} \frac{1}{2}{\nabla}_{i}^{2} \ ,
  \label{eq:chQC_ham_tot_op_c} \\
& \hat{V}_{NN} := \frac{1}{2}\sum_{\alpha \ne \beta} \frac{Z_{\alpha}Z_{\beta}}{R_{\alpha\beta}} \ ,
  \label{eq:chQC_ham_tot_op_d} \\
& \hat{V}_{eN} := - \sum_{i=1}^{N}\sum_{\alpha=1}^{N_{N}} \frac{Z_{\alpha}}{R_{\alpha i}} \ ,
  \label{eq:chQC_ham_tot_op_e} \\
& \hat{V}_{ee} := \frac{1}{2}\sum_{i \ne j} \frac{1}{r_{ij}} \ .
  \label{eq:chQC_ham_tot_op_f}
\end{align}
\end{subequations}

\section[The Born-Oppenheimer approximation]
         {The Born-Oppenheimer approximation}
\label{sec:QC_Born-Oppenheimer}

To think of a macromolecule as a set of quantum objects described by a
\emph{wavefunction} $\Psi(X_{1},\ldots,X_{N_{N}},x_{1},\ldots,x_{N})$
dependent on the spatial and \label{page:foot:spin_convenient}
spin\footnote{\label{foot:spin_convenient} One convenient way of
thinking about functions that depend on spin-like variables is as an
$m$-tuple of ordinary $\mathbb{R}^{3N}$ functions, where $m$ is the
finite number of possible values of the spin. In the case of a
one-particular wavefunction describing an electron, for example,
$\sigma$ can take two values (say, $-1/2$ and $1/2$) in such a way
that one may picture any general \emph{spin-orbital} $\Psi_{i}(x)$ as
a 2-tuple $\big(\Phi^{-1/2}_{i}({\bm r}),\Phi^{1/2}_{i}({\bm
r})\big)$. Of course, another valid way of imagining $\Psi_{i}(x)$ is
simply as a function of four variables, three real and one discrete.}
degrees of freedom, \mbox{$x_{i}:=({\bm r}_{i},\sigma_{i})$}, of the
electrons and on those of the nuclei, $X_{\alpha}:=({\bm
R}_{\alpha},\Sigma_{\alpha})$, would be too much for the imagination
of physicists and chemists. All the language of chemistry would have
to be remade and simple sentences in textbooks, such as ``rotation
about this single bond allows the molecule to avoid steric clashes
between atoms'' or even ``a polymer is a long chain-like molecule
composed of repeating monomer units'', would have to be translated
into long and counter-intuitive statements involving probability and
`quantum jargon'. Conscious or not, we think of molecules as classical
objects.

More precisely, we are ready to accept that electrons are quantum (we
know of the interference experiments, electrons are light, we are
accustomed to draw atomic `orbitals', etc.), however, we are reluctant
to concede the same status to nuclei. Nuclei are heavier than
electrons (at least $\sim 2000$ times heavier, in the case of the
single proton nucleus of hydrogen) and we picture them in our
imagination as `classical things' that move, bond to each other,
rotate around bonds and are at precise points at precise times. We
imagine nuclei `slowly moving' in the field of the electrons, which,
for each position of the first, immediately `adjust their quantum
state'.

The formalization of these ideas is called {\it
  \underline{B}orn-\underline{O}ppenheimer} (BO) {\it approximation}
\cite{Bor1954BOOK,Bor1927APL} and the confirmation of its being good
for many relevant problems is a fact that supports our intuitions
about the topic and that lies at the foundations of the vast majority
of the images, the concepts and the research in quantum
chemistry\footnote{\label{foot:bo_spark} There are many phenomena,
  however, in which the Born-Oppenheimer approximation is broken.  For
  example, in striking a flint to create a spark, mechanical motion of
  the nuclei excites electrons into a plasma that then emits light
  \cite{Mar2000BOOK}.}.

Like any approximation, the Born-Oppenheimer one may be either
\emph{derived} from the exact problem (in this case, the entangled
behaviour of electrons and nuclei as the same quantum object) or
simply \emph{proposed} on the basis of physical intuition, and later
confirmed to be good enough (or not) by comparison with the exact
theory or with the experiment. Of course, if it is possible, the first
way should be preferred, since it allows to develop a deeper insight
about the terms we are neglecting and the specific details that we
will miss. However, although in virtually every quantum chemistry book
\cite{Shi2006BOOK,Cra2002BOOK,Jen1998BOOK,Sza1996BOOK,Par1989BOOK}
hand-waving derivations up to different levels of detail are performed
and the Born-Oppenheimer approximation is typically presented as
unproblematic, it seems that the fine mathematical details on which
these `standard' approaches are based are far from clear
\cite{Sut2005PCCP,Sut1997AQC,Sut1993JCSFT}. This state of affairs does
not imply that the final equations that will need to be solved are
ill-defined or that the numerical methods based on the theory are
unstable; in fact, it is just the contrary (see the discussion below),
because the problems are related only to the precise relation between
the concepts in the whole theory and those in its simplified
version. Nevertheless, the many subtleties involved in a
\emph{derivation} of the Born-Oppenheimer approximation scheme from
the exact equations suggest that the second way, that of
\emph{proposing} the approximation, be taken. Hence, in the following
paragraphs, an \emph{axiomatic} presentation of the main expressions,
aimed mostly to fix the notation and to introduce the language, will
be performed.

First of all, if we examine the Hamiltonian operator in
eq.~(\ref{eq:chQC_ham_tot_au}), we see that the term~$\hat{V}_{eN}$
prevents the problem from being separable in the nuclear and
electronic coordinates, i.e., if we define
$\underbar{x}:=(x_{1},\ldots,x_{N})$ as the set of all electronic
coordinates (spatial and spin-like) and do likewise with the nuclear
coordinates $\underbar{X}$, the term $\hat{V}_{eN}$ prevents any
wavefunction $\Psi(\underbar{X},\underbar{x})$ solution of the
\emph{time-independent Schr\"odinger equation},

\begin{equation}
\label{eq:chQC_schrodinger}
\hat{H}\,\Psi(\underbar{X},\underbar{x})=
\left(\hat{T}_{N}+\hat{T}_{e}+\hat{V}_{NN}+\hat{V}_{eN}+\hat{V}_{ee}\right)
\Psi(\underbar{X},\underbar{x})= E\,\Psi(\underbar{X},\underbar{x}) \ ,
\end{equation}

from being written as a product,
$\Psi(\underbar{X},\underbar{x})=\Psi_{N}(\underbar{X})\Psi_{e}(\underbar{x})$,
of an electronic wavefunction and a nuclear one. If this were the
case, the problem would still be difficult (because of the Coulomb
terms $\hat{V}_{NN}$ and $\hat{V}_{ee}$), but we would be able to
focus on the electrons and on the nuclei separately.

The starting point for the Born-Oppenheimer approximation consists in
assuming that a less strict separability is achieved, in such a way
that, for a pair of suitably chosen $\Psi_{N}(\underbar{X})$ and
$\Psi_{e}(\underbar{x};\underbar{X})$, any wavefunction solution of
eq.~(\ref{eq:chQC_schrodinger}) (or at least those in which we are
interested; for example, the eigenstates corresponding to the lowest
lying eigenvalues) can be expressed as

\begin{equation}
\label{eq:chQC_bo_wf}
\Psi(\underbar{X},\underbar{x})=\Psi_{N}(\underbar{X})
  \Psi_{e}(\underbar{x};\underbar{X}) \ ,
\end{equation}

where we have used a `;' to separate the two sets of variables in the
electronic part of the wavefunction in order to indicate that, in what
follows, it is convenient to use the image that `from the point of
view of the electrons, the nuclear degrees of freedom are fixed', so
that the electronic wavefunction depends `parametrically' on them. In
other words, that the $\underbar{X}$ are not quantum variables in
eq.~(\ref{eq:chQC_clamped}) below. Of course, it is just a `semantic'
semicolon; if anyone feels uncomfortable about it, she may drop it and
write a normal comma.

Notably, in ref.~\cite{Hun1975IJQC}, Hunter showed that any solution
of the Schr\"odinger equation can in fact be written exactly in the
form of eq.~(\ref{eq:chQC_bo_wf}), and that the two functions,
$\Psi_{N}(\underbar{X})$ and $\Psi_{e}(\underbar{x};\underbar{X})$,
into which $\Psi(\underbar{X},\underbar{x})$ is split may be
interpreted as marginal and conditional probability amplitudes
respectively. However, despite the insight that is gained from this
treatment, it is of no practical value, since the knowledge of the
exact solution $\Psi(\underbar{X},\underbar{x})$ is required in order
to compute $\Psi_{N}(\underbar{X})$ and
$\Psi_{e}(\underbar{x};\underbar{X})$ in Hunter's approach.

In the Born-Oppenheimer scheme, an additional assumption is made in
order to avoid this drawback: \emph{the equations obeyed by the
electronic and nuclear parts of the wavefunction are supposed to be
known}. Hence, $\Psi_{e}(\underbar{x};\underbar{X})$ is assumed to be
a solution of the time-independent \emph{clamped nuclei Schr\"odinger
equation},

\begin{equation}
\label{eq:chQC_clamped}
\left(\hat{T}_{e}+\hat{V}_{eN}(\underbar{r};\underbar{R})+
  \hat{V}_{ee}(\underbar{r})\right)
  \Psi_{e}(\underbar{x};\underbar{R}) := \hat{H}_{e}(\underbar{R})\,
    \Psi_{e}(\underbar{x};\underbar{R}) =
    E_{e}(\underbar{R})\,{\Psi_{e}(\underbar{x};\underbar{R})} \ ,
\end{equation}

where the \emph{electronic Hamiltonian operator}
$\hat{H}_{e}(\underbar{R})$ and the \emph{electronic energy}
$E_{e}(\underbar{R})$ (both dependent on the nuclei positions) have
been defined, and, since the nuclear spins do not enter the
expression, we have explicitly indicated that $\Psi_{e}$ depends
parametrically on~$\underbar{R}$ and not on $\underbar{X}$.

The common interpretation of the clamped nuclei equation is, as we
have advanced at the beginning of the section, that the nuclei are
much `slower' than the electrons and, therefore, the latter can
automatically adjust their quantum state to the instantaneous
positions of the former. Physically, eq.~(\ref{eq:chQC_clamped}) is
just the time-independent Schr\"odinger equation of $N$ particles (the
electrons) of mass $m_{e}$ and charge $-e$ in the \emph{external
electric field} of $N_{N}$~\emph{point charges} (the nuclei) of size
$eZ_{\alpha}$ at locations ${\bm R}_{\alpha}$. Mathematically, it is
an eigenvalue problem that has been thoroughly studied in the
literature and whose properties are well-known
\cite{Yse2003TR,Sim2000JMP,Hun2000JMP,Rus1992LNP,Hun1966HPA,VWi1964MFSDVS}. In
particular, it can be shown that, in the case of neutral or positively
charged molecules (i.e., with $Z:=\sum_{\alpha}Z_{\alpha}\ge N$), the
clamped nuclei equation has an infinite number of normalizable
solutions in the discrete spectrum of $\hat{H}_{e}(\underbar{R})$
(\emph{bound-states}) for every value of $\underbar{R}$
\cite{Fri2003ARMA,Zhi1960TMMO}.

These solutions must be regarded as the different \emph{electronic
energy levels}, and a further approximation that is typically made
consists in, not only accepting that the electrons immediately
`follow' nuclear motion, but also that, for each value of the nuclear
positions~$\underbar{R}$, they are in the electronic
ground-state\footnote{\label{foot:fundamental_e} This is customarily
assumed in the literature and it is supported by the general fact that
electronic degrees of freedom are typically more difficult to excite
than nuclear ones. Hence, in the vast majority of the numerical
implementations of the theory, only the electronic ground-state is
sought. We will see this in the forecoming sections.}, i.e., the one
with the lower $E_{e}(\underbar{R})$.

Consequently, we define

\begin{equation}
\label{eq:chQC_bo_Eeff}
E_{e}^{\mathrm{eff}}(\underbar{R}):=E_{e}^{0}(\underbar{R}) \ .
\end{equation}

to be the {\it effective electronic field} in which the nuclei move,
in such a way that, once we have solved the problem in
eq.~(\ref{eq:chQC_clamped}) and know $E_{e}^{0}(\underbar{R})$, the
time-independent \emph{nuclear Schr\"odinger equation} obeyed by
$\Psi_{N}(\underbar{X})$ is:

\begin{equation}
\label{eq:chQC_bo_N_ham}
\left ( \hat{T}_{N}+\hat{V}_{NN}(\underbar{R})+
  E_{e}^{\mathrm{eff}}(\underbar{R})\right )
 \Psi_{N}(\underbar{X}) := \hat{H}_{N} \Psi_{N}(\underbar{X}) = 
  E_{N}\Psi_{N}(\underbar{X}) \ ,
\end{equation}

where the \emph{effective nuclear Hamiltonian} $\hat{H}_{N}$ has been
implicitly defined.

Now, to close the section, we put together the main expressions of the
Born-Oppenheimer approximation for quick reference and we discuss them
in some more detail:

\begin{subequations}
\label{eq:chQC_bo_summary}
\begin{align}
& \hat{H}_{e}(\underbar{R})\,\Psi_{e}(\underbar{x};\underbar{R}) :=
  \left ( \hat{T}_{e} + \hat{V}_{eN}(\underbar{R}) + \hat{V}_{ee} \right )
  \Psi_{e}(\underbar{x};\underbar{R}) =
  E_{e}(\underbar{R})\,\Psi_{e}(\underbar{x};\underbar{R}) \  ,
  \label{eq:chQC_bo_summary_a} \\
& E_{e}^{\mathrm{eff}}(\underbar{R}):=E_{e}^{0}(\underbar{R}) \  ,
  \label{eq:chQC_bo_summary_b} \\
& \hat{H}_{N} \Psi_{N}(\underbar{X}) :=
  \left ( \hat{T}_{N}+V_{NN}(\underbar{R})+
          E_{e}^{\mathrm{eff}}(\underbar{R}) \right )
  \Psi_{N}(\underbar{X}) = E_{N}\Psi_{N}(\underbar{X}) \ ,
  \label{eq:chQC_bo_summary_c} \\
& \Psi(\underbar{x},\underbar{X}) \simeq
      \Psi_{e}^{0}(\underbar{x};\underbar{R})\,\Psi_{N}(\underbar{X})
    \ , \qquad  E \simeq E_{N} \ . \label{eq:chQC_bo_summary_d}
\end{align}
\end{subequations}

To start, note that the above equations are written in the logical
order in which they are imagined and used in any numerical
calculation. First, we assume the nuclei fixed at $\underbar{R}$ and
we (hopefully) solve the clamped nuclei electronic Schr\"odinger
equation (eq.~(\ref{eq:chQC_bo_summary_a})), obtaining the electronic
ground-state $\Psi_{e}^{0}(\underbar{x},\underbar{R})$ with its
corresponding energy $E_{e}^{0}(R)$. Next, we repeat this procedure
for all possible values\footnote{\label{foot:only_grid} Of course,
this cannot be done in practice. Due to the finite character of
available computational resources, what is customarily done is to
define a `grid' in $\underbar{R}$-space and compute
$E_{e}^{0}(\underbar{R})$ in a finite number of points.} of
$\underbar{R}$ and end up with an hyper-surface
$E_{e}^{0}(\underbar{R})$ in $\underbar{R}$-space. Finally, we add
this function to the analytical and easily computable
$V_{NN}(\underbar{R})$ and find the \emph{effective potential} that
determines the nuclear motion:

\begin{equation}
\label{eq:chQC_V_eff}
V_{N}^{\mathrm{eff}}(\underbar{R}) := 
     V_{NN}(\underbar{R}) + E_{e}^{0}(\underbar{R}) \ .
\end{equation}

It is, precisely, this effective potential that is called
\emph{\underline{P}otential \underline{E}nergy \underline{S}urface}
(PES) (or, more generally, \emph{\underline{P}otential
\underline{E}nergy \underline{H}yper-\underline{S}urface} (PEHS)) in
quantum chemistry and that is the central object through which
scientists picture chemical reactions or conformational changes of
macromolecules \cite{Hra2005BOOK}. In fact, the concept is so
appealing and the classical image so strong that, after `going
quantum', we can `go classical' back again and think of nuclei as
perfectly classical particles that move in the classical potential
$V_{N}^{\mathrm{eff}}(\underbar{R})$. In such a case, we would not
have to solve eq.~(\ref{eq:chQC_bo_summary_c}) but, instead, integrate
the Newtonian equations of motion. This is the basic assumption of
every typical force field used for classical ground-state
\emph{molecular dynamics}, such as the ones in the popular CHARMM
\cite{Bro1983JCC,Mac1998BOOK}, AMBER
\cite{Pea1995CPC,Pon2003APC,Che2001BP} or OPLS \cite{Jor1988JACS}
packages.

Finally, we would like to remind the reader that, despite the
hand-waving character of the arguments presented, up to this point,
every computational step has a clear description and
eqs.~(\ref{eq:chQC_bo_summary_a}) through~(\ref{eq:chQC_bo_summary_c})
could be considered as \emph{definitions} involving a certain degree
of notational abuse. To assume that the quantities obtained through
this process are close to those that proceed from a rigorous solution
of the time independent Schr\"odinger equation
(eq.~(\ref{eq:chQC_schrodinger})) is where the approximation really
lies. Hence, the more accurate eqs.~(\ref{eq:chQC_bo_summary_d}) are,
the better the Born-Oppenheimer guess is, and, like any other one, if
one does not trust in the heuristic grounds on which the final
equations stand, they may be taken as axiomatic and judged a
posteriori according to their results in particular
cases\footnote{\label{foot:approximations} Until now, two
approximations have been done: the non-relativistic character of the
objects studied and the Born-Oppenheimer approximation. In the
forecoming, many more will be done. The a priori quantification of
their goodness in large molecules is a formidable task and, despite
the efforts in this direction, in the end, the comparison with
experimental data is the only sound method for validation.}.

In quantum chemistry, the Born-Oppenheimer approximation is assumed in
a great fraction of the studies and it allows the central concept of
potential energy surface to be well-defined, apart from considerably
simplifying the calculations. The same decision is taken in this work.

\section[The variational method]
         {The variational method}
\label{sec:QC_variational}

There exists a mathematically appealing way of deriving the time
independent Schr\"odinger equation (eq.~(\ref{eq:chQC_schrodinger}))
from an extremal principle. To this end, we first define the
functional (see appendix~A) that corresponds to the expected value of
the energy,

\begin{equation}
\label{eq:chQC_functional_F}
\mathcal{F}[\Psi] := \langle\Psi|\,\hat{H}\,|\Psi\rangle \ ,
\end{equation}

where the traditional \emph{bra} and \emph{ket} notation is read as

\begin{equation}
\label{eq:chQC_braket}
\langle\Psi|\,\hat{O}\,|\Psi\rangle := \int \Psi^* (x) \hat{O} \Psi (x)
\,\mathrm{d}x \ ,
\end{equation}

$\hat{O}$ being any operator in the space of wavefunctions and $x$ a
dummy variable representing all possible coordinates on which $\Psi$
depends. The norm of $\Psi$, in this notation is expressed as $\langle
\Psi|\Psi\rangle = \int |\Psi (x)|^2 \,\mathrm{d}x$, and we shall say
that $\Psi$ is \emph{normalized} if $\langle \Psi|\Psi\rangle = 1$.  

If we want to optimize the energy functional above restricting the
search space to the normalized wavefunctions, the
constrained-extremals problem that results can be solved via the
\emph{Lagrange multipliers method} (see appendix~B) by constructing
the \emph{associated functional} $\widetilde{\mathcal{F}}[\Psi]$,
where we introduce a \emph{Lagrange multiplier} $\lambda$ to force
normalization:

\begin{equation}
\label{eq:chQC_functional_Ftilde}
\widetilde{\mathcal{F}}[\Psi] := \mathcal{F}[\Psi] + 
 \lambda \,\Big( \langle \Psi|\Psi\rangle - 1 \Big) \ .
\end{equation}

If we now ask that the functional derivative of
$\widetilde{\mathcal{F}}[\Psi]$ with respect to the complex conjugate
$\Psi^{*}$ of the wave
\label{page:foot:deriving_conjugate}
function\footnote{\label{foot:deriving_conjugate} A function of a
complex variable $z$ (or, analogously, a functional on a space of
complex functions) may be regarded as depending on two different sets
of independent variables: either $\mathrm{Re}(z)$ and $\mathrm{Im}(z)$
or $z$ and $z^{*}$. The choice frequently depending on technical
issues.} be zero, i.e., we look for the stationary points of
$\mathcal{F}[\Psi]$ conditioned by \mbox{$\langle\Psi|\Psi\rangle =
1$}, we obtain the eigenvalues equation for $\hat{H}$, i.e., the
time-independent Schr\"odinger equation. Additionally, it can be
shown, first, that, due to the self-adjointedness of $\hat{H}$, the
equation obtained from the stationarity condition with respect to
$\Psi$ (not $\Psi^{*}$) is just the complex conjugate and adds no new
information.

Moreover, one can see that the reverse implication is also true
\cite{Coh1977BOOK}, so that, if a given normalized wavefunction~$\Psi$
is a solution of the eigenvalue problem and belongs to the discrete
spectrum of $\hat{H}$, then the functional in
eq.~(\ref{eq:chQC_functional_Ftilde}) is stationary with respect to
$\Psi^{*}$:

\begin{equation}
\label{eq:chQC_extremal_principle}
\frac{\delta\widetilde{\mathcal{F}}[\Psi]}{\delta\Psi^{*}} = 0 \quad
 \Longleftrightarrow \quad
 \hat{H}\,\Psi=-\lambda\,\Psi:=E\,\Psi \quad \mathrm{and}
 \quad \langle\Psi|\Psi\rangle = 1 \ .
\end{equation}

This result, despite its conceptual interest, is of little practical
use, because it does not indicate an operative way to solve the
Schr\"odinger equation different from the ones that we already
knew. The equivalence above simply illustrates that mathematical
variational principles are over-arching theoretical statements from
which the differential equations that actually contain the details of
physical systems can be extracted. Nevertheless, using similar ideas,
we will derive another simple theorem which is indeed powerfully
practical: the \emph{Variational Theorem}.

Let $\{\Psi_{n}\}$ be a basis of eigenstates of the Hamiltonian
operator $\hat{H}$ and $\{E_{n}\}$ their corresponding
eigenvalues. Since~$\hat{H}$ is self-adjoint, the eigenstates
$\Psi_{n}$ can be chosen to be orthonormal (i.e.,
$\langle\Psi_{m}|\Psi_{n}\rangle=\delta_{mn}$) and any normalized
wavefunction $\Psi$ in the Hilbert space can be written as a linear
combination of them\footnote{\label{foot:discrete_H} We assume here,
for the sake of simplicity and in order to highlight the relevant
concepts, that $\hat{H}$ has only discrete spectrum. The ideas
involved in a general derivation are the same, but the technical
details and the notation are more complicated \cite{Dir1929PRSL}.}:

\begin{equation}
\label{eq:chQC_decomposition}
|\Psi\rangle = \sum_{n}C_{n}|\Psi_{n}\rangle \quad
 \mbox{provided that} \quad \sum_{n}|C_{n}|^{2}=1 \ .
\end{equation}

If we now denote by $E_{0}$ the lowest $E_{n}$ (i.e., the energy of
the ground-state)\footnote{\label{foot:sometimes_nofundamental} Its
existence is not guaranteed: it depends on the particular potential in
$\hat{H}$. However, for the physically relevant cases, there is indeed
a minimum energy in the set $\{E_{n}\}$.} and calculate the expected
value of the energy on an arbitrary state $\Psi$ such as the one in
eq.~(\ref{eq:chQC_decomposition}), we obtain

\begin{eqnarray}
\label{eq:chQC_variational_th}
\langle\Psi|\,\hat{H}\,|\Psi\rangle &=& \sum_{m,n}C_{m}^{*}C_{n}
   \langle\Psi_{m}|\,\hat{H}\,|\Psi_{n}\rangle = 
   \sum_{m,n}C_{m}^{*}C_{n}E_{n}\langle\Psi_{m}|\Psi_{n}\rangle \nonumber \\
& = & \sum_{m,n}C_{m}^{*}C_{n}E_{n}\delta_{mn}=
   \sum_{n}|C_{n}|^{2}E_{n} \geq \sum_{n}|C_{n}|^{2}E_{0} = E_{0} \ .
\end{eqnarray}

This simple relation is the \emph{Variational Theorem} and it states
that any wavefunction of the Hilbert space has an energy larger than
the one of the ground-state (the equality can only be achieved if
$\Psi=\Psi_{0}$). However trivial this fact may appear, it allows a
very fruitful `everything-goes' strategy when trying to approximate
the ground-state in a difficult problem. If one has a procedure for
finding a promising guess wavefunction (called \emph{variational
ansatz}), no matter how heuristic, semi-empirical or intuitive it may
be, one may expect that the lower the corresponding energy, the closer
to the ground-state it is\footnote{\label{foot:variational_ruggedness}
Of course, this not necessarily so (and, in any case, it depends on
the definition of `closer'), since it could happen that the
$\langle\Psi|\hat{H}|\Psi\rangle$ landscape in the constrained subset
of the Hilbert space in which the search is performed be `rugged'. In
such a case, we may have very different wavefunctions (say, in the
sense of the $L^{2}$-norm) with similar energies
$\langle\Psi|\hat{H}|\Psi\rangle$. The only `direction' in which one
can be sure that the situation improves when using the variational
procedure is the (very important) energetic one. That one is also
moving towards better values of any other observable is, in general,
no more than a \emph{bona fide} assumption.}.  This provides a
systematic strategy for improving the test wavefunction which may take
a number of particular forms.

One example of the application of the Variational Theorem is to
propose a family of normalized wavefunctions $\Psi_{\theta}$
parametrically dependent on a number $\theta$ and calculate the
$\theta$-dependent expected value of the
energy\footnote{\label{foot:overlap} Note that, if the functions
$\Psi_{\theta}$ were not normalized, then we should deal with the
constrained problem as in~(\ref{eq:chQC_functional_Ftilde}), or,
equivalently, we could include a dividing \emph{overlap term}
$\langle\Psi_{\theta}|\Psi_{\theta}\rangle$
in~(\ref{eq:chQC_E_theta}).}:

\begin{equation}
\label{eq:chQC_E_theta}
E(\theta) := \langle\Psi_{\theta}|\,\hat{H}\,|\Psi_{\theta}\rangle \ .
\end{equation}

Then, one may use the typical tools of one-variable calculus to find
the minimum of~$E(\theta)$ and thus make the best guess of the energy
$E_{0}$ constrained to the family $\Psi_{\theta}$. If the ansatz is
cleverly chosen, this estimate could be rather accurate, however, for
large systems that lack symmetry, it is very difficult to write a good
enough form for $\Psi_{\theta}$.

When dealing with a large number of particles, there exists another
protocol based on the Variational Theorem that will permit us to
derive the Hartree and Hartree-Fock equations for the electronic
wavefunction $\Psi_{e}$ (see secs.~\ref{sec:QC_Hartree}
and~\ref{sec:QC_Hartree-Fock}, respectively). The first step is to
devise a restricted way (a function $f$ with no free parameters) to
express $\Psi_{e}$ in terms of \emph{one-electron wavefunctions}, also
called \emph{orbitals} and denoted\footnote{\label{foot:a_eq_i} In
principle, there could be more orbitals than electrons, however, in
both the Hartree and Hartree-Fock applications of this formalism, the
index $a$ runs, just like $i$, from 1 to $N$.} by $\{\psi_{a}(x)\}$,
thus reducing the search space to a (typically small) subset of the
whole Hilbert space:

\begin{equation}
\label{eq:chQC_constraint_total}
\Psi_{e}(x_{1},\ldots,x_{N})=f\Big(\{\psi_{a}(x_{i})\}\Big) \ .
\end{equation}

The second step consists in establishing a (possibly infinite) number
of constraints on the one-electron
functions\footnote{\label{foot:both_are_constraints} Actually, both
restrictions (the one at the level of the total wavefunction in
eq.~(\ref{eq:chQC_constraint_total}) and the one involving the
one-particle ones in eq.~(\ref{eq:chQC_constraint_mono})) are simply
constraints (see appendix~B).  The distinction is not fundamental but
operative, and it also helps us to devise variational ansatzs
separating the two conceptual playgrounds.},

\begin{equation}
\label{eq:chQC_constraint_mono}
L_{k}\Big(\{\psi_{a}(x_{i})\}\Big)=0 \ .
\end{equation}

With these two ingredients, we can now write the Lagrange functional
that describes the constrained problem in terms of the orbitals
$\psi_{a}$ (see eq.~(\ref{eq:chQC_functional_Ftilde})):

\begin{equation}
\label{eq:chQC_constrained_functional}
\widetilde{\mathcal{F}}\,\big[\{\psi_{a}\}\big] =
   \Big\langle f\big(\{\psi_{a}(x_{i})\}\big) \,\Big|\, \hat{H}\, \Big|\,
   f\big(\{\psi_{a}(x_{i})\}\big) \Big\rangle + \sum_{k} \lambda_{k}
   L_{k}\Big(\{\psi_{a}(x_{i})\}\Big) \ .
\end{equation}

Finally, we take the derivatives of
$\widetilde{\mathcal{F}}[\{\psi_{a}\}]$ with respect to every
$\psi_{a}(x)$ (normally, with respect to the complex conjugate
$\psi_{a}^{*}(x)$, see footnote~\ref{foot:deriving_conjugate} in
page~\pageref{page:foot:deriving_conjugate}) and we ask each one to be
zero (see appendix~A). This produces the final equations that must be
solved in order to find the stationary one-electron orbitals.

Of course, these final equations may have multiple solutions. In the
cases discussed in this work, there exist procedures to check that a
particular solution (found computationally) is, not only stationary,
but also minimal \cite{See1977JCP}. However, to assure that it is, not
only locally minimal, but also globally (i.e., that is
\emph{optimal}), could be, in general, as difficult as for any other
multi-dimensional optimization problem
\cite{Mar1992EPL,Cer1985JOTA,Kir1983SCI}. In the Hartree and
Hartree-Fock cases, discussed in secs.~\ref{sec:QC_Hartree}
and~\ref{sec:QC_Hartree-Fock} respectively, the aufbau principle and a
clever choice of the starting guess constitute particular techniques
intended to alleviate this problem.

\section[Statement of the problem]
         {Statement of the problem}
\label{sec:QC_statement}

Assuming the Born-Oppenheimer approximation (see
sec.~\ref{sec:QC_Born-Oppenheimer} and
eqs.~(\ref{eq:chQC_bo_summary})), the central problem that one must
solve in quantum chemistry is \emph{to find the ground-state of the
electronic Hamiltonian for a fixed position} $\underbar{R}$ \emph{of
the nuclei}\footnote{\label{foot:non-e_notation} Since, from now on,
we will only be dealing with the `electronic problem', the notation
has been made simpler by dropping superfluous subindices $e$ where
there is no possible ambiguity. As a consequence, for example, the
electronic Hamiltonian is now denoted by $\hat{H}$, the electronic
kinetic energy by $\hat{T}$ and the electronic wavefunction by
$\Psi(\underbar{x})$ (dropping the parametric dependence on
$\underbar{R}$ in the same spirit).}:

\begin{equation}
\label{eq:chQC_el_ham_1}
 \hat{H} := \hat{T} + \hat{V}_{eN} + \hat{V}_{ee} :=
  - \sum_{i=1}^{N} \frac{1}{2}{\nabla}_{i}^{2}
  - \sum_{i=1}^{N}\sum_{\alpha=1}^{N_{N}} \frac{Z_{\alpha}}{R_{\alpha i}}
  + \frac{1}{2}\sum_{i \ne j} \frac{1}{r_{ij}} \  .
\end{equation}

As already remarked in sec.~\ref{sec:QC_Born-Oppenheimer}, this
problem is well posed for neutral and positively charged molecules,
and, in the same way in which the term $\hat{V}_{eN}$ prevented the
total wavefunction to be a product of an electronic and a nuclear
part, the term $\hat{V}_{ee}$ in the expression above breaks the
separability in the one-electron variables $x_{i}$ of the electronic
time-independent Schr\"odinger equation associated to $\hat{H}$.
Hence, a general solution $\Psi(\underbar{x})$ cannot be a product of
orbitals and the search must be a priori performed in the whole
Hilbert space. However, this is a much too big place to look for
$\Psi(\underbar{x})$, since the computational requirements to solve
the Schr\"odinger equation grow exponentially on the number of
electrons.

Partially recognizing this situation, in the first days of quantum
mechanics, Dirac wrote that,

\begin{quote}
The underlying physical laws necessary for the mathematical theory of
a large part of physics and the whole of chemistry are thus completely
known, and the difficulty is only that the exact application of these
equations leads to equations much too complicated to be soluble. It
therefore becomes desirable that approximate practical methods of
applying quantum mechanics should be developed, which can lead to an
explanation of the main features of complex atomic systems without too
much computation \cite{Dir1929PRSL}.
\end{quote}

The description of the most popular approximate methods, which the
great physicist envisaged to be necessary, will be the objective of
the following sections. Two basic points responsible of the relative
success of such an enterprise are the severe reduction of the space in
where the ground-state is sought (which, of course, leads to only an
approximation of it) and the availability of computers unimaginably
faster than anything that could be foreseen in times of Dirac.

\section[The Hartree approximation]
        {The Hartree approximation}
\label{sec:QC_Hartree}

\index{Hartree!approximation}
\index{Pauli exclusion principle}

One of the first and most simple approximations aimed to solve the
problem posed in the previous section is due to Hartree in 1927
\cite{Har1927PCPS} (although the way in which the Hartree equations
will be derived here, using the Variational Theorem, is due to Slater
\cite{Sla1930PR}). In this approximation, the total wavefunction is
constrained to be a product (typically referred to as \emph{Hartree
product}) of $N$ one-electron orbitals (see
eq.~(\ref{eq:chQC_constraint_total})), where the spin of the electrons
and the antisymmetry (i.e., the Pauli exclusion principle) are not
taken into account\footnote{\label{foot:orbitals_spinorbitals} We
shall denote with capital Greek letters the wavefunctions depending on
all the electronic variables, and with lowercase Greek letters the
one-electron orbitals. In addition, by $\Psi$ (or $\psi$), we shall
indicate wavefunctions containing spin part (called
\emph{spin-orbitals}) and, by $\Phi$ (or $\phi$), those that depend
only on spatial variables.}:

\begin{equation}
\label{eq:chQC_constraint_total_H}
\Phi({\bm r}_{1},\ldots,{\bm r}_{N})=\prod_{i=1}^{N}\phi_{i}({\bm r}_{i}) \ ,
\end{equation}

where the $a$ index in the orbitals has been substituted by $i$ due to
the fact that each function is paired to a specific set of electron
coordinates, consequently being the same number of both of them.

Also, the additional requirement that the one-particle wavefunctions be
normalized is imposed (see eq.~(\ref{eq:chQC_constraint_mono})):

\begin{equation}
\label{eq:chQC_constraint_mono_H}
\langle\phi_{i}|\phi_{i}\rangle=1 \ ,\qquad i=1,\ldots,N \ .
\end{equation}

With these two ingredients, we can construct the auxiliary functional
whose zero-derivative condition produces the solution of the
constrained stationary points problem (see
eq.~(\ref{eq:chQC_constrained_functional})). To this effect, we
introduce $N$ Lagrange multipliers $\varepsilon_i$ that force the
normalization constraints\footnote{\label{foot:H_total_normalized}
Note that the normalization of the total wavefunction is a consequence
of the normalization of the one-electron ones and needs not to be
explicitly asked.}:

\begin{equation}
\label{eq:chQC_constrained_functional_H}
\widetilde{\mathcal{F}}\,\big[\{\phi_{i}\}\big] =
   \left \langle \prod_{i=1}^{N}\phi_{i}({\bm r}_{i}) \right | \hat{H}
   \left | \prod_{i=1}^{N}\phi_{i}({\bm r}_{i}) \right \rangle -
   \sum_{i=1}^{N} \varepsilon_{i}
   \Big( \langle\phi_{i}|\phi_{i}\rangle-1 \Big) \ ,
\end{equation}

where the minus sign in the Lagrange multipliers term is chosen in
order to get to the most common form of the final equations without
having to define new quantities

This functional may be considered to depend on $2N$ independent
functions: the $N$ one-electron~$\phi_{i}$ and their $N$ complex
conjugates (see footnote~\ref{foot:deriving_conjugate} in
page~\pageref{page:foot:deriving_conjugate}). The \emph{Hartree
equations} are then obtained by imposing that the functional
derivative of $\widetilde{\mathcal{F}}$ with respect to $\phi_{k}^{*}$
be zero for $k=1,\ldots,N$. In order to obtain them and as an
appetizer for the slightly more complicated process in the more used
Hartree-Fock approximation, the functional derivative will be here
computed in detail following the steps indicated in appendix~A.

First, we write out\footnote{\label{foot:H_drop_limits} The limits in
sums and products are dropped if there is no possible ambiguity.} the
first term in the right-hand side of
eq.~(\ref{eq:chQC_constrained_functional_H}):

\begin{eqnarray}
\label{eq:chQC_H_step_1}
\lefteqn{ \left \langle \prod_{i}\phi_{i}({\bm r}_{i}) \right | \hat{H}
   \left | \prod_{i}\phi_{i}({\bm r}_{i}) \right \rangle = } \nonumber \\
&& \mbox{}
   -\frac{1}{2}\sum_{i}
   \left(\prod_{j \ne i}\langle\phi_{j}|\phi_{j}\rangle\right)
   \int\phi_{i}^{*}({\bm r})\nabla^{2}\phi_{i}({\bm r})\mathrm{d}{\bm r}
   \nonumber \\
&& \mbox{}
   - \sum_{i}\left(\prod_{j \ne i}\langle\phi_{j}|\phi_{j}\rangle\right)
   \int\phi_{i}^{*}({\bm r})\,\phi_{i}({\bm r})
   \left ( \sum_{A=1}^{N_{N}}\frac{Z_{A}}{|{\bm r}-{\bm R}_{A}|} \right )
   \mathrm{d}{\bm r} \nonumber \\
&& \mbox{}
   + \frac{1}{2}\sum_{i}\sum_{j \ne i} \left(\prod_{k \ne i,j}
     \langle\phi_{k}|\phi_{k}\rangle\right)
     \int\!\!\!\!\int
     \frac{\phi_{i}^{*}({\bm r})\,\phi_{i}({\bm r})\,\phi_{j}^{*}({\bm r}\, ')
                                   \,\phi_{j}({\bm r}\, ')}
	 {|{\bm r}-{\bm r}\, '|}\,\mathrm{d}{\bm r}\, '
   \mathrm{d}{\bm r} \ ,
\end{eqnarray}

where $\mathrm{d}{\bm r}$ denotes the Euclidean $\mathbb{R}^{3}$
volume element $\mathrm{d}x\,\mathrm{d}y\,\mathrm{d}z$.

Now, we realize that the products outside the integrals can be dropped
using the constraints in eq.~(\ref{eq:chQC_constraint_mono_H}) (see
the last paragraphs of appendix~B for a justification that this can be
done before taking the derivative). Then, using the previous
expression and conveniently rearranging the order of the integrals and
sums, we write out the first term in the numerator of the left-hand
side of eq.~(\ref{eq:def_funct_der}) that corresponds to an
infinitesimal variation of the function $\phi_{k}^{*}$:

\begin{eqnarray}
\label{eq:chQC_H_step_2}
\lefteqn{\widetilde{\mathcal{F}}\,\big[\phi_{k}^{*}+
         \epsilon \delta \phi_{k}^{*}\big]:= } \nonumber \\
\lefteqn{
   \widetilde{\mathcal{F}}\,\big[\phi_{1},\phi_{1}^{*},\ldots,
        \phi_{k},\phi_{k}^{*}+
        \epsilon \delta \phi_{k}^{*},\ldots,
        \phi_{N},\phi_{N}^{*}\big] =} \nonumber \\
&& \mbox{}
   -\frac{1}{2}\sum_{i}
   \int\phi_{i}^{*}({\bm r})\nabla^{2}\phi_{i}({\bm r})\,\mathrm{d}{\bm r}
   - \sum_{i}
   \int\big|\phi_{i}({\bm r})\big|^{2}
   \left ( \sum_{A=1}^{N_{N}}\frac{Z_{A}}{|{\bm r}-{\bm R}_{A}|} \right )
   \mathrm{d}{\bm r} \nonumber \\
&& \mbox{}
    + \frac{1}{2}\sum_{i}\int\big|\phi_{i}({\bm r})\big|^{2}
   \left ( \int\frac{\sum_{j \ne i}\big|\phi_{j}({\bm r}\, ')\big|^{2}}
	 {|{\bm r}-{\bm r}\, '|}\,\mathrm{d}{\bm r}\, ' \right )
   \mathrm{d}{\bm r} 
   - \sum_{i}\varepsilon_{i} \left ( \int\big|\phi_{i}({\bm r})\big|^{2}
   \mathrm{d}{\bm r} - 1 \right ) \nonumber \\
&& \mbox{}
    -\frac{1}{2}\epsilon\int\delta\phi_{k}^{*}({\bm r})\nabla^{2}
    \phi_{k}({\bm r})\,\mathrm{d}{\bm r} -
    \epsilon\int\delta\phi_{k}^{*}({\bm r})\,\phi_{k}({\bm r})
   \left ( \sum_{A=1}^{N_{N}}\frac{Z_{A}}{|{\bm r}-{\bm R}_{A}|} \right )
   \mathrm{d}{\bm r} \nonumber \\
&& \mbox{}
   + \epsilon\int\delta\phi_{k}^{*}({\bm r})\,\phi_{k}({\bm r})
   \left ( \int\frac{\sum_{i \ne k}\big|\phi_{i}({\bm r}\, ')\big|^{2}}
	 {|{\bm r}-{\bm r}\, '|}\mathrm{d}{\bm r}\, ' \right )
   \mathrm{d}{\bm r} 
  -\epsilon\,\varepsilon_{k}\int\delta\phi_{k}^{*}({\bm r})\,\phi_{k}({\bm r})
   \,\mathrm{d}{\bm r} \ .
\end{eqnarray}

We subtract from this expression the quantity
$\widetilde{\mathcal{F}}\,\big[\{\phi_{i}({\bm r}_{i})\}\big]$, so
that the first four terms cancel, and we can write

\begin{eqnarray}
\label{eq:chQC_H_step_3}
\lefteqn{ \lim_{\epsilon \rightarrow 0}
  \frac{\widetilde{\mathcal{F}}\,\big[\phi_{k}^{*}+\epsilon\delta \phi_{k}^{*}\big] -
        \widetilde{\mathcal{F}}\,\big[\phi_{k}^{*}\big]}{\epsilon} =
       } \nonumber \\
&& \int \Bigg{[} -\frac{1}{2} \nabla^{2} \phi_{k}({\bm r}) -
     \left ( \sum_{A=1}^{N_{N}}\frac{Z_{A}}
             {|{\bm r}-{\bm R}_{A}|} \right ) \phi_{k}({\bm r}) \nonumber \\
&& \mbox{}
  + \left ( \int\frac{\sum_{i \ne k}|\phi_{i}({\bm r}\, ')|^{2}}
	     {|{\bm r}-{\bm r}\, '|}\,\mathrm{d}{\bm r}\, ' \right )
            \phi_{k}({\bm r}) - \varepsilon_{k}\phi_{k}({\bm r}) \Bigg{]}
      \,\delta\phi_{k}^{*}({\bm r}) \,\mathrm{d}{\bm r} \ .
\end{eqnarray}

Now, by simple inspection of the right-hand side, we see that the
functional derivative (see eq.~(\ref{eq:def_funct_der})) is the part
enclosed by square brackets:

\begin{equation}
\label{eq:chQC_funct_der_H}
\frac{\delta\widetilde{\mathcal{F}}\,\big[\{\phi_{i}\}\big]}
     {\delta \phi_{k}^{*}}=
 \left ( - \frac{1}{2}\nabla^{2} + \hat{V}_{e}({\bm r}) +
         \hat{V}_{e}^{k}({\bm r}) - \varepsilon_{k} \right )
       \phi_{k}({\bm r}) \ ,
\end{equation}

where the {\it nuclear potential energy} and the {\it electronic
potential energy} have been respectively defined
as\footnote{\label{foot:change_e} Compare the notation with the one in
eqs.~(\ref{eq:chQC_ham_tot_op}), here a subindex $e$ has been dropped
to distinguish the new objects defined.}

\begin{subequations}
\label{eq:chQC_potentials_H}
\begin{align}
& \hat{V}_{N}({\bm r}):=-\sum_{A=1}^{N_{N}}\frac{Z_{A}}
     {|{\bm r}-{\bm R}_{A}|} \ , \label{eq:chQC_potentials_H_a} \\
& \hat{V}_{e}^{k}({\bm r}):=\int\frac{\sum_{i\ne k}|\phi_{i}({\bm r}\, ')|^{2}}
	     {|{\bm r}-{\bm r}\, '|}\,\mathrm{d}{\bm r}\, '
  \ . \label{eq:chQC_potentials_H_b}
\end{align}
\end{subequations}

Finally, if we ask the functional derivative to be zero for
$k=1,\ldots,N$, we arrive to the equations that the stationary points
must satisfy, the {\it Hartree equations}:

\begin{equation}
\label{eq:chQC_Hartree_eqs}
\hat{\mathcal{H}}_{k}[\phi] \, \phi_{k}({\bm r}) :=
\left ( - \frac{1}{2}\nabla^{2} + \hat{V}_{N}({\bm r}) +
          \hat{V}_{e}^{k}({\bm r}) \right ) \phi_{k}({\bm r}) = 
          \varepsilon_{k}\,\phi_{k}({\bm r})\ ,
\end{equation}

for all $k=1,\ldots,N$.

Let us note that, despite the fact that the object
$\hat{\mathcal{H}}_{k}[\phi]$ defined above is not a operator strictly
speaking, since, as the notation emphasizes, it depends on the
orbitals $\phi_{i \ne k}$, we will stick to the name \emph{Hartree
operator} for it, in order to be consistent with most of the
literature.

Now, some remarks related to the Hartree equations are worth
making. First, it can be shown that, if the variational ansatz in
eq.~(\ref{eq:chQC_constraint_total_H}) included the spin degrees of
freedom of the electrons, all the expressions above would be kept,
simply changing the orbitals $\phi_{i}({\bm r}_i)$ by the
spin-orbitals $\psi_{i}({\bm r}_i,\sigma_{i})$.

Secondly, and moving into more conceptual playgrounds, we note that
the special structure of $\hat{V}_{e}^{k}({\bm r})$ in
eq.~(\ref{eq:chQC_potentials_H_b}) makes it mandatory to interpret the
Hartree scheme as one in which each electron `feels' only the average
effect of the rest. In fact, if the \emph{quantum charge density}
$\rho_{i}({\bm r}):=|\phi_{i}({\bm r})|^{2}$ is regarded for a moment
as a classical continuum distribution, then the potential produced by
all the electrons but the $k$-th is precisely the one in
eq.~(\ref{eq:chQC_potentials_H_b}). Supporting this image, note also
the fact that, if we write the joint probability density of electron 1
being at the point ${\bm r}_{1}$, electron $2$ being at the point
${\bm r}_{2}$ and so on (simply squaring
eq.~(\ref{eq:chQC_constraint_total_H})),

\begin{equation}
\label{eq:chQC_Hartree_rho}
\rho({\bm r}_{1},\ldots,{\bm r}_{N}):=
 \big| \Phi({\bm r}_{1},\ldots,{\bm r}_{N}) \big|^{\,2} =
 \prod_{i=1}^{N}\big|\phi_{i}({\bm r}_{i})\big|^{\,2} =
 \prod_{i=1}^{N}\rho_{i}({\bm r}_{i}) \ ,
\end{equation}

we see that, \emph{in a probabilistic sense}, the electrons are
independent (they could not be independent in a physical, complete
sense, since we have already said that they `see' each other in an
average way).

Anyway, despite these appealing images and also despite the fact that,
disguised under the misleading (albeit common) notation, these
equations seem `one-particle', they are rather complicated from a
mathematical point of view.  On the one hand, it is true that, whereas
the original electronic Schr\"odinger equation
in~(\ref{eq:chQC_bo_summary_a}) depended on $3N$ spatial variables,
the expressions above only depend on 3. This is what we have gained
from drastically reducing the search space to the set of Hartree
products in eq.~(\ref{eq:chQC_constraint_total_H}) and what renders
the approximation tractable. On the other hand, however, we have paid
the price of greatly increasing the mathematical complexity of the
expressions, so that, while the electronic Schr\"odinger equation was
one linear differential equation, the Hartree ones
in~(\ref{eq:chQC_Hartree_eqs}) are $N$ coupled non-linear
integro-differential equations \cite{Can2003BOOK}.

This complexity precludes any analytical approach to the problem and
forces us to look for the solutions using less reliable iterative
methods. Typically, in computational studies, one proposes a
\emph{starting guess} for the set of $N$ orbitals $\{\phi^0_k\}$; with
them, the Hartree operator $\hat{\mathcal{H}}_{k}[\phi^0]$ in the
left-hand side of eq.~(\ref{eq:chQC_Hartree_eqs}) is constructed for
every $k$ and the $N$ equations are solved as simple eigenvalue
problems. For each $k$, the $\phi^1_k$ that corresponds to the lowest
$\varepsilon^1_k$ is selected and a new Hartree operator
$\hat{\mathcal{H}}_{k}[\phi^1]$ is constructed with
the~$\{\phi^1_k\}$.  The process is iterated until (hopefully) the
$n$-th set of solutions $\{\phi^{n}_k\}$ differ from the
\mbox{$(n-1)$-th} one $\{\phi^{n-1}_k\}$ less than a reasonably small
amount.

Many technical issues exist that raise doubts about the possible
success of such an approach. The most important ones being related to
the fact that a proper definition of the \emph{Hartree problem} should
be: \emph{find the global minimum of the energy functional $\langle
\Phi | \,\hat{H}\, | \Phi \rangle$ under the constraint that the
wavefunction $\Phi$ be a Hartree product} and not: \emph{solve the
Hartree equations}~(\ref{eq:chQC_Hartree_eqs}), whose solutions indeed
include the global minimum sought but also all the rest of stationary
points.

While the possibility that a found solution be a maximum or a saddle
point can be typically ruled out \cite{Can2003BOOK,See1977JCP}, as we
remarked in sec.~\ref{sec:QC_variational} and due to the fact that
there are an infinite number of solutions to the Hartree equations
\cite{Lio1987CMP}, to be sure that any found minimum is the global one
is impossible in a general case. There exists, however, one way,
related to a theorem by Simon and Lieb \cite{Lie1974JCP,Lie1977CMP},
of hopefully biasing a particular found solution of the Hartree
equations to be the global minimum that we are looking for. They
showed that, first, for neutral or positively charged molecules ($Z
\ge N$), the Hartree global minimization problem has a solution (its
uniqueness is not established yet \cite{Can2003BOOK}) and, second,
that the minimizing orbitals $\{\phi_k\}$ correspond to the lowest
eigenvalues of the~$\hat{\mathcal{H}}_{k}[\phi]$ operators
self-consistently constructed with
them\footnote{\label{foot:hartreeOp_k} In quantum chemistry, where the
number of electrons considered is typically small, the version of the
Hartree equations that is used is the one derived here, with the
Hartree operators depending on the index~$k$ in a non-trivial
way. However, if the number of electrons is large enough (such as in
condensed matter applications), is customary to add to the effective
electronic repulsion in eq.~(\ref{eq:chQC_potentials_H_b}) the
self-interaction of electron $k$ with himself. In such a case, the
Hartree operator is independent of $k$ so that, after having achieved
self-consistency, the orbitals $\phi_k$ turn out to be eigenstates
corresponding to different eigenvalues of the same Hermitian operator,
$\hat{\mathcal{H}}[\phi]$, and, therefore, mutually orthogonal.}. Now,
although \emph{the reverse of the second part of the theorem is not
true in general} (i.e., from the fact that a particular set of
orbitals are the eigenstates corresponding to the lowest eigenvalues
of the associated Hartree operators, does not necessarily follow that
they are the ones that minimize the energy) \cite{Can2003BOOK}, in
practice, the insight provided by Lieb and Simon's result is invoked
to build each successive state in the iterative procedure described
above, choosing the lowest lying eigenstates each time. In this way,
although one cannot be sure that the global minimum has been reached,
the fact that the found one has a property that the former also
presents is regarded as a strong hint that it must be so (see also the
discussion for the Hartree-Fock case in the next section).

This drawback and all the problems arising from the fact that an
iterative procedure such as the one described above could converge to
a fixed point, oscillate eternally or even diverge, are circumvented
in practice by a clever choice of the starting guess
orbitals~$\{\phi^0_k\}$. If they are extracted, for example, from a
slightly less accurate theory, one may expect that they could be `in
the basin of attraction' of the true Hartree minimum (so that the
stationary point found will be the correct one) and close to it (so
that the iterative procedure will converge). This kind of wishful
thinking combined with large amounts of heuristic protocols born from
many decades of trial-and-error-derived knowledge pervade and make
possible the whole quantum chemistry discipline.

\section[The Hartree-Fock approximation]
        {The Hartree-Fock approximation}
\label{sec:QC_Hartree-Fock}

The Hartree theory discussed in the previous section is not much used
in quantum chemistry and many textbooks on the subject do not even
mention it. Although it contains the seed of almost every concept
underlying the \emph{Hartree-Fock approximation} discussed in this
section, it lacks an ingredient that turns out to be essential to
correctly describe the behaviour of molecular species: \emph{the
indistinguishability of the electrons}. This was noticed independently
by Fock \cite{Foc1930ZP} and Slater \cite{Sla1930PR} in 1930, and it
was corrected by proposing a variational ansatz for the total
wavefunction that takes the form of a so-called \emph{Slater
determinant} (see eq.~(\ref{eq:chQC_constraint_total_SD_GHF}) below).

The most important mathematical consequence of the
indistinguishability among a set of $N$ quantum objects of the same
type is the requirement that the total $N$-particle wavefunction must
either remain unchanged (\emph{symmetric}) or change sign
(\emph{antisymmetric}) when any pair of coordinates, $x_{i}$ and
$x_{j}$, are swapped. In the first case, the particles are called
\emph{bosons} and must have integer spin, while in the second case,
they are called \emph{fermions} and have semi-integer spin. Electrons
are fermions, so the total wavefunction must be antisymmetric under
the exchange of any pair of one-electron coordinates. This is a
property that is certainly not met by the single Hartree product in
eq.~(\ref{eq:chQC_constraint_total_H}) but that can be easily
implemented by forming linear combinations of many of them. The trick
is to add all the possible Hartree products that are obtained from
eq.~(\ref{eq:chQC_constraint_total_H}) changing the order of the
orbitals labels while keeping the order of the coordinates
ones\footnote{\label{foot:coordinates_or_labels} It is immaterial
whether the orbitals labels are kept and the coordinates ones changed
or vice versa.}, and assigning to each term the sign of the
permutation $p$ needed to go from the natural order $1,\ldots,N$ to
the corresponding one $p(1),\ldots,p(N)$. The sign of a permutation
$p$ is 1 if $p$ can be written as a composition of an even number of
two-element transpositions, and it is $-1$ if the number of
transpositions needed is odd. Therefore, we define $\mathcal{T}(p)$ as
the minimum number\footnote{\label{foot:transpositions} It can be
shown that the parity of all decompositions of $p$ into products of
elementary transpositions is the same. We have chosen the minimum only
for $\mathcal{T}(p)$ to be well defined.} of transpositions needed to
perform the permutation $p$, and we write the sign of~$p$ as
$(-1)^{\mathcal{T}(p)}$.

Using this, an antisymmetric wavefunction constructed from Hartree
products of $N$ different orbitals may be written as

\begin{equation}
\label{eq:chQC_constraint_total_GHF}
\Psi(x_{1},\ldots,x_{N})=\frac{1}{\sqrt{N!}}\sum_{p \,\in S_{N}}
  (-1)^{\mathcal{T}(p)}\psi_{p(1)}(x_{1}) \cdots \psi_{p(N)}(x_{N})\ ,
\end{equation}

where the factor $1/\sqrt{N!}$ enforces normalization of the total
wavefunction $\Psi$ (if we use the constraints in
eq.~(\ref{eq:chQC_constraint_mono_GHF})) and $S_{N}$ denotes the
\emph{symmetric group} of order $N$, i.e., the set of all permutations
of $N$ elements (with a certain multiplication rule).

The above expression is more convenient to perform the calculations
that lead to the Hartree-Fock equations, however, there is also a
compact way of rewriting eq.~(\ref{eq:chQC_constraint_total_GHF})
which is commonly found in the literature and that is useful to
illustrate some particular properties of the problem. It is the
\emph{Slater determinant}:

\begin{equation}
\label{eq:chQC_constraint_total_SD_GHF}
\Psi(x_{1},\ldots,x_{N})=\frac{1}{\sqrt{N!}} \left |
\begin{array}{cccc}
\psi_1(x_1) & \psi_2(x_1) & \cdots & \psi_N(x_1) \\
\psi_1(x_2) & \psi_2(x_2) & \cdots & \psi_N(x_2) \\
   \vdots   &   \vdots    & \ddots &   \vdots    \\
\psi_1(x_N) & \psi_2(x_N) & \cdots & \psi_N(x_N)
\end{array} \right | \ .
\end{equation}

Now, having established the constraints on the form of the total
wavefunction, we ask the Hartree-Fock one-electron orbitals to be, not
only normalized, like we did in the Hartree case, but also mutually
orthogonal:

\begin{equation}
\label{eq:chQC_constraint_mono_GHF}
\langle\psi_{i}|\psi_{j}\rangle=\delta_{ij}\ , \qquad i,j=1,\ldots,N \ .
\end{equation}

Additionally note that, contrarily to what we did in the previous
section, we have now used one-electron wavefunctions $\psi_i$
dependent also on the spin $\sigma$ (i.e., spin-orbitals) to construct
the variational ansatz. A \emph{general}
spin-orbital\footnote{\label{foot:general_spin-orbital} Note that, if
we had not included the spin degrees of freedom, the search space
would have been half as large, since, where we now have $2N$ functions
of ${\bm r}$ (i.e., $\varphi_{i}^{\alpha}({\bm r})$ and
$\varphi_{i}^{\beta}({\bm r})$, with $i=1,\ldots,N$), we would have
had just $N$ (the $\phi_{i}({\bm r})$).} may be written as (see also
footnote~\ref{foot:spin_convenient} in
page~\pageref{page:foot:spin_convenient})

\begin{equation}
\label{eq:chQC_general_so}
\psi(x)=\varphi^{\alpha}({\bm r})\,\alpha(\sigma)+
        \varphi^{\beta}({\bm r})\,\beta(\sigma) \ ,
\end{equation}

where the functions $\alpha$ and $\beta$ correspond to the
\emph{spin-up} and \emph{spin-down} eigenstates of the operator
associated to the $z$-component of the one-electron spin. They are
defined as

\begin{equation}
\label{eq:chQC_alpha_beta}
\begin{array}{r@{\hspace{4pt}}c@{\hspace{4pt}}l@{\hspace{20pt}}r@{\hspace{4pt}}c@{\hspace{4pt}}l}
\alpha(-1/2)&=&0 & \beta(-1/2)&=&1 \\
\alpha(1/2)&=&1 & \beta(1/2)&=&0 \ .
\end{array}
\end{equation}

The formalism obtained when these general spin-orbitals are used
is accordingly called \emph{\underline{G}eneral
\underline{H}artree-\underline{F}ock}~(GHF). The first part of the
mathematical treatment in the following paragraphs shall be performed
assuming this situation. The advantage of such a choice is that, later
on, by imposing additional constraints to the spin part of the
one-electron orbitals, we will be able to derive the basic equations
for some other flavours of the Hartree-Fock theory, such as UHF, RHF
and ROHF, in a very direct way.

Now, to calculate the expected value of the energy in a state such as
the one in eqs.~(\ref{eq:chQC_constraint_total_GHF})
and~(\ref{eq:chQC_constraint_total_SD_GHF}), let us denote the
one-particle part (that operates on the $i$-th coordinates) of the
total electronic Hamiltonian $\hat{H}$ in eq.~(\ref{eq:chQC_el_ham_1})
by

\begin{equation}
\label{eq:chQC_hi}
\hat{h}_{i}:= - \frac{{\nabla}_{i}^{2}}{2} -
          \sum_{\alpha=1}^{N_{N}}
            \frac{Z_{\alpha}}{|{\bm R}_{\alpha}-{\bm r}_{i}|} \ ,
\end{equation}

in such a way that,

\begin{equation}
\label{eq:chQC_expected_H_GHF}
\langle\Psi|\,\hat{H}\,|\Psi\rangle =
  \sum_{i}\langle\Psi|\,\hat{h}_{i}\,|\Psi\rangle + \frac{1}{2}
  \sum_{i \ne j} \langle\Psi|\,\frac{1}{r_{ij}}\,|\Psi\rangle \ ,
\end{equation}

where $r_{ij}:=|{\bm r}_{j}-{\bm r}_{i}|$.

\label{page:one_e_calculation}

We shall compute separately each one of the sums in the expression
above. Let us start now with the first one: For a given $i$ in the
sum,  and due to the structure of the electronic wavefunction
in~(\ref{eq:chQC_constraint_total_GHF}), the expected value
$\langle\Psi|\,\hat{h}_{i}\,|\Psi\rangle$ is a sum of $(N!)^2$ terms
of the form

\begin{equation}
\label{eq:chQC_calculation_hi_1}
\frac{1}{N!}\,(-1)^{\mathcal{T}(p)+\mathcal{T}(p^{\prime})}
  \big\langle \psi_{p(1)}(x_{1}) \cdots \psi_{p(N)}(x_{N})
  \,\big|\, \hat{h}_{i} \,\big|\,
  \psi_{p^{\prime}(1)}(x_{1}) \cdots \psi_{p^{\prime}(N)}(x_{N})
  \big\rangle \ ,
\end{equation}

but, since $\hat{h}_{i}$ operates only on $x_i$ and due to the
orthogonality of the spin-orbitals with different indices, we have
that the only non-zero terms are those with $p=p^{\prime}$. Taking
this into account, all permutations $p$ appear still as terms of the
sum, and we see that every orbital $\psi_j$ occurs depending on every
coordinate $x_i$. Given a particular pair $i$ and $j$, this
happens in the terms for which $p(i)=j$ and one of such terms may be
expressed as

\begin{equation}
\label{eq:chQC_calculation_hi_2}
\frac{1}{N!}
  \left(\prod_{k \ne j} \langle \psi_k |  \psi_k \rangle \right)
  \langle \psi_j |\,\hat{h}\,| \psi_j \rangle \ ,
\end{equation}

where we have used that $(-1)^{2\,\mathcal{T}(p)}=1$, and we have
dropped the index $i$ from $\hat{h}_{i}$ noticing that the integration
variables in
$\langle\psi_j(x_{i})|\,\hat{h}_{i}\,|\psi_j(x_{i})\rangle$ are
actually dummy.

Next, we use again the one-electron wavefunctions constraints in
eq.~(\ref{eq:chQC_constraint_mono_GHF}) to remove the product of norms
in brackets, and we realize that, for each $j$, there are as many
terms like the one in the expression above as permutations of the
remaining $N-1$ orbital indices (i.e., $(N-1)!$). In addition, we
recall that every $j$ must appear and perform the first sum in
eq.~(\ref{eq:chQC_expected_H_GHF}), yielding

\begin{eqnarray}
\label{eq:chQC_calculation_hi_3}
\lefteqn{\sum_{i}\langle\Psi|\,\hat{h}_{i}\,|\Psi\rangle =
\sum_{i}(N-1)!\sum_{j}\frac{1}{N!}\langle\psi_j|\,\hat{h}\,|\psi_j\rangle = }
 \nonumber \\
&& N(N-1)!\sum_{j}\frac{1}{N!}\langle\psi_j|\,\hat{h}\,|\psi_j\rangle =
\sum_{j}\langle\psi_j|\,\hat{h}\,|\psi_j\rangle \ ,
\end{eqnarray}

where all factorial terms have canceled out.

The next step is to calculate the second sum in
eq.~(\ref{eq:chQC_expected_H_GHF}). Again, we have that, for each
pair~$(i,j)$, $\langle\Psi|\,1/r_{ij}\,|\Psi\rangle$ is a sum of
$(N!)^2$ terms like

\begin{equation}
\label{eq:chQC_calculation_rij_1}
\frac{1}{N!}\,(-1)^{\mathcal{T}(p)+\mathcal{T}(p^{\prime})}
  \big\langle \psi_{p(1)}(x_{1}) \cdots \psi_{p(N)}(x_{N})
  \,\big|\, \frac{1}{r_{ij}} \,\big|\,
  \psi_{p^{\prime}(1)}(x_{1}) \cdots \psi_{p^{\prime}(N)}(x_{N})
  \big\rangle \ .
\end{equation}

\label{page:two_e_calculation}

For this expected value, contrarily to the case of $\hat{h}_{i}$ and
due to the two-body nature of the operator $1/r_{ij}$, not only do the
terms with $p=p^{\prime}$ survive, but also those in which $p$ and
$p^{\prime}$ differ over only a pair of values $i$ and $j$, i.e.,
those for which $p(i)=p^{\prime}(j)$, $p(j)=p^{\prime}(i)$ and
$p(k)=p^{\prime}(k), \forall k \ne i,j$. The reason for this is
that, even if $p(i) \ne p^\prime(i)$ and $p(j) \ne p^\prime(j)$, the
integral $\langle\psi_{p(i)}(x_i)\psi_{p(j)}(x_j)|\,1/r_{ij}\,|
\psi_{p^\prime(i)}(x_i)\psi_{p^\prime(j)}(x_j)\rangle$ does not
vanish.

Now, using that $1/r_{ij}$ operates only on $x_i$ and $x_j$, the
orthonormality conditions in eq.~(\ref{eq:chQC_constraint_mono_GHF})
and the fact that $(-1)^{2\,\mathcal{T}(p)}=1$, we have that, when
$\psi_k$ depends on $x_i$ and $\psi_l$ depends on~$x_j$, the
$p=p^{\prime}$ part of the corresponding terms in
eq.~(\ref{eq:chQC_calculation_rij_1}) reads

\begin{equation}
\label{eq:chQC_calculation_rij_2}
\frac{1}{N!}
  \langle \psi_k\psi_l |\,\frac{1}{r}\,| \psi_k\psi_l \rangle \ ,
\end{equation}

where we have defined

\begin{equation}
\label{eq:chQC_two_e_1r}
\langle \psi_i\psi_j |\,\frac{1}{r}\,| \psi_k\psi_l \rangle :=
\sum_{\sigma,\,\sigma^{\prime}}\int\!\!\!\!\int
     \frac{\psi_{i}^{*}(x)\,\psi_{j}^{*}(x^{\prime})\,
           \psi_{k}(x)\,\psi_{l}(x^{\prime})}
	 {|{\bm r}-{\bm r}^{\prime}|}\,\mathrm{d}{\bm r}
   \,\mathrm{d}{\bm r}^{\prime} \ .
\end{equation}

Next, we see that, for each pair $(k,l)$, and keeping
$p=p^\prime$, we can make $(N-2)!$ permutations among the $N-2$
indices of the orbitals on which $1/r_{ij}$ does not operate and still
find the same expression~(\ref{eq:chQC_calculation_rij_2}).
Therefore, for each pair $(k,l)$, we have a sum of $(N-2)!$ identical
terms. In addition, if we perform the sum on $i$ and $j$ in
eq.~(\ref{eq:chQC_expected_H_GHF}) and remark that the term in
eq.~(\ref{eq:chQC_calculation_rij_2}) does not depend on the pair
$(i,j)$ (which is obvious from the suggestive notation above), we have
that the $p=p^{\prime}$ part of the second sum in
eq.~(\ref{eq:chQC_expected_H_GHF}), which is typically called
\emph{Coulomb energy}, reads

\begin{equation}
\label{eq:chQC_calculation_rij_3}
\frac{1}{2}\sum_{i \ne j}(N-2)!\sum_{k \ne l}\frac{1}{N!}
  \langle\psi_k\psi_l|\,\frac{1}{r}\,|\psi_k\psi_l\rangle=
\frac{1}{2}\sum_{k \ne l}
  \langle\psi_k\psi_l|\,\frac{1}{r}\,|\psi_k\psi_l\rangle \ ,
\end{equation}

where we have used that the sum $\sum_{i \ne j}$ is performed on
$N(N-1)$ identical terms which do not depend neither on $i$ nor on
$j$.

On the other hand, in the case in which $p$ and $p^{\prime}$ only
differ in that the indices of the orbitals that depend on $x_{i}$ and
$x_{j}$ are swapped, all the derivation above applies except for the
facts that, first, $(-1)^{\mathcal{T}(p)+\mathcal{T}(p^{\prime})}=-1$
and, second, the indices $k$ and $l$ must be \emph{exchanged} in
eq.~(\ref{eq:chQC_calculation_rij_3}) (it is immaterial if they are
exchanged in the bra or in the ket, since the indices are summed over
and are dummy). Henceforth, the remaining part of the second sum in
eq.~(\ref{eq:chQC_expected_H_GHF}), typically termed \emph{exchange
energy}, may be written as

\begin{equation}
\label{eq:chQC_calculation_rij_4}
-\frac{1}{2}\sum_{k \ne l}
  \langle\psi_k\psi_l|\,\frac{1}{r}\,|\psi_l\psi_k\rangle \ .
\end{equation}

Finally, the expected value of the energy in the GHF variational state
$\Psi$ turns out to be

\begin{eqnarray}
\label{eq:chQC_expected_H_2_GHF}
\lefteqn{E^{\mathrm{GHF}}:=\langle\Psi|\,\hat{H}\,|\Psi\rangle =}\nonumber\\
&&  \sum_{i}\underbrace{\langle\psi_{i}|\,\hat{h}\,|\psi_{i}\rangle
                      }_{\displaystyle h_i} + \frac{1}{2}
  \sum_{i, j} \bigg( \underbrace{
     \langle\psi_i\psi_j|\,\frac{1}{r}\,|\psi_i\psi_j\rangle
                                 }_{\displaystyle J_{ij}}
  -  \underbrace{
     \langle\psi_i\psi_j|\,\frac{1}{r}\,|\psi_j\psi_i\rangle
                 }_{\displaystyle K_{ij}} \bigg) \ ,
\end{eqnarray}

where the \emph{one-electron integrals} $h_{i}$ have been defined
together with the \emph{two-electron integrals}, $J_{ij}$ and
$K_{ij}$, and the fact that $J_{ii}=K_{ii},\forall i$ has been used to
include the diagonal terms in the second sum.

Now, the energy functional above is the quantity that we want to
minimize under the orthonormality constraints in
eq.~(\ref{eq:chQC_constraint_mono_GHF}). So we are prepared to write
the auxiliary functional $\widetilde{\mathcal{F}}$, introducing $N^2$
Lagrange multipliers $\lambda_{ij}$ (see
eq.~(\ref{eq:chQC_constrained_functional}) and compare with the
Hartree example in the previous section):

\begin{equation}
\label{eq:chQC_constrained_functional_GHF}
\widetilde{\mathcal{F}}\,\big[\{\psi_{i}\}\big] =
  \sum_i h_i + \frac{1}{2} \sum_{i, j} (J_{ij}-K_{ij}) -
 \sum_{i, j} \lambda_{ij}\,
   \Big( \langle\psi_{i}|\psi_{j}\rangle-\delta_{ij} \Big) \ .
\end{equation}

In order to get to the \emph{Hartree-Fock equations} that the
stationary orbitals $\psi_k$ must satisfy, we impose that the
functional derivative of
$\widetilde{\mathcal{F}}\,\big[\{\psi_{i}\}\big]$ with respect to
$\psi^{*}_k$ be zero. To calculate $\delta \widetilde{\mathcal{F}} /
\delta \psi^{*}_k$, we follow the procedure described in appendix~A,
using the same notation as in eq.~(\ref{eq:chQC_H_step_2}). The
variation with respect to each $\psi^{*}_k$ shall yield the
Hartree-Fock equations for the unconjugated $\psi_i$. The equations
for the $\psi^*_i$ are obtained either by differentiating
$\widetilde{\mathcal{F}}\,\big[\{\psi_{i}\}\big]$ with respect to each
$\psi_k$ or, if the $\lambda_{ij}$ matrix is Hermitian (which will
turn out to be the case), by simply taking the complex conjugate of
both sides of the final equations in~(\ref{eq:chQC_HF_eqs_v1_GHF}).

Now,

\begin{eqnarray}
\label{eq:chQC_limit_GHF}
\lefteqn{ \lim_{\epsilon \rightarrow 0}
  \frac{\widetilde{\mathcal{F}}\,
  \big[\psi_{k}^{*}+\epsilon\delta \psi_{k}^{*}\big] -
        \widetilde{\mathcal{F}}\,\big[\psi_{k}^{*}\big]}{\epsilon} =
       } \nonumber \\
&& \langle\delta\psi_{k}|\,\hat{h}\,|\psi_{k}\rangle + 
     \sum_j\bigg(\langle\delta\psi_k\psi_j|\,\frac{1}{r}\,|\psi_k\psi_j\rangle
  -  \langle\delta\psi_k\psi_j|\,\frac{1}{r}\,|\psi_j\psi_k\rangle\bigg)
  -  \sum_j \lambda_{kj} \langle\delta\psi_{k}|\psi_{j}\rangle = 
  \nonumber \\
&& \int \Bigg{[} \hat{h}\,\psi_{k}(x)
  + \sum_j \left ( \psi_k(x) \!\! \int
    \frac{|\psi_{j}(x^{\prime})|^{2}}
	 {|{\bm r}-{\bm r}^{\prime}|}\,\mathrm{d}x^{\prime}
  - \psi_j(x) \!\! \int
    \frac{\psi^{*}_{j}(x^{\prime})\,\psi_{k}(x^{\prime})}
	 {|{\bm r}-{\bm r}^{\prime}|}\,\mathrm{d}x^{\prime} \right)
 \nonumber \\
&& \mbox{}
  - \sum_j \lambda_{kj}\,\psi_{j}(x) \Bigg{]}
      \,\delta\psi_{k}^{*}(x) \,\mathrm{d}x \ ,
\end{eqnarray}

where we have used the more compact notation $\int\mathrm{d}x$ instead of
$\sum_\sigma\int\mathrm{d}{\bm r}$.

Then, like in the previous section, by simple inspection of the
right-hand side, we see that the functional derivative is the part
enclosed by square brackets (see eq.~(\ref{eq:def_funct_der})):

\begin{equation}
\label{eq:chQC_funct_der_GHF}
\frac{\delta\widetilde{\mathcal{F}}\,\big[\{\psi_{i}\}\big]}
     {\delta \psi_{k}^{*}}=
 \left[\hat{h} + \sum_j \Big(\hat{J}_{j}[\psi]-
        \hat{K}_{j}[\psi]\Big)\right]\psi_{k}(x)-
        \sum_j\lambda_{kj}\psi_{j}(x) \ ,
\end{equation}

where the \emph{Coulomb} and \emph{exchange operators} are
respectively defined by their action on an arbitrary function
$\varphi(x)$ as follows\footnote{\label{foot:again_operator} Like in
the Hartree case in the previous section, the word \emph{operator} is
a common notational abuse if they act upon the very $\psi_i$ on which
they depend. This is again made explicit in the notation.}:

\begin{subequations}
\label{eq:chQC_operators_GHF}
\begin{align}
& \hat{J}_{j}[\psi]\,\varphi(x):=\left(\int\frac{|\psi_{j}(x^{\prime})|^{2}}
	     {|{\bm r}-{\bm r}^{\prime}|}\,\mathrm{d}x^{\prime} \right)
             \varphi(x)\ , \label{eq:chQC_operators_GHF_a} \\
& \hat{K}_{j}[\psi]\,\varphi(x):= \left(
             \int\frac{\psi^{*}_{j}(x^{\prime})\,\varphi(x^{\prime})}
	     {|{\bm r}-{\bm r}^{\prime}|}\,\mathrm{d}x^{\prime} \right)
             \psi_{j}(x)\ . \label{eq:chQC_operators_GHF_b} 
\end{align}
\end{subequations}

Therefore, if we define the GHF \emph{Fock operator} as

\begin{equation}
\label{eq:chQC_Fock_op_GHF}
\hat{F}^\mathrm{GHF}[\psi]:=
\hat{h}+\sum_j \Big(\hat{J}_{j}[\psi]-\hat{K}_{j}[\psi]\Big) \ ,
\end{equation}

we arrive to a first version of the \emph{Hartree-Fock equations} by
asking that the functional derivative in
eq.~(\ref{eq:chQC_funct_der_GHF}) be zero:

\begin{equation}
\label{eq:chQC_HF_eqs_v1_GHF}
\hat{F}^\mathrm{GHF}[\psi]\,\psi_{i}(x)=\sum_j\lambda_{ij}\psi_{j}(x)\ ,
 \qquad i=1,\ldots,N\ .
\end{equation}

Now, in order to obtain a simpler version of them, we shall take
profit from the fact that the whole problem is invariant under a
unitary transformation among the one-electron orbitals.

If we repeat the calculation in eq.~(\ref{eq:chQC_limit_GHF}) but varying
$\psi_k$ this time, instead of $\psi^{*}_{k}$, and use the following
relation:

\begin{equation}
\label{eq:chQC_cc_GHF}
\int \psi^*_i(x)\hat{h}\psi_j(x)\mathrm{d}x =
\int \big[\hat{h}\psi^*_i(x)\big]\psi_j(x)\mathrm{d}x
  \ ,
\end{equation}

we arrive to the GHF equations for the conjugated spin-orbitals:

\begin{equation}
\label{eq:chQC_HF_eqs_v1_GHF_conj}
\hat{F}^\mathrm{GHF}[\psi]\,\psi^*_{i}(x)=\sum_j\lambda_{ji}\psi^*_{j}(x)\ .
 \qquad i=1,\ldots,N\ .
\end{equation}

Then, we may subtract the complex conjugate of
eq.~(\ref{eq:chQC_HF_eqs_v1_GHF_conj}) from eq.~(\ref{eq:chQC_HF_eqs_v1_GHF})
yielding

\begin{equation}
\label{eq:chQC_lambdakjjk_GHF}
\sum_j \big(\lambda_{ij}-\lambda^{*}_{ji}\big)\,
 \psi_j(x) = 0 \ , \qquad k=1,\ldots,N \ .
\end{equation}

Therefore, since the set of the $\psi_j$ is orthogonal and hence
linearly independent, we have that the $N \times N$ matrix
$\Lambda:=(\lambda_{ij})$ of Lagrange multipliers is Hermitian:

\begin{equation}
\label{eq:chQC_lambdakjjk_GHF_herm}
\lambda_{ij}=\lambda^{*}_{ji}\ , \qquad k,j=1,\ldots,N \ .
\end{equation}

This actually means that we have a set of three equations that the
stationary spin-orbitals satisfy, but only two of them are
independent. These equations are the GHF equations for the $\psi_i$
and $\psi^*_i$, in~(\ref{eq:chQC_HF_eqs_v1_GHF})
and~(\ref{eq:chQC_HF_eqs_v1_GHF_conj}), respectively,
and~(\ref{eq:chQC_lambdakjjk_GHF_herm}). Any pair of them could be in
principle be chosen as the basic equations, however, in common
practice the first and the last one of them are typically picked.

In any case, due to~(\ref{eq:chQC_lambdakjjk_GHF_herm}), a unitary
matrix $U$ exists that \emph{diagonalizes} $\Lambda$; in the sense
that $\varepsilon:=U^{-1}\Lambda U=U^{+}\Lambda U$ is a diagonal
matrix, i.e., $\varepsilon_{ij}=\delta_{ij}\varepsilon_{i}$. Using
this unitary matrix $U$, we can transform the set of orbitals
$\{\psi_i\}$ into a new one $\{\psi^{\prime}_i\}$:

\begin{equation}
\label{eq:chQC_transformation_phi_GHF}
\psi_k(x)=\sum_j U_{kj}\,\psi^{\prime}_j(x) \ .
\end{equation}

This transformation is physically legitimate since it only changes the
$N$-electron wavefunction~$\Psi$ in an unmeasurable phase
$e^{i\phi}$. To see this, let us denote by $S_{ij}$ the $(ij)$-element
of the matrix inside the Slater determinant in
eq.~(\ref{eq:chQC_constraint_total_SD_GHF}), i.e.,
$S_{ij}:=\psi_{j}(x_{i})$. Then, after using the expression above, the
$(ij)$-element of the new matrix $S^{\prime}$ can be related to the
old ones via $S_{ki}=\sum_{j}U_{kj}S^{\prime}_{ij}$, in such a way
that $S=S^{\prime}U^{T}$ and the desired result follows:

\begin{equation}
\label{eq:chQC_Phi_invariant_GHF}
\Psi\big(\{\psi_{i}\}\big)=\frac{\det S}{\sqrt{N!}}=
\frac{\det \big(S^{\prime}U^{T}\big)}{\sqrt{N!}}=
\frac{\det S^{\prime} \det U^{T}}{\sqrt{N!}}=
e^{i\phi}\Psi\big(\{\psi^{\prime}_{i}\}\big) \ .
\end{equation}

Now, we insert eq.~(\ref{eq:chQC_transformation_phi_GHF}) into the
first version of the Hartree-Fock equations
in~(\ref{eq:chQC_HF_eqs_v1_GHF}):

\begin{equation}
\label{eq:chQC_HF_eqs_v2_GHF}
\hat{F}^\mathrm{GHF}[U\psi^{\prime}]\,\left(\sum_j 
  U_{ij}\psi^{\prime}_{j}(x)\right)=
 \sum_{j,k}\lambda_{ij}U_{jk}\psi^{\prime}_{k}(x)\ ,
 \qquad i=1,\ldots,N\ .
\end{equation}

Next, we multiply by $U^{-1}_{li}$ each one of the $N$ expressions and
sum in $i$:

\begin{eqnarray}
\label{eq:chQC_HF_eqs_v3_GHF}
\lefteqn{\hat{F}^\mathrm{GHF}[U\psi^{\prime}]\,\left(\sum_{i,j}
\underbrace{U^{-1}_{li}U_{ij}}_{\displaystyle \delta_{lj}}
                                \psi^{\prime}_{j}(x)\right)=
 \sum_{i,j,k}\underbrace{U^{-1}_{li}\lambda_{ij}U_{jk}
    }_{\displaystyle \varepsilon_{lk}=\delta_{lk}\varepsilon_{l}}
 \psi^{\prime}_{k}(x)} \nonumber \\[6pt]
&&\Longrightarrow\quad
\hat{F}^\mathrm{GHF}[U\psi^{\prime}]\,\psi^{\prime}_{l}(x)=
 \varepsilon_{l}\psi^{\prime}_{l}(x)\ ,
 \qquad l=1,\ldots,N\ .
\end{eqnarray}

Although this new version of the Hartree-Fock equations can be readily
seen as a \emph{pseudo-eigenvalue problem} and solved by the customary
iterative methods, we can go a step further and show that, like the
$N$-particle wavefunction $\Psi$ (see
eq.~(\ref{eq:chQC_Phi_invariant_GHF})), the Fock operator
$\hat{F}^\mathrm{GHF}[\psi]$, as a function of the one-electron
orbitals, is invariant under a unitary transformation such as the one
in eq.~(\ref{eq:chQC_transformation_phi_GHF}). In fact, this is true
for each one of the sums of Coulomb and exchange operators in
eq.~(\ref{eq:chQC_Fock_op_GHF}) separately:

\begin{eqnarray}
\label{eq:chQC_opJ_invariant_GHF}
\lefteqn{\sum_j \hat{J}_{j}[U\psi^{\prime}]\,\varphi(x)=
\sum_j \Bigg(\int\frac{ \left |\sum_k U_{jk}
           \,\psi^{\prime}_{k}(x^{\prime}) \right|^{2}}
{|{\bm r}-{\bm r}^{\prime}|}\,\mathrm{d}x^{\prime} \Bigg)\,\varphi(x) =
} \nonumber \\
&&\sum_j \Bigg(\int\frac{ \sum_{k,l} 
U^{*}_{jk}U_{jl}
           \,\psi^{\prime *}_{k}(x^{\prime})
           \,\psi^{\prime}_{l}(x^{\prime})}
{|{\bm r}-{\bm r}^{\prime}|}\,\mathrm{d}x^{\prime} \Bigg)
             \,\varphi(x)=
\nonumber \\
&& \Bigg(\int\frac{ \sum_{j,k,l} U^{-1}_{kj}U_{jl}
           \,\psi^{\prime *}_{k}(x^{\prime})
           \,\psi^{\prime}_{l}(x^{\prime})}
{|{\bm r}-{\bm r}^{\prime}|}\,\mathrm{d}x^{\prime} \Bigg)\,\varphi(x)=
\nonumber \\
&& \sum_k \Bigg(\int\frac{ |\psi^{\prime}_{k}(x^{\prime})|^{2}}
 {|{\bm r}-{\bm r}^{\prime}|}\,\mathrm{d}x^{\prime} \Bigg)\,\varphi(x)=
 \sum_j \hat{J}_{j}[\psi^{\prime}]\,\varphi(x) \ , \qquad
 \forall \varphi(x) \ ,
\end{eqnarray}

where, in the step before the last, we have summed on $j$ and $l$,
using that \mbox{$\sum_j U^{-1}_{kj}U_{jl} = \delta_{kl}$}.

Performing very similar calculations, one can show that

\begin{equation}
\label{eq:chQC_opK_invariant_GHF}
\sum_j \hat{K}_{j}[U\psi^{\prime}]\,\varphi(x) =
\sum_j \hat{K}_{j}[\psi^{\prime}]\,\varphi(x)\ ,\qquad \forall \varphi(x) \ ,
\end{equation}

and therefore, that
$F^\mathrm{GHF}[U\psi^{\prime}]=F^\mathrm{GHF}[\psi^{\prime}]$. In
such a way that any unitary transformation on a set of orbitals that
constitute a solution of the Hartree-Fock equations
in~(\ref{eq:chQC_HF_eqs_v1_GHF}) yields a different set that is also a
solution of \emph{the same} equations. For computational and
conceptual reasons (see, for example, Koopmans' Theorem below), it
turns out to be convenient to use this freedom and choose the
matrix~$U$ in such a way that the Lagrange multipliers matrix is
diagonalized (see eq.~(\ref{eq:chQC_lambdakjjk_GHF}) and the paragraph
below it). The particular set of one-electron orbitals
$\{\psi^{\prime}_i\}$ obtained with this~$U$ are called
\emph{canonical orbitals} and their use is so prevalent that we will
circumscribe the forecoming discussion to them and drop the prime from
the notation.

Using the canonical orbitals, the \emph{Hartree-Fock equations} can be
written as

\begin{equation}
\label{eq:chQC_HF_eqs_vFinal_GHF}
\hat{F}^\mathrm{GHF}[\psi]\,\psi_{i}(x)=\varepsilon_{i}\psi_{i}(x)\ ,
 \qquad i=1,\ldots,N\ .
\end{equation}

Many of the remarks related to these equations are similar to those
made about the Hartree ones in~(\ref{eq:chQC_Hartree_eqs}), although
there exist important differences due to the inclusion of the
indistinguishability of the electrons in the variational ansatz. This
is clearly illustrated if we calculate the joint probability density,
associated to a wavefunction like the one in
eq.~(\ref{eq:chQC_constraint_total_GHF}), of the coordinates with
label 1 taking the value $x_{1}$, the coordinates with label $2$
taking the value $x_{2}$, and so on:

\begin{eqnarray}
\label{eq:chQC_rho_GHF}
\lefteqn{\rho^\mathrm{GHF}(x_1,\ldots,x_N)=
  \big|\Psi(x_1,\ldots,x_N)\big|^{2}=}\nonumber \\
&& \!\!\!\!\!\!\!\!\!\!\!\! \frac{1}{N!}\sum_{p,p^{\prime} \,\in S_{N}}
  (-1)^{\mathcal{T}(p)+\mathcal{T}(p^{\prime})}
  \psi^{*}_{p(1)}(x_{1}) \cdots \psi^{*}_{p(N)}(x_{N})\,
  \psi_{p^{\prime}(1)}(x_{1}) \cdots \psi_{p^{\prime}(N)}(x_{N}) 
  \ .
\end{eqnarray}

If we compare this expression with eq.~(\ref{eq:chQC_Hartree_rho}), we
see that the antisymmetry of $\Psi$ has completely spoiled the
statistical independence among the one-electron coordinates. However,
there is a weaker quasi-independence that may be recovered: If, using
the same reasoning about permutations that took us to the one-electron
part $\sum_{i}\langle\Psi|\,\hat{h}_{i}\,|\Psi\rangle$ of the energy
functional in page~\pageref{page:one_e_calculation}, we calculate the
marginal probability density of the $i$-th coordinates taking the
value $x_i$, we find

\begin{equation}
\label{eq:chQC_rhoi_GHF}
\rho^\mathrm{GHF}_i(x_i):=\int \left(\prod_{k \ne i} \mathrm{d}x_k \right)
\rho^\mathrm{GHF}(x_1,\ldots,x_N)=\frac{1}{N}\sum_j \big|\psi_j(x_i)\big|^{2} \ .
\end{equation}

Now, since the coordinates indices are just immaterial labels, the
actual probability density of finding \emph{any} electron with
coordinates $x$ is given by

\begin{equation}
\label{eq:chQC_qrhoi_GHF}
\rho^\mathrm{GHF}(x):=\sum_i \rho^\mathrm{GHF}_i(x)=\sum_i \big|\psi_i(x)\big|^{2} \ ,
\end{equation}

which can be interpreted as a \emph{charge density} (except for the
sign), as, in atomic units, the charge of the electron is $e=-1$. The
picture being consistent with the fact that $\rho^\mathrm{GHF}(x)$ is
normalized to the number of electrons $N$:

\begin{equation}
\label{eq:chQC_qrhoi_norm_GHF}
\int \rho^\mathrm{GHF}(x)\,\mathrm{d}x=N \ .
\end{equation}

Additionally, if we perform the same type of calculations that allowed
to calculate the two-electron part of the energy functional in
page~\pageref{page:two_e_calculation}, we have that the two-body
probability density of the $i$-th coordinates taking the value~$x_i$
and of the $j$-th coordinates taking the value $x_j$ reads

\begin{eqnarray}
\label{eq:chQC_rhoij_GHF}
\lefteqn{\rho^\mathrm{GHF}_{ij}(x_i,x_j):=\int \left(\prod_{k \ne i,j}\mathrm{d}x_k \right)
          \rho^\mathrm{GHF}(x_1,\ldots,x_N)=} \nonumber \\
&&  \frac{1}{N(N-1)} \left(\sum_{k,l} \big|\psi_k(x_i)\big|^{2}
                            \big|\psi_l(x_j)\big|^{2} -
  \sum_{k,l} \psi^{*}_k(x_i)\,\psi^{*}_l(x_j)\,\psi_l(x_i)\,\psi_k(x_j)\right)
  \ ,
\end{eqnarray}

and, if we reason in the same way as in the case of
$\rho^\mathrm{GHF}_i(x_i)$, in order to get to the probability density
of finding \emph{any} electron with coordinates $x$ at the same time
that \emph{any other} electron has coordinates $x^{\prime}$, we must
multiply the function above by $N(N-1)/2$, which is the number of
immaterial $(i,j)$-labelings, taking into account that the distinction
between $x$ and $x^{\prime}$ is also irrelevant:

\begin{eqnarray}
\label{eq:chQC_qrhoij_GHF}
\lefteqn{\rho^\mathrm{GHF}(x,x^{\prime}):=\frac{N(N-1)}{2}\,\rho^\mathrm{GHF}_{ij}(x,x^{\prime})=}
\nonumber \\ && \frac{1}{2} \left(\sum_k \big|\psi_k(x)\big|^{2} \sum_l
\big|\psi_l(x^{\prime})\big|^{2} - \sum_{k,l}
\psi^{*}_k(x)\,\psi^{*}_l(x^{\prime})\,\psi_l(x)\,\psi_k(x^{\prime})\right)
\ .
\end{eqnarray}

Finally, taking eq.~(\ref{eq:chQC_qrhoi_GHF}) to this one, we have

\begin{equation}
\label{eq:chQC_qrhoij_qrhoi_GHF}
\rho^\mathrm{GHF}(x,x^{\prime})=\frac{1}{2}\left(
  \rho^\mathrm{GHF}(x)\,\rho^\mathrm{GHF}(x^{\prime})-
  \sum_{k,l} \psi^{*}_k(x)\,\psi^{*}_l(x^{\prime})\,
             \psi_l(x)\,\psi_k(x^{\prime})\right) \ ,
\end{equation}

where the first term corresponds to independent electrons and the
second one, called the \emph{interference term}, could be
interpreted as an exchange correction.

Although, in general, this is the furthest one may go, when additional
constraints are imposed on the spin part of the one-electron
wavefunctions (see the discussion about Restricted Hartree-Fock in the
following pages, for example), the exchange correction in
eq.~(\ref{eq:chQC_qrhoij_qrhoi_GHF}) above vanishes for electrons of
opposite spin, i.e., electrons of opposite spin turn out to be
pairwise independent. However, whereas it is true that more
correlation could be added to the Hartree-Fock results by going to
higher levels of the theory and, in this sense, Hartree-Fock could be
considered the first step in the `correlation ladder', one should not
regard it as an `uncorrelated' approximation, since, even in the
simplest case of RHF (see below), Hartree-Fock electrons (of the same
spin) are statistically correlated. All of this has its roots in
the \emph{Pauli principle}, which states that no pair of electrons can
share all the quantum numbers.

Let us now point out that, like in the Hartree case, the left-hand
side of the Hartree-Fock equations
in~(\ref{eq:chQC_HF_eqs_vFinal_GHF}) is a complicated, non-linear
function of the orbitals~$\{\psi_i\}$ and the notation chosen is
intended only to emphasize the nature of the iterative protocol that
is typically used to solve the problem. However, note that, while the
Hartree operator $\hat{\mathcal{H}}_{k}[\phi]$ depended on the index
of the orbital $\phi_k$ on which it acted, the Fock operator in
eq.~(\ref{eq:chQC_HF_eqs_vFinal_GHF}) is the same for all the
spin-orbitals $\psi_i$. This is due to the inclusion of the $i=j$
terms in the sum of the Coulomb and exchange two-electron integrals in
eq.~(\ref{eq:chQC_expected_H_2_GHF}) and it allows to perform the
iterative procedure solving only one eigenvalue problem at each step,
instead of $N$ of them like in the Hartree case (see however the
UHF and ROHF versions of the Hartree-Fock problem in what follows).

The one-particle appearance of
eqs.~(\ref{eq:chQC_HF_eqs_vFinal_GHF}) is again strong and, whereas
the `eigenvalues' $\varepsilon_i$ are not the energies of the
individual electrons, they are called \emph{orbital energies} due to
the physical meaning they receive via the well-known \emph{Koopmans'
Theorem} \cite{Koo1934PHY}.

To get to this result, let us multiply
eq.~(\ref{eq:chQC_HF_eqs_vFinal_GHF}) from the left by $\psi_i(x)$,
for a given $i$, and then integrate over $x$.  Using the definition of
the Fock operator in eq.~(\ref{eq:chQC_Fock_op_GHF}) together with the
Coulomb and exchange ones in eqs.~(\ref{eq:chQC_operators_GHF}), we
obtain

\begin{equation}
\label{eq:chQC_Koopmans1_GHF}
\langle \psi_i |\,\hat{F}^\mathrm{GHF}\,|\psi_i \rangle=
  h_{i} + \sum_j \Big( J_{ij} - K_{ij} \Big)=
  \varepsilon_{i}\ ,
 \qquad i=1,\ldots,N\ ,
\end{equation}

where we have used the same notation as in
eq.~(\ref{eq:chQC_expected_H_2_GHF}) and the fact that the
one-electron orbitals are normalized.

If we next sum on $i$ and compare the result with the expression in
eq.~(\ref{eq:chQC_expected_H_2_GHF}), we found that the relation of
the eigenvalues $\varepsilon_i$ with the actual Hartree-Fock energy is
given by

\begin{equation}
\label{eq:chQC_Koopmans2_GHF}
E^\mathrm{GHF}=\sum_i\varepsilon_i - \frac{1}{2} \sum_{i,j} \Big( J_{ij} - K_{ij} \Big)\ .
\end{equation}

\label{page:Koopman}

Finally, if we assume that upon `removal of an electron from the
$k$-th orbital' the rest of the orbitals will remain unmodified, we
can calculate the \emph{ionization energy} using the expression
in~(\ref{eq:chQC_expected_H_2_GHF}) together with the equations above:

\begin{eqnarray}
\label{eq:chQC_Koopmans_GHF}
&& \Delta E^\mathrm{GHF} := E^\mathrm{GHF}_{N-1} - E^\mathrm{GHF}_{N} = \sum_{i \ne k} h_i - \sum_i h_i +
   \frac{1}{2} \sum_{i,j \ne k} \Big( J_{ij} - K_{ij} \Big)
  \nonumber \\
&& - \frac{1}{2} \sum_{i,j} \Big( J_{ij} - K_{ij} \Big)=
 - h_k - \sum_j \Big( J_{kj} - K_{kj} \Big) = -\varepsilon_k \ ,
\end{eqnarray}

and this is Koopmans' Theorem, namely, that \emph{the $k$-th
ionization energy in the frozen-orbitals approximation is
$\varepsilon_k$}.

Moving now to the issue about the solution of the Hartree-Fock
equations in~(\ref{eq:chQC_HF_eqs_vFinal_GHF}), we must remark that
the necessity of using the relatively unreliable iterative approach to
tackle them stems again from their complicated mathematical form. Like
in the Hartree case, we have managed to largely reduce the dimension
of the space on which the basic equations are defined: from $3N$ in
the electronic Schr\"odinger equation in~(\ref{eq:chQC_bo_summary_a})
to 3 in the Hartree-Fock ones. However, to have this, we have payed
the price of dramatically increasing their complexity
\cite{Can2003BOOK}, since, while the electronic Schr\"odinger equation
was one linear differential equation, the Hartree-Fock ones
in~(\ref{eq:chQC_HF_eqs_vFinal_GHF}) are $N$ coupled non-linear
integro-differential equations, thus precluding any analytical
approach to their solution.

A typical iterative procedure\footnote{\label{foot:iterative_outline}
The process described in this paragraph must be taken only as an
outline of the one that is performed in practice. It is impossible to
deal in a computer with a general function as it is (a non-countable
infinite set of numbers), and the problem must be discretized in some
way. The truncation of the one-electron Hilbert space using a finite
basis set, described in secs.~\ref{sec:QC_Roothaan}
and~\ref{sec:QC_basis_sets}, is the most common way of doing
this. Moreover, the iterative procedure is normally performed
using not the spin-orbitals but the spatial ones. In this sense, the
restricted versions of the Hartree-Fock problem, discussed below, are
closer to the actual implementation of the theory in computer
applications.} begins by proposing a \emph{starting guess} for the
set of $N$ spin-orbitals $\{\psi^0_i\}$. With them, the Fock operator
$\hat{F}^\mathrm{GHF}[\psi^{0}]$ in the left-hand side of
eq.~(\ref{eq:chQC_HF_eqs_vFinal_GHF}) is constructed and the set of
$N$ equations is solved as one simple eigenvalue problem. Then, the
$\{\psi^1_i\}$ that correspond to the $N$ lowest eigenvalues
$\varepsilon^{1}_i$ are selected (see the discussion of the aufbau
principle below) and a new Fock operator
$\hat{F}^\mathrm{GHF}[\psi^{1}]$ is constructed with them. The process
is iterated until (hopefully) the $n$-th set of solutions
$\{\psi^{n}_i\}$ differs from the $(n-1)$-th one $\{\psi^{n-1}_i\}$
less than a reasonably small amount (defining the distance among
solutions in some suitable way typically combined with a convergence
criterium related to the associated energy change). When this occurs,
the procedure is said to have converged and the solution orbitals are
called \emph{self-consistent}; also, a calculation of this kind is
commonly termed \emph{\underline{s}elf-\underline{c}onsistent
\underline{f}ield} (SCF).

Again, like in the Hartree case, many issues exist that raise doubts
about the possible success of such an approach. The most important
ones are related to the fact that a proper definition of the
\emph{Hartree-Fock problem} should be: \emph{find the global minimum
of the energy functional $\langle \Psi | \,\hat{H}\, | \Psi \rangle$
under the constraint that the wavefunction $\Psi$ be a Slater
determinant of one-electron spin-orbitals}, and not: \emph{solve the
Hartree-Fock equations}~(\ref{eq:chQC_HF_eqs_vFinal_GHF}). The
solutions of the latter are all the stationary points of the
constrained energy functional, while we are interested only in the
particular one that is the global minimum. Even ruling out the
possibility that a found solution may be a maximum or a saddle point
(which can be done \cite{Can2003BOOK,See1977JCP}), one can never be
sure that it is the global minimum and not a local one.

There exists, however, one way, related to the Hartree-Fock version of
the theorem by Simon and Lieb \cite{Lie1974JCP,Lie1977CMP} mentioned
in the previous section, of hopefully biasing a particular found
solution of eqs.~(\ref{eq:chQC_HF_eqs_vFinal_GHF}) to be the global
minimum that we are looking for. They showed, first, that for neutral
or positively charged molecules ($Z \ge N$), the Hartree-Fock global
minimization problem has a solution (its uniqueness is not established
yet \cite{Can2003BOOK}) and, second, that the minimizing orbitals
$\{\psi_i\}$ correspond to the $N$ lowest eigenvalues of the Fock
operator $\hat{F}^\mathrm{GHF}[\psi]$ that is self-consistently
constructed with them. Therefore, although \emph{the reverse is not
true in general} \cite{Can2003BOOK} (i.e., from the fact that a
particular set of orbitals are the eigenstates corresponding to the
lowest eigenvalues of the associated Fock operator, does not
necessarily follow that they are the optimal ones), the information
contained in Simon and Lieb's result is typically invoked to build
each successive state in the iterative procedure described above by
keeping only the $N$ orbitals that correspond to the $N$ lowest
eigenvalues $\varepsilon_i$. Indeed, by doing that, one is effectively
constraining the solutions to have a property that the true solution
does have, so that, in the worst case, the space in which one is
searching is of the same size as the original one, and, in the best
case (even playing with the possibility that the reverse of Simon and
Lieb's theorem be true, though not proved), the space of solutions is
reduced to the correct global minimum alone.  This wishful-thinking
way of proceeding is termed the \emph{aufbau principle}
\cite{Can2003BOOK}, and, together with a clever choice of the
starting-guess set of orbitals \cite{Sch1991BOOK} (typically extracted
from a slightly less accurate theory, so that one may expect that it
could be `in the basin of attraction' of the true Hartree-Fock
minimum), constitute one of the many heuristic strategies that make
possible that the aforementioned drawbacks (and also those related to
the convergence of iterative procedures) be circumvented in real
cases, so that, in practice, most of SCF calculations performed in the
field of quantum chemistry do converge to the true solution of
eqs.~(\ref{eq:chQC_HF_eqs_vFinal_GHF}) in spite of the theoretical
notes of caution.

Now, to close this GHF part, let us discuss some points regarding the
imposition of constraints as a justification for subsequently
introducing three commonly used forms of the Hartree-Fock theory
that involve additional restrictions on the variational ansatz (apart
from those in eqs.~(\ref{eq:chQC_constraint_total_GHF}) and
(\ref{eq:chQC_constraint_mono_GHF})).

In principle, the target systems in which we are interested in our
group and to which the theory developed in this work is meant to be
applied are rather complex (short peptides, small ligands, etc.). They
have many degrees of freedom and the different interactions that drive
their behaviour typically compete with one another, thus producing
complicated, `frustrated' energy landscapes (see
refs.~\cite{Onu2004COSB,Plo2002QRB,Dil1999PSC,Dob1998ACIE,Bry1995PRO,Bry1987PNAS},
but note, however, that we do not need to think about macromolecules;
a small molecule like CO$_2$ already has 22 electrons). This state of
affairs renders the a priori assessment of the accuracy of any
approximation to the exact equations an impossible task. As
researchers calculate more and more properties of molecular species
using quantum chemistry and the results are compared to higher-level
theories or to experimental data, much empirical knowledge about `how
good is Theory~A for calculating Property~X' is being
gathered. However, if the characterization of a completely new
molecule that is not closely related to any one that has been
previously studied is tackled with, say, the Hartree-Fock
approximation, it would be very unwise not to `ask for a second
opinion'.

All of this also applies, word by word, to the choice of the
constraints on the wavefunction in variational approaches like the one
discussed in this section: For example, it is impossible to know a
priori what will be the loss of accuracy due to the requirement that
the $N$-particle wavefunction $\Psi$ be a Slater determinant as in
eq.~(\ref{eq:chQC_constraint_total_GHF}). However, in the context of
the Hartree-Fock approximation, there exists a way of proceeding,
again, partly based on wishful thinking and partly confirmed by actual
calculations in particular cases, that is almost unanimously used to
choose additional constraints which are expected to yield more
\emph{efficient} theories. It consists of imposing constraints to the
variational wavefunction that are \emph{properties that the exact
solution to the problem does have}. In such a way that the obvious
loss of accuracy due to the reduction of the search space is expected
to be minimized, while the decrease in computational cost could be
considerable.

This way of thinking is clearly illustrated by the question of whether
or not one should allow that the one-electron spin-orbitals $\psi_i$
(and therefore the total wavefunction $\Psi$) be complex
valued. Indeed, due to the fact that the electronic Hamiltonian in
eq.~(\ref{eq:chQC_el_ham_1}) is self-adjoint, the real and imaginary
parts of any complex eigenfunction solution of the time independent
Schr\"odinger equation in~(\ref{eq:chQC_bo_summary_a}) are also
solutions of it \cite{Can2003BOOK}. Therefore, the ground-tate, which
is the exact solution of the problem that we are trying to solve, may
be chosen to be real valued. Nevertheless, the exact minimum will not
be achieved, in general, in the smaller space defined by the
Hartree-Fock constraints in eqs.~(\ref{eq:chQC_constraint_total_GHF})
and (\ref{eq:chQC_constraint_mono_GHF}), so that there is no a priori
reason to believe that allowing the Hartree-Fock wavefunction to take
complex values would not improve the results by finding a lower
minimum. In fact, in some cases, this happens
\cite{Sch1991BOOK}. Nevertheless, if one constrains the search to real
orbitals, the computational cost is reduced by a factor two, and,
after all, the whole formalism discussed in this work profits
from the imposition of constraints (starting by the consideration of
only one Slater determinant), all of which save some computational
effort at the expense of a reduction in the accuracy. The search for
the most efficient of these approximations constitutes the main part
of the quantum chemistry field.

Apart from these `complex vs. real' considerations, there exist
three further restrictions that are commonly found in the literature
and that affect the spin part of the one-electron orbitals
$\psi_i$. The $N$-electron wavefunction $\Psi$ of the GHF
approximation (which is the one discussed up to now) is not an
eigenstate of the total-spin operator, $\hat{S}^2$, nor of the
$z$-component of it, $\hat{S}_z$ \cite{Sch1991BOOK}. However, since
both of them commute with the electronic Hamiltonian in
eq.~(\ref{eq:chQC_el_ham_1}), the true ground-state of the exact
problem can be chosen to be an eigenstate of both operators
simultaneously. Therefore three additional types of constraints
on the spin part of the GHF wavefunction
in~(\ref{eq:chQC_constraint_total_GHF}) are typically made that force
the variational ansatz to satisfy these ground-state properties and
that should be seen in the light of the above discussion, i.e., as
reducing the search space, thus yielding an intrinsically less
accurate theory, but also as being good candidates to hope that the
computational savings will pay for this.

\begin{figure}
\centerline{
\epsfxsize=7cm
\epsfbox{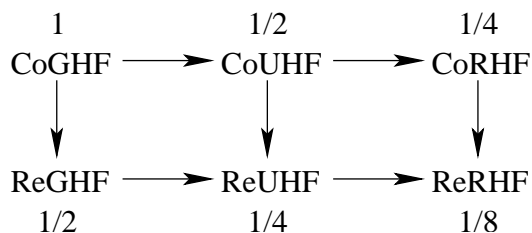}
}\vspace{0.2cm}
\caption{\label{fig:HF_computational_cost} Schematic relation map among
  six types of Hartree-Fock methods discussed in the text:
  \emph{\underline{G}eneral \underline{H}artree-\underline{F}ock} (GHF),
  \emph{\underline{U}nrestricted \underline{H}artree-\underline{F}ock} (UHF)
  and \emph{\underline{R}estricted \underline{H}artree-\underline{F}ock}
  (RHF), in both their \emph{complex} (Co) and \emph{real} (Re) versions. The
  arrows indicate imposition of constraints; horizontally, in the spin part of
  the orbitals, and, vertically, from complex- to real-valued wavefunctions.
  Next to each method, the size of the search space relative to that in CoGHF
  is shown.}
\end{figure}

The first approximation to GHF (in a logical sense) is called
\emph{\underline{U}nrestricted \underline{H}artree-\underline{F}ock}
(UHF) and it consists of asking the orbitals $\psi_i$ to be a product
of a part $\phi_i({\bm r})$ depending on the positions ${\bm r}$ times
a spin eigenstate of the one-electron $\hat{s}_z$ operator, i.e.,
either $\alpha(\sigma)$ or $\beta(\sigma)$ (see
eq.~(\ref{eq:chQC_alpha_beta})). This is denoted by
$\psi_i(x):=\phi_i({\bm r})\gamma_i(\sigma)$, where $\gamma_i$ is
either the $\alpha$ or the $\beta$ function. Now, if we call
$N_\alpha$ and $N_\beta$ the number of spin-orbitals of each type, we
have that, differently from the GHF one, the UHF \mbox{$N$-particle}
wavefunction~$\Psi$ is an eigenstate of the $\hat{S}_z$ operator with
eigenvalue $(1/2)(N_\alpha-N_\beta)$ (in atomic units, see
sec.~\ref{sec:QC_molecular_H}). However, it is not an eigenstate
of $\hat{S}^{2}$ and, when this deviation results into a poor
description of the observables in which we are interested, we talk
about \emph{spin contamination} \cite{Bal1999RCC}. Although the UHF
wavefunction can be projected into pure $\hat{S}^{2}$-states, the
result is multideterminantal \cite{Sch1991BOOK} and will not be
considered here since it spoils many of the properties that render
Hartree-Fock methods a low-cost choice.

Regarding the computational cost of the UHF approximation, it is
certainly lower than that of GHF, since the search space is half as
large: In the latter case, we had to consider $2N$ (complex or real)
functions of $\mathbb{R}^{3}$ (the $\varphi^\alpha_i({\bm r})$ and
the~$\varphi^\beta_i({\bm r})$, see eq.~(\ref{eq:chQC_general_so})),
while in UHF we only have to deal with $N$ of them: the $\phi_i({\bm
r})$.

Now, if we introduce into the expression for the GHF constrained
functional in~(\ref{eq:chQC_constrained_functional_GHF}) the following
relations that hold for the UHF spin- and spatial
orbitals\footnote{\label{foot:weird_deltas1} Of course, the average
values at both sides of the expressions are taken over different
variables: over $x$ and $x^\prime$ on the left-hand side, and over
${\bm r}$ and ${\bm r}^\prime$ on the right-hand side. Also, let us
remark that, although placing \emph{functions} as arguments of the
Kronecker's delta $\delta_{\gamma_i \gamma_j}$ is a bit unorthodox
mathematically, it constitutes an intuitive (and common) notation.}

\begin{subequations}
\label{eq:chQC_help_relations_UHF}
\begin{align}
 \langle \psi_i |\,\hat{h}\,| \psi_i \rangle &=
  \langle \phi_i |\,\hat{h}\,| \phi_i \rangle
  \ , \label{eq:chQC_help_relations_UHF_a} \\
 \langle \psi_i\psi_j |\,\frac{1}{r}\,| \psi_i\psi_j \rangle &=
  \langle \phi_i\phi_j |\,\frac{1}{r}\,| \phi_i\phi_j \rangle
  \ , \label{eq:chQC_help_relations_UHF_b} \\
 \langle \psi_i\psi_j |\,\frac{1}{r}\,| \psi_j\psi_i \rangle &=
  \delta_{\gamma_i \gamma_j} 
  \langle \phi_i\phi_j |\,\frac{1}{r}\,| \phi_j\phi_i \rangle
  \ , \label{eq:chQC_help_relations_UHF_c}
\end{align}
\end{subequations}

we may perform a derivation analogous to the one performed for the GHF
case, and get to a first version of the \emph{UHF equations}

\begin{subequations}
\label{eq:chQC_HF_eqs_vFinal_UHF}
\begin{align}
& \hat{F}^{\mathrm{UHF}}_{i}[\phi]\,\phi_{i}({\bm r}):=
\left(\hat{h}+\sum^N_j \hat{J}_j[\phi] - \sum^N_{j}\delta_{\gamma_i \gamma_j}
      \hat{K}_j[\phi] \right)\,\phi_{i}({\bm r})=
\sum_j \lambda_{ij}\phi_{j}({\bm r})\ , \label{eq:chQC_HF_eqs_vFinal_UHF_a}\\
& \hat{F}^{\mathrm{UHF}}_{i}[\phi]\,\phi^*_{i}({\bm r})=
\sum_j \lambda_{ji}\phi^*_{j}({\bm r})\ , \label{eq:chQC_HF_eqs_vFinal_UHF_b}
\end{align}
\end{subequations}

for all $i=1,\ldots,N$ and where the Coulomb and exchange operators
dependent on the spatial orbitals $\phi_i$ are defined by their action
on an arbitrary function $\varphi({\bm r})$ as follows:

\begin{subequations}
\label{eq:chQC_operators_UHF}
\begin{align}
& \hat{J}_{j}[\phi]\,\varphi({\bm r}):=
             \left( \int\frac{|\phi_{j}({\bm r}^{\prime})|^{2}}
	   {|{\bm r}-{\bm r}^{\prime}|}\,\mathrm{d}{\bm r}^{\prime} \right)
             \varphi({\bm r})\ , \label{eq:chQC_operators_UHF_a} \\
& \hat{K}_{j}[\phi]\,\varphi({\bm r}):= \left(
      \int\frac{\phi^{*}_{j}({\bm r}^{\prime})\,\varphi({\bm r}^{\prime})}
	{|{\bm r}-{\bm r}^{\prime}|}\,\mathrm{d}{\bm r}^{\prime} \right)
             \phi_{j}({\bm r})\ . \label{eq:chQC_operators_UHF_b} 
\end{align}
\end{subequations}

Now, one must note that, differently from the GHF case, due to the
fact that the exchange interaction only takes place between orbitals
`of the same spin', the \emph{UHF Fock operator}
$\hat{F}^{\mathrm{UHF}}_{i}[\phi]$ depends on the index $i$. This
precludes the solution of UHF as a \emph{single} pseudoeigenvalue
problem (c.f. eq.~(\ref{eq:chQC_HF_eqs_vFinal_GHF})) and makes
necessary some further considerations in order to arrive to a more
directly applicable form of the expressions:

First, although the equations for $\phi_i$ and $\phi_i^*$
in~(\ref{eq:chQC_HF_eqs_vFinal_UHF}) can be combined in the same way
as in the GHF case to yield the Hermiticity conditions
$\lambda_{ij}=\lambda^*_{ji}$, there are fewer Lagrange multipliers in
the UHF scheme than in the previous derivation. To see this, one must
notice that the orthogonality constraints must be imposed \emph{on the
spin-orbitals, not on the spatial orbitals} (see
eq.~(\ref{eq:chQC_constraint_mono_GHF})). Therefore, since two UHF
spin-orbitals $\psi_i=\phi_i\alpha$ and $\psi_j=\phi_j\beta$ are
orthogonal no matter the value of $\langle \phi_i | \phi_j \rangle$
due to the different spin parts, the corresponding Lagrange multiplier
$\lambda_{ij}$ needs not to be included in the constrained functional
from which the UHF equations come. This may be incorporated into the
formalism by simply using that the matrix $\Lambda:=(\lambda_{ij})$ in
eqs.~(\ref{eq:chQC_HF_eqs_vFinal_UHF}) presents the following
block-diagonal form:

\begin{equation}
\label{eq:chQC_lambda_UHF}
\Lambda^{\mathrm{UHF}}:=
\left( \begin{array}{c@{\hspace{5pt}}c}
\Lambda^{\alpha}  & 0 \\[5pt]
0 & \Lambda^\beta
\end{array} \right)
\ ,
\end{equation}

where we have assumed (without loss of generality) that the UHF
spin-orbitals are ordered in such a way that the $\alpha$ ones occur
first, and $0$ indicates a block of zeros of the appropriate size. (Of
course, redundant constraints may be imposed on the orbitals by, for
example in this case, including matrix terms in $\Lambda$ that connect
the $\alpha$ and $\beta$ spaces. However, in order to know the exact
freedom we have in the choice of the constraints, it is convenient to
use the minimal number of Lagrange multipliers. If this approach were
not followed, for example, the discussion below about the
diagonalization of $\Lambda^{\alpha}$ and $\Lambda^\beta$ would become
much less direct.)

The next step consists in noticing that, despite the dependence of
$\hat{F}^{\mathrm{UHF}}_{i}$ on the orbital index $i$ in
eqs.~(\ref{eq:chQC_HF_eqs_vFinal_UHF}), there are actually only
\emph{two} different Fock operators: one for the $\alpha$ orbitals and
one for the $\beta$ ones. Defining the sets of indices $A:=\{i|1\leq i
\leq N_\alpha\}$ and $B:=\{i|N_\alpha + 1 \leq i \leq N_\alpha +
N_\beta = N\}$, we can write these two $\alpha$ and $\beta$ operators:

\begin{subequations}
\label{eq:chQC_alpha_beta_F_UHF}
\begin{align}
&  \hat{F}^{\mathrm{UHF}}_\alpha[\phi]=\hat{h}+\sum^{N}_{j=1}
      \hat{J}_j[\phi] - \sum_{j \in A}
      \hat{K}_j[\phi] \ , \label{eq:chQC_alpha_beta_F_UHF_a} \\
&  \hat{F}^{\mathrm{UHF}}_\beta[\phi]=\hat{h}+\sum^{N}_{j=1}
      \hat{J}_j[\phi] - \sum_{j \in B}
      \hat{K}_j[\phi] \ . \label{eq:chQC_alpha_beta_F_UHF_b} 
\end{align}
\end{subequations}

With them, and using the particular structure of the matrix
$\Lambda^{\mathrm{UHF}}$ in~(\ref{eq:chQC_lambda_UHF}), the original
UHF equations in~(\ref{eq:chQC_HF_eqs_vFinal_UHF}) are split into two
disjoint sets of expressions that are only coupled through the Fock
operators on the left hand sides, plus the Hermiticity condition:

\begin{subequations}
\label{eq:chQC_HF_eqs_vFinal_1_UHF}
\begin{align}
&  \hat{F}^{\mathrm{UHF}}_\alpha[\phi]\phi_i({\bm r}) =
   \sum_{j \in A} \lambda_{ij}^\alpha \phi_j({\bm r}) \ , \quad
   \mathrm{if} \ i \in A \ , \label{eq:chQC_HF_eqs_vFinal_1_UHF_a} \\
&  \hat{F}^{\mathrm{UHF}}_\beta[\phi]\phi_i({\bm r}) =
   \sum_{j \in B} \lambda_{ij}^\beta \phi_j({\bm r}) \ , \quad
   \mathrm{if} \ i \in B \ , \label{eq:chQC_HF_eqs_vFinal_1_UHF_b} \\
&  \lambda_{ij} = \lambda_{ji}^* \ . \label{eq:chQC_HF_eqs_vFinal_1_UHF_c}
\end{align}
\end{subequations}

The last step needed to arrive to the final form of the UHF equations
(found by Pople and Nesbet \cite{Pop1954JCP} and named after them) is
the diagonalization of both the $\Lambda^\alpha$ and $\Lambda^\beta$
Hermitian matrices above. In order to achieve this, the orbitals
$\phi_i$ must be transformed similarly to the GHF case. However, this
is trickier than it was then, since not only the $N$-electron
wavefunction $\Psi$ and the Fock operators must remain invariant under
the sought transformation, but also the UHF constraints must be kept.

As we saw before, any unitary transformation $U$ in the set of
spin-orbitals $\psi_i$ like the one in
eq.~(\ref{eq:chQC_transformation_phi_GHF}) is physically legitimate,
since it changes the Slater determinant by only an unmeasurable phase
and leave the Fock operators invariant. Then, if we write each
spin-orbital as in~(\ref{eq:chQC_general_so}):

\begin{equation}
\label{eq:chQC_general_so_i}
\psi_i(x)=\varphi_i^{\alpha}({\bm r})\,\alpha(\sigma)+
        \varphi_i^{\beta}({\bm r})\,\beta(\sigma) \ ,
\end{equation}

we can make use of~(\ref{eq:chQC_transformation_phi_GHF}), to obtain

\begin{equation}
\label{eq:chQC_transformation_psi_i}
\varphi_i^{\alpha}({\bm r})\,\alpha(\sigma)+
        \varphi_i^{\beta}({\bm r})\,\beta(\sigma)=
\sum_j U_{ij}
\big[ \varphi^{\prime\alpha}_j({\bm r})\,\alpha(\sigma)+
        \varphi^{\prime\beta}_j({\bm r})\,\beta(\sigma)\big] \ .
\end{equation}

By setting $\sigma=-1/2$ and $\sigma=1/2$ in this expression, we see
that any transformation $U$ in the spin-orbitals $\psi_i$ induces
exactly the same transformation in their spatial components
$\varphi_i^{\alpha}$ and $\varphi_i^{\beta}$,

\begin{equation}
\label{eq:chQC_transformation_phi_i}
\varphi_i^{\gamma}({\bm r})=
\sum_j U_{ij}\,\varphi^{\prime\gamma}_j({\bm r})\ ,
\qquad \gamma=\alpha,\beta \ .
\end{equation}

Now, if we order the sets of spin and spatial orbitals: ${\bm
\psi}^T:=(\psi_1,\ldots,\psi_N)$ and $({\bm
\varphi}^{\gamma})^T:=(\varphi^{\gamma}_1,\ldots,\varphi^{\gamma}_N)$,
with $\gamma=\alpha,\beta$, we can express the \emph{UHF constraints}
by saying that the ${\bm \varphi}^{\gamma}$ must have the form

\begin{subequations}
\label{eq:chQC_constraints_UHF}
\begin{align}
&  ({\bm \varphi}^\alpha)^T=
    (\phi_1,\ldots,\phi_{N_\alpha},0,\ldots,0)
    \ , \label{eq:chQC_constraints_UHF_a} \\
&  ({\bm \varphi}^\beta)^T=
    (0,\ldots,0,\phi_{N_\alpha + 1},\ldots,
     \phi_{N_\alpha + N_\beta})
   \ . \label{eq:chQC_constraints_UHF_b} 
\end{align}
\end{subequations}

Then, since the fact that the $\alpha$ orbitals appear first
constitutes no loss of generality, we must ask the transformed
$\varphi_i^{\prime \gamma}$, with $\gamma=\alpha,\beta$, to have also
the structure in~(\ref{eq:chQC_constraints_UHF}) if we want to remain
inside the UHF scheme.  As a consequence, and due to the linear
independence among the orbitals, not every unitary matrix $U$ is
allowed, but only those of the form

\begin{equation}
\label{eq:chQC_U_UHF}
U^{\mathrm{UHF}}:=
\left( \begin{array}{c@{\hspace{5pt}}c}
U^{\alpha}  & 0 \\[5pt]
0 & U^\beta
\end{array} \right)
\ ,
\end{equation}

using the same notation as in eq.~(\ref{eq:chQC_lambda_UHF}).

\label{eq:QC_ROHF_argument} This can be easily proved by focusing on a
particular value for $i$ and $\gamma$ in
eq.~(\ref{eq:chQC_transformation_phi_i}), say, $i \in A$ and $\gamma =
\beta$.  Due to the UHF constraints
in~(\ref{eq:chQC_constraints_UHF}), we know that the left-hand side of
such an expression is zero and that only spatial orbitals with $j \in
B$ appear in the sum on the right-hand side, yielding the relation
$0=\sum_{j \in B}U_{ij}\phi^\prime_j({\bm r})$. But the
$\phi^\prime_j({\bm r})$, with $j \in B$, form an orthonormal, and
therefore linearly independent set, so that the only possibility that
such a relation can hold is that all coefficients $U_{ij}$ be zero. By
repeating this for all $i \in A$ and, then, for $\gamma = \alpha$, the
result follows.

Fortunately, this limited freedom in the choice of $U$ is still enough
to independently diagonalize $\Lambda^\alpha$ and $\Lambda^\beta$ in
eq.~(\ref{eq:chQC_HF_eqs_vFinal_1_UHF}) (which are both Hermitian) by
suitably choosing the unitary submatrices $U^\alpha$ and $U^\beta$
respectively.

This takes us to the final, diagonal form of the UHF equations, the
\emph{Pople-Nesbet equations} \cite{Pop1954JCP}:

\begin{subequations}
\label{eq:chQC_HF_eqs_vFinal_2_UHF}
\begin{align}
&  \hat{F}^{\mathrm{UHF}}_\alpha[\phi]\phi_i({\bm r}) =
   \varepsilon_{i}^\alpha \phi_i({\bm r}) \ , \quad
   \mathrm{if} \ i \in A \ , \label{eq:chQC_HF_eqs_vFinal_2_UHF_a} \\
&  \hat{F}^{\mathrm{UHF}}_\beta[\phi]\phi_i({\bm r}) =
   \varepsilon_{i}^\beta \phi_i({\bm r}) \ , \quad
   \mathrm{if} \ i \in B \ . \label{eq:chQC_HF_eqs_vFinal_2_UHF_b}
\end{align}
\end{subequations}

Although these equations are coupled through the Coulomb term in the
Fock operators on the left-hand side, at each step of the iterative
SCF procedure, they can be solved as two independent eigenvalue
problems. This has allowed to implement them in most quantum chemical
packages and it has made UHF calculations now routine.

To close the UHF discussion, we shall now study the statistical
properties of the probability densities appearing in this model. If
we introduce the special form of the UHF orbitals
in~(\ref{eq:chQC_constraints_UHF}) into the general expression
in~(\ref{eq:chQC_qrhoij_GHF}), we can calculate the two-body
probability density of finding \emph{any} electron with coordinates
$x$ at the same time that \emph{any other} electron has coordinates
$x^{\prime}$:

\begin{eqnarray}
\label{eq:chQC_qrhoij_UHF}
\lefteqn{\rho^{\mathrm{UHF}}(x,x^{\prime})=\frac{1}{2}\,
        \Bigg( \sum_{k,l}
        \big|\phi_k({\bm r})\big|^{2}\big|\phi_l({\bm r}^{\prime})\big|^{2}
        \gamma_k(\sigma)\,\gamma_l(\sigma^{\prime})}\nonumber \\
 && \mbox{} - \sum_{k,l}
 \phi^{*}_k({\bm r})\,\phi^{*}_l({\bm r}^{\prime})\,\phi_l({\bm r})\,
 \phi_k({\bm r}^{\prime})\,\gamma_k(\sigma)\,\gamma_l(\sigma^{\prime})
 \,\gamma_k(\sigma^{\prime})\,\gamma_l(\sigma) \Bigg)\ .
\end{eqnarray}

If we compute this probability density for `electrons of the same
spin', i.e., for $\sigma=\sigma^{\prime}$, we
obtain\footnote{\label{foot:weird_deltas2} Placing \emph{a function
and a coordinate} as arguments of the Kronecker's delta
$\delta_{\gamma_k \sigma}$ is even more unorthodox mathematically than
placing two functions (in fact, $\delta_{\gamma_k \sigma}$ is exactly
the same as $\gamma_k(\sigma)$), however, the intuitive character of
the notation compensates again for this.}

\begin{eqnarray}
\label{eq:chQC_qrhoij_samespins_UHF}
\lefteqn{\rho^{\mathrm{UHF}}({\bm r},{\bm r}^{\prime};
              \sigma=\sigma^{\prime})=} \nonumber \\[8pt]
&& \frac{1}{2} \sum_{k,l \in I_\sigma} \left(
        \big|\phi_k({\bm r})\big|^{2}\big|\phi_l({\bm r}^{\prime})\big|^{2}
- \phi^{*}_k({\bm r})\,\phi^{*}_l({\bm r}^{\prime})\,\phi_l({\bm r})\,
  \phi_k({\bm r}^{\prime}) \right) = \nonumber \\
&& \frac{1}{2}\left(\rho^{\mathrm{UHF}}({\bm r},\sigma)\,\rho^{\mathrm{UHF}}({\bm r}^{\prime},\sigma)
 - \sum_{k,l}
  \delta_{\gamma_k \sigma}\delta_{\gamma_l \sigma}
  \phi^{*}_k({\bm r})\,
  \phi^{*}_l({\bm r}^{\prime})\,\phi_l({\bm r})\,
  \phi_k({\bm r}^{\prime})\right) \ ,
\end{eqnarray}

where the following expression for the one-electron charge density has
been used:

\begin{equation}
\label{eq:chQC_qrhoi_UHF}
\rho^{\mathrm{UHF}}({\bm r},\sigma)=\sum_{k}
   \big|\phi_k({\bm r})\big|^{2}\gamma_k(\sigma) \ .
\end{equation}

At this point, note that eq.~(\ref{eq:chQC_qrhoij_samespins_UHF})
contains, like in the GHF case, the exchange correction to the first
(independent electrons) term. Nevertheless, as we advanced, if we
calculate the two-body $\rho^{\mathrm{UHF}}$ for `electrons of
opposite spin', i.e., for $\sigma \ne \sigma^{\prime}$, we have that

\begin{equation}
\label{eq:chQC_qrhoij_diffspins_UHF}
\rho^{\mathrm{UHF}}({\bm r},{\bm r}^{\prime};
              \sigma \ne \sigma^{\prime})=\frac{1}{2}\,
       \rho^{\mathrm{UHF}}({\bm r},\sigma)\,\rho^{\mathrm{UHF}}({\bm r}^{\prime},\sigma^{\prime}) \ ,
\end{equation}

i.e., that \emph{UHF electrons of opposite spin are statistically
pairwise independent}.

There is another common approximation to GHF that is more restrictive
than UHF and is accordingly called \emph{\underline{R}estricted
\underline{H}artree-\underline{F}ock} (RHF). Apart from asking the
orbitals $\psi_i$ to be a product of a spatial part times a spin
eigenstate of the one-electron $\hat{s}_z$ operator (like in the UHF
case), in RHF, the number of `spin-up' and `spin-down' orbitals is the
same, $N_\alpha=N_\beta$ (note that this means that RHF may only be
used with molecules containing an even number of electrons), and each
spatial wavefunction occurs twice: once multiplied by $\alpha(\sigma)$
and the other time by $\beta(\sigma)$. This is typically referred to
as a \emph{closed-shell} situation and we shall denote it by writing
$\psi_i(x):=\phi_i({\bm r})\,\alpha(\sigma)$ if $i \le N/2$, and
$\psi_i(x):=\phi_{i-N/2}({\bm r})\,\beta(\sigma)$ if $i > N/2$; in
such a way that there are $N/2$ different spatial orbitals denoted by
$\phi_{I}({\bm r})$, with, $I=1,\ldots,N/2$.  Using the same
notation as in~(\ref{eq:chQC_constraints_UHF}), the \emph{RHF
constraints} on the spin-orbitals are

\begin{subequations}
\label{eq:chQC_constraints_RHF}
\begin{align}
&  ({\bm \varphi}^\alpha)^T=
    (\phi_1,\ldots,\phi_{N/2},0,\ldots,0)
    \ , \label{eq:chQC_constraints_RHF_a} \\
&  ({\bm \varphi}^\beta)^T=
    (0,\ldots,0,\phi_1,\ldots,
     \phi_{N/2})
   \ . \label{eq:chQC_constraints_RHF_b} 
\end{align}
\end{subequations}

Due to these additional restrictions, we have that, differently from
the GHF and UHF ones, the RHF \mbox{$N$-particle} wavefunction~$\Psi$
is an eigenstate of both the $\hat{S}^2$ and the $\hat{S}_z$
operators, with zero eigenvalue in both cases
\cite{Sza1996BOOK,Sch1991BOOK}, just like the ground-state of the
exact problem. I.e., \emph{there is no spin-contamination in
RHF}.

Regarding the computational cost of the RHF approximation, it is even
lower than that of UHF, since the size of the search space has been
reduced to one quarter that of GHF: In the latter case, we had to
consider $2N$ (complex or real) functions of $\mathbb{R}^{3}$ (the
$\varphi^\alpha_i({\bm r})$ and the~$\varphi^\beta_i({\bm r})$, see
eq.~(\ref{eq:chQC_general_so})), while in RHF we only have to deal
with $N/2$ of them: the $\phi_I({\bm r})$ (see
fig.~\ref{fig:HF_computational_cost}).

Now, using again the relations in~(\ref{eq:chQC_help_relations_UHF}),
we can derive a first version of the \emph{RHF equations}:

\begin{equation}
\label{eq:chQC_HF_eqs_vFinal_1_RHF}
\hat{F}^{\mathrm{RHF}}[\phi]\,\phi_{I}({\bm r}):=
\left[ \hat{h}+\sum^{N/2}_J \bigg( 2\hat{J}_J[\phi] - 
      \hat{K}_J[\phi] \bigg) \right]\phi_{I}({\bm r})=
\sum_{J}^{N/2}\lambda_{IJ}\phi_{J}({\bm r})\ ,
\end{equation}

for all $I=1,\ldots,N/2$, where we have used that the \emph{RHF Fock
operator} $\hat{F}^{\mathrm{RHF}}[\phi]$, differently from the UHF
case, does not depend on the index $I$ of the orbital on which it
operates, and following the same steps as before, the minimal Lagrange
multipliers matrix needed to enforce the orthogonality constraints
among the \emph{spin}orbitals in the RHF case is

\begin{equation}
\label{eq:chQC_lambda_RHF}
\Lambda^{\mathrm{RHF}}:=
\left( \begin{array}{c@{\hspace{5pt}}c}
\Lambda^{(N/2)}/2  & 0 \\[5pt]
0 & \Lambda^{(N/2)}/2
\end{array} \right)
\ ,
\end{equation}

where, this time, $\Lambda^{(N/2)}:=(\lambda_{IJ})$ is an arbitrary
$N/2 \times N/2$ Hermitian matrix, and the $1/2$ has been included in
order to get to the classical RHF equations
in~(\ref{eq:chQC_HF_eqs_vFinal_1_RHF}) without irrelevant numerical
factors.

Using the same arguments as for UHF, it is clear that, in order to
`remain inside RHF' upon an unitary transformation of the
spin-orbitals $\psi_i$, not every unitary matrix $U$ is allowed, but
only those of the form

\begin{equation}
\label{eq:chQC_U_RHF}
U^{\mathrm{RHF}}:=
\left( \begin{array}{c@{\hspace{5pt}}c}
U^{(N/2)}  & 0 \\[5pt]
0  & U^{(N/2)}
\end{array} \right)
\ .
\end{equation}

Finally, by suitable choosing the $N/2 \times N/2$ unitary block
$U^{(N/2)}$, the Hermitian matrix $\Lambda^{(N/2)}$ can be
diagonalized and the final, diagonal form of the \emph{RHF equations}
can be written:

\begin{equation}
\label{eq:chQC_HF_eqs_vFinal_RHF}
\hat{F}^{\mathrm{RHF}}[\phi]\,\phi_{I}({\bm r}):=
\left[ \hat{h}+\sum^{N/2}_J \bigg( 2\hat{J}_J[\phi] - 
      \hat{K}_J[\phi] \bigg) \right]\phi_{I}({\bm r})=
\varepsilon_{I}\phi_{I}({\bm r})\ ,
\end{equation}

with $I=1,\ldots,N/2$.

As we can see, this version of the Hartree-Fock theory can be
numerically solved as a \emph{single} pseudoeigenvalue problem. This,
together with the aforementioned small size of the RHF space, has made
the RHF approximation (in its real-valued version) the first one cast
into a computationally manageable form \cite{Roo1951RMP,HalPRSLA1951}
and the most used one in recent literature
\cite{Var2002JPCA,Lan2005PSFB,Per2003JCC,Yu2001JMS,Els2001CP,Bal2000JMS,Rod1998JMS,Csa1999PBMB,Bea1997JACS,Fre1992JACS}
(for molecules with an even number of electrons).

Now, in order to investigate the statistical features of RHF, if we
introduce the special form of the orbitals
in~(\ref{eq:chQC_constraints_RHF}) into the general expression
in~(\ref{eq:chQC_qrhoij_GHF}), we can calculate the RHF two-body
probability density of finding \emph{any} electron with coordinates
$x$ at the same time that \emph{any other} electron has coordinates
$x^{\prime}$:

\begin{eqnarray}
\label{eq:chQC_qrhoij_RHF}
\rho^{\mathrm{RHF}}(x,x^{\prime})&=&\frac{1}{2}
        \left( \sum^{N/2}_{K,L}
        \big|\phi_K({\bm r})\big|^{2}\big|\phi_L({\bm r}^{\prime})\big|^{2}
        - \delta_{\sigma\sigma^{\prime}}\sum^{N/2}_{K,L} 
     \phi^{*}_K({\bm r})\,\phi^{*}_L({\bm r}^{\prime})\,\phi_L({\bm r})\,
        \phi_K({\bm r}^{\prime}) \right) = \nonumber \\
 &&  \frac{1}{2}\left(\rho({\bm r},\sigma)\,
          \rho({\bm r}^{\prime},\sigma^{\prime})
        - \delta_{\sigma\sigma^{\prime}}\sum^{N/2}_{K,L} 
        \phi^{*}_K({\bm r})\,\phi^{*}_L({\bm r}^{\prime})\,\phi_L({\bm r})\,
        \phi_K({\bm r}^{\prime})\right) \ .
\end{eqnarray}

where the following expression for the RHF one-electron charge density
has been used:

\begin{equation}
\label{eq:chQC_rhoi_RHF}
\rho^{\mathrm{RHF}}({\bm r},\sigma)=
   \sum^{N/2}_{K}\big|\phi_K({\bm r})\big|^{2} \ .
\end{equation}

We notice that the situation is the same as in the UHF case: For RHF
electrons with equal spin, there exists an exchange term in
$\rho^{\mathrm{RHF}}(x,x^{\prime})$ that corrects the `independent'
part, whereas \emph{RHF electrons of opposite spin are statistically
pairwise independent}.

Finally, if we follow the same steps as for GHF, in
page~\pageref{page:Koopman}, we can relate the \emph{RHF energy} to
the eigenvalues $\varepsilon_I$ and the two-electron spatial
integrals:

\begin{equation}
\label{eq:chQC_energy_RHF}
E=2\sum_I^{N/2} \varepsilon_I - \sum_{I,J}^{N/2}
  \Bigg( 2\langle\phi_I\phi_J|\,\frac{1}{r}\,|\phi_I\phi_J\rangle  -
        \langle\phi_I\phi_J|\,\frac{1}{r}\,|\phi_J\phi_I\rangle \Bigg)\ .
\end{equation}

To close this section, we shall discuss a fourth flavour of
Hartree-Fock which is called \emph{\underline{R}estricted
\underline{O}pen-shell \underline{H}artree-\underline{F}ock}
(ROHF). Compared to the rest of variants, ROHF is the most difficult
to derive theoretically and it shall be described here only in an
introductory manner. The ROHF wavefunction is somewhat between the RHF
and the UHF ones, and both of them can be obtained as particular cases
of the ROHF scheme (the RHF one provided that the molecule has an even
number of electrons). In the (monodeterminantal) ROHF case, the
one-electron spin orbitals $\psi_i$ are constrained to be of two
different types: $2N_D$ of them are \emph{doubly occupied}, like in
the RHF case, in such a way that they are formed by only $N_D$ spatial
orbitals $\phi_a$, each one of them appearing once multiplied by
$\alpha(\sigma)$ and once by $\beta(\sigma)$. The associated $2N_D$
spin-orbitals are said to belong to the \emph{closed shell} part of
the wavefunction. The remaining $N_S:=N-2N_D$ ones are \emph{singly
occupied}, like in the UHF case, and are said to belong to the
\emph{open shell}. Among them, $N_\alpha$ are multiplied by an
$\alpha(\sigma)$ spin part, and $N_\beta$ by $\beta(\sigma)$.

If we number the whole set of ROHF spatial orbitals $\phi_a$, with
$a=1,\ldots,N_D+N_\alpha+N_\beta$, in such a way that the doubly
occupied ones occur first, with $a \in D:=\{a|1\leq a \leq N_D\}$,
then the alpha ones, with $a \in A:=\{a|N_D + 1 \leq a \leq N_D +
N_\alpha\}$, and finally the beta ones, with $a \in B:=\{a|N_D +
N_\alpha + 1 \leq a \leq N_D + N_\alpha + N_\beta\}$, we can express
the \emph{ROHF constraints} on the spin-orbitals using the same
notation as in~(\ref{eq:chQC_constraints_UHF})
and~(\ref{eq:chQC_constraints_RHF}):

\begin{subequations}
\label{eq:chQC_constraints_ROHF}
\begin{align}
&  ({\bm \varphi}^\alpha)^T=
    (\underbrace{\phi_1,\ldots,\phi_{N_D}}_{N_D},
     \underbrace{\phi_{N_D + 1},\ldots,\phi_{N_D + N_\alpha}}_{N_\alpha},
     \underbrace{0,\ldots,0}_{N_D + N_\beta})
    \ , \label{eq:chQC_constraints_ROHF_a} \\
&  ({\bm \varphi}^\beta)^T=
    (\underbrace{0,\ldots,0}_{N_D + N_\alpha},
     \underbrace{\phi_1,\ldots,\phi_{N_D}}_{N_D},
     \underbrace{\phi_{N_D + N_\alpha + 1},\ldots,
      \phi_{N_D + N_\alpha + N_\beta}}_{N_\beta})
    \ ,
    \label{eq:chQC_constraints_ROHF_b} 
\end{align}
\end{subequations}

where the particular ordering has been chosen in order to facilitate
the forecoming calculations.

The general monodeterminantal ROHF wavefunction considered here and
constructed using the above constraints has the same spin properties
as the UHF one, i.e., it is an eigenstate of the the $\hat{S}_z$
operator with eigenvalue $(1/2)(N_\alpha-N_\beta)$, but it is not an
eigenstate of $\hat{S}^{2}$.  However, in the particular (and common)
case in which all the $N_S$ open shell orbitals are constrained to
present parallel spin parts (either all $\alpha$ or all $\beta$), the
ROHF wavefunction becomes an eigenstate of the $\hat{S}^{2}$ operator
too, with eigenvalue $\frac{N_S}{2}(\frac{N_S}{2}+1)$, thus avoiding
the problem of spin contamination \cite{Coo2005BOOK}. In order to
construct wavefunctions with the same $S^2$ but lower $S_z$, several
ROHF Slater determinants must be linearly combined. The subtleties
arising from such a procedure are beyond the scope of this review; the
interested reader may want to check references
\cite{Coo2005BOOK,Car1977BOOK,Hur1976BOOK,McW1992BOOK}, which discuss
this topic.

Regarding the size of the variational space in ROHF, it is somewhere
between UHF and RHF, depending on the $2N_D/(N_\alpha+N_\beta)$ ratio.

Now, if the ROHF constraints in~(\ref{eq:chQC_constraints_ROHF}) are
imposed on the GHF energy in eq.~(\ref{eq:chQC_expected_H_2_GHF}), the
ROHF analogue in terms of the spatial orbitals $\phi_a$ can be
calculated:

\begin{eqnarray}
\label{eq:chQC_energy_ROHF}
\lefteqn{E^{\mathrm{ROHF}}\big[\{\phi_a\}\big]=
   \sum_{a}f_a\langle\phi_{a}|\hat{h}|\phi_{a}\rangle} \nonumber \\
&& \mbox{} + \frac{1}{2} \sum_{a, b} \bigg(f_a f_b
 \langle\phi_a\phi_b|\frac{1}{r}|\phi_a\phi_b\rangle
 -  g_{ab}\langle\phi_a\phi_b|\frac{1}{r}|\phi_b\phi_a\rangle
    \bigg) \ ,
\end{eqnarray}

where the $f_a$ are a sort of `occupation numbers' that take the value
$f_a=2$ when $a \in D$ (i.e., when it corresponds to a closed shell
orbital) and $f_a=1$ otherwise. The matrix $g:=(g_{ab})$ is defined as

\begin{equation}
\label{eq:chQC_g_ROHF}
g:=
\left( \begin{array}{c@{\hspace{3pt}}|@{\hspace{3pt}}c}
2^D  & 1 \\[5pt]
\hline\\[-8pt]
1 & 
\begin{array}{c@{\hspace{5pt}}c}
1^\alpha & 0 \\[5pt]
0 & 1^\beta
\end{array}
\end{array} \right)
\ ,
\end{equation}

being $2^D$ a $N_D \times N_D$ box of 2's, $1^\alpha$ and $1^\beta$,
$N_\alpha \times N_\alpha$ and $N_\beta \times N_\beta$ boxes of 1's
respectively. The off-diagonal blocks contain in all elements the
number indicated and are of the appropriate size.

Then, as we did for GHF, in order to derive the ROHF equations, we
construct the functional for the conditioned stationary values
problem, adding to the energy in~(\ref{eq:chQC_energy_ROHF}) the
Lagrange multipliers terms needed to enforce the orthonormality
constraints:

\begin{equation}
\label{eq:chQC_constrained_functional_ROHF}
\widetilde{\mathcal{F}}\,\big[\{\phi_a\}\big] =
   E^{\mathrm{ROHF}}\big[\{\phi_a\}\big] -
 \sum_{a, b} \lambda_{ab}\,
   \Big( \langle\phi_a|\phi_b\rangle-\delta_{ab} \Big) \ .
\end{equation}

Now, like in all restricted HF cases, special attention must be payed
to the structure of the matrix
$\Lambda^{\mathrm{ROHF}}:=(\lambda_{ab})$, since the requirement is
that all \emph{spin-orbitals} be orthogonal, not the spatial
orbitals. In the ROHF case, this leads to explicitly impose the
orthonormality conditions (of course) inside the three sets of
$\phi_a$: the doubly occupied, the alpha and the beta ones; but also
between the doubly occupied and the alpha ones, and between the doubly
occupied and the beta ones. The orthormality between the alpha and
beta sets, however, needs not to be enforced, since the associated
spin-orbitals are already orthogonal due to the different spin parts.

These considerations lead to the following form for the minimal
Lagrange multipliers matrix:

\begin{equation}
\label{eq:chQC_lambda_ROHF}
\Lambda^{\mathrm{ROHF}}:=
\left( \begin{array}{c@{\hspace{5pt}}c@{\hspace{5pt}}c}
\Lambda^D & \Lambda^{D\alpha} & \Lambda^{D\beta} \\[5pt]
(\Lambda^{D\alpha})^+ & \Lambda^\alpha & 0 \\[5pt]
(\Lambda^{D\beta})^+ & 0 & \Lambda^\beta
\end{array} \right)
\ ,
\end{equation}

where the notation used for the different blocks connecting the doubly
occupied, alpha and beta shells is self-explanatory, and the fact
that $\Lambda^\mathrm{ROHF}$ is Hermitian after reaching stationarity
has been advanced.

Next, we impose the condition of zero functional derivative on the
functional in~(\ref{eq:chQC_constrained_functional_ROHF}), and obtain
a first version of the \emph{ROHF equations}:

\begin{subequations}
\label{eq:chQC_HF_eqs_v1_ROHF}
\begin{align}
& \hat{F}_a^{\mathrm{ROHF}}[\phi]\phi_{a}({\bm r})=
   \sum_{b}\lambda_{ab}\phi_{b}({\bm r})\ , \qquad
   a = 1,\ldots,N_D+N_\alpha+N_\beta \ ,
\label{eq:chQC_HF_eqs_v1_ROHF_a} \\
& \hat{F}_a^{\mathrm{ROHF}}[\phi]\,\phi^*_{a}({\bm r})=
   \sum_{b}\lambda_{ba}\phi^*_{b}({\bm r})\ , \qquad
   a = 1,\ldots,N_D+N_\alpha+N_\beta \ ,
\label{eq:chQC_HF_eqs_v1_ROHF_b}
\end{align}
\end{subequations}

where the \emph{ROHF operator} $\hat{F}_a^{\mathrm{ROHF}}[\phi]$ is
defined as

\begin{equation}
\label{eq:chQC_ROHF_op}
\hat{F}_a^{\mathrm{ROHF}}[\phi]:=
 f_a \hat{h}+ \sum_b \bigg( f_a f_b \hat{J}_b[\phi] - 
      g_{ab}\hat{K}_b[\phi] \bigg) \ ,
\end{equation}

and the Hermiticity property of the Lagrange multipliers matrix
$\Lambda^{\mathrm{ROHF}}$ follows from conjugation and subtraction in
eqs.~(\ref{eq:chQC_HF_eqs_v1_ROHF}).

Again, although the Fock operator depends on the index $a$ of the
orbital upon which it operates, this dependence presents a very
particular structure, yielding \emph{only three} different types of
operators:

\begin{subequations}
\label{eq:chQC_ROHF_ops}
\begin{align}
& \hat{F}_D^{\mathrm{ROHF}}[\phi]:=
 2 \hat{h}+ \sum_b \bigg( 2 f_b \hat{J}_b[\phi] - 
      f_{b}\hat{K}_b[\phi] \bigg) \ , \qquad
   a \in D \ ,
\label{eq:chQC_HF_ROHF_ops_a} \\
& \hat{F}_\alpha^{\mathrm{ROHF}}[\phi]:=
 \hat{h}+ \sum_{b \in (D \cup A)} \bigg(\hat{J}_b[\phi] - 
      \hat{K}_b[\phi] \bigg) \ , \qquad
   a \in A \ ,
\label{eq:chQC_HF_ROHF_ops_b} \\
& \hat{F}_\beta^{\mathrm{ROHF}}[\phi]:=
 \hat{h}+ \sum_{b \in (D \cup B)} \bigg( \hat{J}_b[\phi] - 
      \hat{K}_b[\phi] \bigg) \ , \qquad
   a \in B \ .
\label{eq:chQC_HF_ROHF_ops_c}
\end{align}
\end{subequations}

Note that, in the particular case that we had no open-shell orbitals,
the operator for the doubly occupied ones does not reduce (as it
should) to the RHF one in~(\ref{eq:chQC_HF_eqs_vFinal_1_RHF}). This is
because we hid a factor 2 in the definition of the Lagrange
multipliers in the RHF derivation.

At this point, it would be desirable to continue with the same program
that we followed in the UHF and RHF cases and diagonalize the matrix
$\Lambda^{\mathrm{ROHF}}$ in order to arrive to a system of three
pseudoeigenvalue equations, coupled only via the Fock
operators. Nevertheless, this is not possible in ROHF, as we shall
show in the following lines, and it is the root of ROHF being the most
tricky flavour in the Hartree-Fock family. The obstruction to achieve
this diagonalization comes from the fact, already used in the UHF
case, that not every unitary transformation of the
\emph{spin-orbitals} is allowed if we want to remain inside the ROHF
scheme, i.e., if we want that the transformed spin-orbitals satisfy
the same ROHF constraints in~(\ref{eq:chQC_constraints_ROHF}) that the
untransformed ones did.

To begin with, if we recall that the doubly occupied spatial orbitals
must be orthogonal to both the alpha and the beta ones, we can use the
same reasoning as in page~\pageref{eq:QC_ROHF_argument} to show that
the alpha-beta connecting parts of the allowed unitary matrix in this
case must be zero. Hence, we can write

\begin{equation}
\label{eq:chQC_U_ROHF}
U^{\mathrm{ROHF}}:=
\left(
\begin{array}{c@{\hspace{3pt}}|@{\hspace{3pt}}c}
\begin{array}{c@{\hspace{5pt}}c}
U^D & U^{D\alpha} \\[5pt]
U^{\prime D\alpha} & U^\alpha
\end{array}
&
0
\\[5pt]
\hline\\[-8pt]
0
&
\begin{array}{c@{\hspace{5pt}}c}
U^{\prime D} & U^{D\beta} \\[5pt]
U^{\prime D\beta} & U^\beta
\end{array}
\end{array} \right) \ ,
\end{equation}

where the size of each block may be easily found
from~(\ref{eq:chQC_constraints_ROHF}) and the notation is again
self-explanatory.

Now, in order to obtain further restrictions to the form of
$U^{\mathrm{ROHF}}$, we write the transformation of the \emph{two}
spin-orbitals in the closed shell that correspond to \emph{the same}
spatial orbital $\phi_a$, with $a \in A$. To this end, we
use~(\ref{eq:chQC_U_ROHF}), (\ref{eq:chQC_constraints_ROHF}), and the
appropriate lines in the first and third lines of blocks
in~(\ref{eq:chQC_U_ROHF}):

\begin{equation}
\label{eq:chQC_ROHF_proof1}
\phi_a({\bm r}) = \sum_{b \in D} U^D_{ab} \phi^\prime_b({\bm r}) +
         \sum_{b \in A} U^{D \alpha}_{ab} \phi^\prime_b({\bm r}) =
         \sum_{b \in D} U^{\prime D}_{ab} \phi^\prime_b({\bm r}) +
         \sum_{b \in B} U^{D \beta}_{ab} \phi^\prime_b({\bm r}) \ .
\end{equation}

Using the orthogonality relations inside and among the three sets of
spatial orbitals, we multiply the above expression by any $\phi_c({\bm
r})$, with $c \in D$ and integrate on ${\bm r}$. From this, the
equality of $U^D$ and $U^{\prime D}$ follows, and the corresponding
sums in~(\ref{eq:chQC_ROHF_proof1}) subtract to zero, yielding

\begin{equation}
\label{eq:chQC_ROHF_proof2}
\sum_{b \in A} U^{D \alpha}_{ab} \phi^\prime_b({\bm r}) =
\sum_{b \in B} U^{D \beta}_{ab} \phi^\prime_b({\bm r}) \ ,
\end{equation}

which equates one vector in the linear space spanned by the alpha
orbitals to another in the linear space spanned by the beta ones.

However, only the zero vector can belong to both spaces if we want the
ROHF assumptions regarding the spin-orbitals to hold. To see this, we
can use a \emph{reductio ad absurdum} type of argument: Assume that
both sides of~(\ref{eq:chQC_ROHF_proof2}) are different from
zero. Then, we may perform a unitary transformation changing only the
alpha and beta spaces, and with no elements connecting the two. This
is allowed, since it does not change neither the $N$-electron
wavefunction, nor the Fock operators, nor the ROHF constraints.  Now,
if we select the partial unitary transformations in the alpha and beta
sets so that they turn the vectors at both sides
of~(\ref{eq:chQC_ROHF_proof2}) into single elements in the bases of
their respective spaces, we have that a single alpha orbital
\emph{equals} a beta one. Although this can happen in particular
cases, we cannot ask it or we would be changing the fundamental
assumptions made in~(\ref{eq:chQC_constraints_ROHF}). Therefore, both
sides of~(\ref{eq:chQC_ROHF_proof2}) must be zero, and, since the
alpha and beta sets are linearly independent, all coefficients must be
zero too.

The proof that $U^{\prime D \alpha}$ and $U^{\prime D \beta}$ are also
zero is performed using similar arguments, and the final form of
$U^\mathrm{ROHF}$ satisfying all the restrictions reads

\begin{equation}
\label{eq:chQC_U_ROHF2}
U^{\mathrm{ROHF}}=
\left(
\begin{array}{c@{\hspace{3pt}}|@{\hspace{3pt}}c}
\begin{array}{c@{\hspace{5pt}}c}
U^D & 0 \\[5pt]
0 & U^\alpha
\end{array}
& 0
\\[5pt]
\hline\\[-8pt]
0
&
\begin{array}{c@{\hspace{5pt}}c}
U^D & 0 \\[5pt]
0 & U^\beta
\end{array}
\end{array} \right) \ .
\end{equation}

Finally, if we write the associated matrix using the $a$ indices,
i.e., operating on the set of spatial orbitals with the doubly
occupied ones \emph{unrepeated}:

\begin{equation}
\label{eq:chQC_U_ROHF3}
\tilde{U}^{\mathrm{ROHF}}:=
\left(
\begin{array}{c@{\hspace{5pt}}c@{\hspace{5pt}}c}
U^D & & 0 \\[5pt]
 & U^\alpha & \\[5pt]
0 & & U^\beta
\end{array} \right) \ ,
\end{equation}

then, the transformed $\Lambda^{\prime \mathrm{ROHF}}$ is related to
the original one in~(\ref{eq:chQC_lambda_ROHF}) through simple matrix
multiplication: $\Lambda^\prime=\tilde{U}^+\Lambda\tilde{U}$ (dropping
the ROHF superindices). It is clear that such a restricted
$\tilde{U}^{\mathrm{ROHF}}$ does not operate on the off-diagonal
blocks of $\Lambda^{\mathrm{ROHF}}$ in~(\ref{eq:chQC_lambda_ROHF})
and, therefore, \emph{the sought diagonalization is not possible}.

Using the fact that, however, the $\Lambda^D$, $\Lambda^\alpha$ and
$\Lambda^\beta$ do allow to be diagonalized, we can write the final,
simplest possible form of the \emph{ROHF equations} (forgetting their
complex conjugate counterparts):

\begin{subequations}
\label{eq:chQC_HF_eqs_vfinal_ROHF}
\begin{align}
& \hat{F}_D^{\mathrm{ROHF}}[\phi]\phi_{a}({\bm r})=
   \varepsilon^D_a \phi_{a}({\bm r}) +
   \sum_{b \in A}\lambda^{D\alpha}_{ab}\phi_{b}({\bm r}) +
   \sum_{b \in B}\lambda^{D\beta}_{ab}\phi_{b}({\bm r})\ , \quad
   a \in D \ ,
\label{eq:chQC_HF_eqs_vfinal_ROHF_a} \\
& \hat{F}_\alpha^{\mathrm{ROHF}}[\phi]\phi_{a}({\bm r})=
   \varepsilon^\alpha_a \phi_{a}({\bm r}) +
   \sum_{b \in D}(\lambda^{D\alpha}_{ba})^*\phi_{b}({\bm r})\ , \quad
   a \in A \ ,
\label{eq:chQC_HF_eqs_vfinal_ROHF_b} \\
& \hat{F}_\beta^{\mathrm{ROHF}}[\phi]\phi_{a}({\bm r})=
   \varepsilon^\beta_a \phi_{a}({\bm r}) +
   \sum_{b \in D}(\lambda^{D\beta}_{ba})^*\phi_{b}({\bm r})\ , \quad
   a \in B \ ,
\label{eq:chQC_HF_eqs_vfinal_ROHF_c}
\end{align}
\end{subequations}

where the superindices in the matrix elements are only written for
visual convenience when comparing with~(\ref{eq:chQC_lambda_ROHF}).

In this final form, it is evident that the off-diagonal elements of
the Lagrange multipliers matrix, i.e., those related to the
orthogonality constraints between the closed and open shells,
introduce a coupling in the right-hand side of the ROHF equations that
completely spoils the possibility of casting them into
pseudoeigenvalue ones. In the previous UHF and RHF versions,
\emph{all} orthogonality constraints were handled by simply choosing a
special basis in the space spanned by the spatial orbitals in which
the Lagrange multipliers matrix was diagonal. However, in the ROHF
scheme, there is no such basis and we must deal with the problem in a
different way, resulting into higher computational and theoretical
difficulty.

Since the pioneering work by Roothaan \cite{Roo1960RMP}, the solution
of the ROHF problem has been attempted by distinct means, ranging from
directly tackling the ROHF equations
in~(\ref{eq:chQC_HF_eqs_vfinal_ROHF}) explicitly forcing the
orthogonality constraints \cite{Bin1974MP,Pet1992SOFT}, to the
construction of a so-called \emph{unified coupling operator}
\cite{Roo1960RMP,Hir1974JCP,Gue1974MP}, which allows to turn the ROHF
scheme into a single pseudoeigenvalue problem at the price of
introducing certain ambiguities in the one-electron orbital energies
\cite{Pla2006JCP,Car1977BOOK}. The details and subtleties involved in
these issues are beyond the scope of a review of the fundamental
topics such as this one. The interested reader may want to check the
more specialized accounts in
\cite{Coo2005BOOK,Car1977BOOK,Hur1976BOOK,McW1992BOOK}.

\section[The Roothaan-Hall equations]{The Roothaan-Hall equations}
\label{sec:QC_Roothaan}

The Hartree-Fock equations in the RHF form in
expression~(\ref{eq:chQC_HF_eqs_vFinal_RHF}) are a set of $N/2$
coupled integro-differential equations. As such, they can be tackled
by finite-differences methods and solved on a discrete grid; this is
known as \emph{numerical Hartree-Fock} \cite{Kob1997AQC}, and,
given the present power of computers, it is only applicable to very
small molecules.

In order to deal with larger systems, such as biological
macromolecules, independently proposed by Roothaan
\cite{Roo1951RMP} and Hall \cite{HalPRSLA1951} in 1951, a
different kind of discretization must be performed, not in
$\mathbb{R}^3$ but in the Hilbert space $\mathcal{H}$ of the
one-electron orbitals. Hence, although the actual dimension of
$\mathcal{H}$ is infinite, we shall approximate any function in it by
a finite linear combination of $M$ different functions
$\chi_a$\footnote{\label{foot:i_vs_a_roothaan} In all this section and
the next one, the indices belonging to the first letters of the
alphabet, $a, b, c, d$, etc., run from~1 to $M$ (the number of
functions in the finite basis set); whereas those named with capital
letters from $I$ towards the end of the alphabet, $I, J, K, L$, etc.,
run from 1 to $N/2$ (the number of spatial wavefunctions $\phi_I$,
also termed the number of \emph{occupied orbitals}).}. In particular,
the one-electron orbitals that make up the RHF wavefunction, shall be
approximated by

\begin{equation}
\label{eq:chQC_LCAO}
\phi_I({\bm r})\simeq
  \sum^{M}_{a}c_{aI}\,\chi_a({\bm r})\ , \qquad I=1,\ldots,N/2\ , 
  \quad M \geq \frac{N}{2} \ .
\end{equation}

In both cases, numerical Hartree-Fock and discretization of the
function space, the correct result can be only be reached
asymptotically; when the grid is very fine, for the former, and when
$M \to \infty$, for the latter. This exact result, which, in the case
of small systems, can be calculated up to several significant digits,
is known as the \emph{Hartree-Fock limit} \cite{Jen2005TCA}.

In practical cases, however, $M$ is finite (often, only about an order
of magnitude larger than $N/2$) and the set $\{\chi_a\}_{a=1}^M$ in
the expression above is called the \emph{basis set}. We shall devote
the next section to discuss its special characteristics, but, for now,
it suffices to say that, in typical applications, the functions
$\chi_a$ are \emph{atom-centred}, i.e., each one of them has
non-negligible value only in the vicinity of a particular
nucleus. Therefore, like all the electronic wavefunctions we have
dealt with in the last sections, they parametrically depend on the
positions $\underbar{R}$ of the nuclei (see
sec.~\ref{sec:QC_Born-Oppenheimer}). This is why, sometimes, the
functions $\chi_a$ are called \emph{\underline{a}tomic
\underline{o}rbitals}\footnote{\label{foot:noAO} Some authors
\cite{Jen1998BOOK} suggest that, being strict, the term
\emph{atomic orbitals} should be reserved for the one-electron
wavefunctions $\phi_I$ that are the solution of the Hartree-Fock
problem (or even to the exact Schr\"odinger equation of the isolated
atom), and that the elements $\chi_a$ in the basis set should be
termed simply \emph{localized functions}. However, it is very common
in the literature not to follow this recommendation and choose the
designation that appear in the text
\cite{Pop1999RMP,Jen1998BOOK,Roo1951RMP}. We shall do the
same for simplicity.}~(AO) (since they are localized at individual
atoms), the $\phi_I$ are referred to as \emph{\underline{m}olecular
\underline{o}rbitals}~(MO) (since they typically have non-negligible
value in the whole space occupied by the molecule), and the
approximation in eq.~(\ref{eq:chQC_LCAO}) is called
\emph{\underline{l}inear \underline{c}ombination of \underline{a}tomic
\underline{o}rbitals}~(LCAO). In addition, since we voluntarily
circumscribe to real-RHF, we assume that both the coefficients
$c_{aI}$ and the functions $\chi_a$ in the above expression are real.

Now, if we introduce the linear combination in
eq.~(\ref{eq:chQC_LCAO}) into the Hartree-Fock equations
in~(\ref{eq:chQC_HF_eqs_vFinal_RHF}), multiply the result from the
left by $\chi_b$ (for a general value of $b$) and integrate on ${\bm
r}$, we obtain

\begin{equation}
\label{eq:chQC_RH_1}
\sum_a F_{ba}c_{aI}=\varepsilon_I\sum_a S_{ba}c_{aI}\ ,\qquad I=1,\ldots,N/2\ ,
\ b=1,\ldots,M \ ,
\end{equation}

where we denote by $F_{ba}$ the $(b,a)$-element of the \emph{Fock
matrix}\footnote{\label{foot:no_RHF} Note that the RHF superindex has
been dropped from $F$.}, and by $S_{ba}$ the one of the \emph{overlap
matrix}, defined as

\begin{equation}
\label{eq:chQC_matrices_1}
F_{ba}:=\langle \chi_b | \, \hat{F}[\phi] \, | \chi_a \rangle
\qquad \mathrm{and} \qquad
S_{ba}:=\langle \chi_b | \chi_a \rangle \ ,
\end{equation}

respectively.

Note that we do not ask the $\chi_a$ in the basis set to be mutually
orthogonal, so that the overlap matrix is not diagonal in general.

Next, if we define the $M\times M$ matrices $F[c]:=(F_{ab})$ and
$S:=(S_{ab})$, together with the (column) $M$-vector $c_I:=(c_{aI})$,
we can write eq.~(\ref{eq:chQC_RH_1}) in matricial form:

\begin{equation}
\label{eq:chQC_RH_2}
F[c] c_I=\varepsilon_I\,Sc_{I}\ .
\end{equation}

Hence, using the LCAO approximation, we have traded the $N/2$ coupled
integro-differential Hartree-Fock equations
in~(\ref{eq:chQC_HF_eqs_vFinal_RHF}) for this system of $N/2$
algebraic equations for the $N/2$ orbital energies $\varepsilon_I$ and
the $M \cdot N/2$ coefficients~$c_{aI}$, which are called
\emph{Roothaan-Hall equations} \cite{Roo1951RMP,HalPRSLA1951}
and which are manageable in a computer.

Now, if we forget for a moment that the Fock matrix depends on the
coefficients $c_{aI}$ (as stressed by the notation~$F[c]$) and also
that we are only looking for $N/2$ vectors $c_I$ while the matrices
$F$ and $S$ are $M \times M$, we may regard the above expression as a
$M$-dimensional \emph{generalized eigenvalue problem}. Many properties
are shared between this kind of problem and a classical eigenvalue
problem (i.e., one in which $S_{ab}=\delta_{ab}$)
\cite{Roo1951RMP}, being the most important one that, due to the
Hermiticity of $F[c]$, one can find an orthonormal set of $M$ vectors
$c_a$ corresponding to real eigenvalues $\varepsilon_a$ (where, of
course, some eigenvalue could be repeated).

In fact, it is using this formalism how most of actual Hartree-Fock
computations are performed, although the reader must also note
that other approaches, in which the orthonormality constraints are
automatically satisfied due to the choice of variables also exist in
the literature \cite{Hea1988JPC}. The general outline of the
iterative procedure is essentially the same as the one discussed in
sec.~\ref{sec:QC_Hartree-Fock}: Choose a \emph{starting-guess} for the
coefficients $c_{aI}$ (let us denote it by~$c^0_{aI}$), construct the
corresponding Fock matrix
$F[c^0]$\footnote{\label{foot:F_onlydependsonoccupied} Note (in
eq.~(\ref{eq:chQC_matrices_2}), for example) that the Fock matrix only
depends on the vectors $c_a$ with $a \le N/2$.} and solve the
generalized eigenvalue problem in eq.~(\ref{eq:chQC_RH_2}). By virtue
of the aufbau principle discussed in the previous section, from the
$M$ eigenvectors $c_a$, keep the $N/2$ ones $c^1_I$ that correspond to
the $N/2$ lowest eigenvalues~$\varepsilon^1_I$, construct the new Fock
matrix $F[c^1]$ and iterate (by convention, the eigenvalues
$\varepsilon_a^n$, for all $n$, are ordered from the lowest to the
largest as $a$ runs from 1 to $M$). This procedure ends when the
$n$-th solution is close enough (in a suitable defined way) to the
$(n-1)$-th one. Also, note that, after convergence has been achieved,
we end up with $M$ orthogonal vectors $c_{a}$. Only the $N/2$ ones
that correspond to the lowest eigenvalues represent real one-electron
solutions and they are called \emph{occupied orbitals}; the $M-N/2$
remaining ones do not enter in the $N$-electron wavefunction (although
they are relevant for calculating corrections to the Hartree-Fock
results) and they are called \emph{virtual orbitals}.

Regarding the mathematical foundations of this procedure, let us
stress, however, that, whereas in the finite-dimensional GHF and UHF
cases it has been proved that the analogous of Lieb and Simon's
theorem (see the previous section) is satisfied, i.e., that the global
minimum of the original optimization problem corresponds to the lowest
eigenvalues of the self-consistent Fock operator, in the RHF and ROHF
cases, contrarily, no proof seems to exist \cite{Can2003BOOK}. Of
course, in practical applications, the positive result is assumed to
hold.

Finally, if we expand $F_{ab}$ in eq.~(\ref{eq:chQC_matrices_1}),
using the shorthand $|a\rangle$ for $|\chi_a \rangle$, we have

\begin{equation}
\label{eq:chQC_matrices_2}
F_{ab}=\langle a|\,\hat{h}\,|b\rangle + \sum_{c,d} \underbrace{\left( \sum_J 
 c_{cJ}c_{dJ} \right)}_{\displaystyle D_{cd}[c]}
 \underbrace{\left( 2 \langle ac|\,\frac{1}{r}\,|bd\rangle -
                    \langle ac|\,\frac{1}{r}\,|db\rangle\right)}_{
                    \displaystyle G^{cd}_{ab}}  \ ,
\end{equation}

where we have introduced the \emph{density matrix} $D_{cd}[c]$, and
also the matrix $G^{cd}_{ab}$, made up by the \emph{two-electron
four-centre integrals} $\langle ac|\,1/r\,|bd\rangle$ (also called
\emph{\underline{e}lectron \underline{r}epulsion
\underline{i}ntegrals} (ERIs)) defined by

\begin{equation}
\label{eq:chQC_twoe_integrals}
\langle ac|\,\frac{1}{r}\,|bd\rangle :=
\int\!\!\!\!\int
 \frac{\chi_{a}({\bm r})\,\chi_{c}({\bm r}^{\prime})\,
           \chi_{b}({\bm r})\,\chi_{d}({\bm r}^{\prime})}
	 {|{\bm r}-{\bm r}^{\prime}|}\,\mathrm{d}{\bm r}
   \,\mathrm{d}{\bm r}^{\prime} \ .
\end{equation}

It is also convenient to introduce the \emph{Coulomb} ($J_{ab}[c]$)
and \emph{exchange} ($K_{ab}[c]$) \emph{matrices} 

\begin{subequations}
\label{eq:chQC_coulomb_exchange}
\begin{align}
& J_{ab}[c] := \sum_{c,d}D_{cd}[c]\langle ac|\,\frac{1}{r}\,|bd\rangle
  \ , \label{eq:chQC_coulomb_exchange_a} \\
& K_{ab}[c] := \sum_{c,d}D_{cd}[c]\langle ac|\,\frac{1}{r}\,|db\rangle
  \ , \label{eq:chQC_coulomb_exchange_b} 
\end{align}
\end{subequations}

in terms of which, the Fock operator in eq.~(\ref{eq:chQC_matrices_2})
may be expressed as 

\begin{equation}
\label{eq:chQC_matrices_3}
F_{ab}=\langle a|\,\hat{h}\,|b\rangle + 2J_{ab}[c] - K_{ab}[c] \ .
\end{equation}

After SCF convergence has been achieved, the \emph{RHF energy} in the
finite-dimensional case can be computed using the discretized version
of eq.~(\ref{eq:chQC_energy_RHF}):

\index{energy!RHF}
\index{RHF!energy}

\begin{eqnarray}
\label{eq:chQC_energy-finite_RHF}
&& E=2\sum_I^{N/2} \varepsilon_I - \sum_{I,J}^{N/2}\sum_{a,b,c,d}
    \Bigg( 2c_{aI}c_{bJ}c_{cI}c_{dJ}
        \langle ab|\,\frac{1}{r}\,|cd\rangle  -
        c_{aI}c_{bJ}c_{cJ}c_{dI}
        \langle ab|\,\frac{1}{r}\,|dc\rangle \Bigg) = \nonumber \\
&& \qquad 2\sum_I^{N/2} \varepsilon_I -\sum_{I,J}^{N/2}\sum_{a,b,c,d}
c_{aI}c_{bJ}c_{cI}c_{dJ}\langle ab|\,\frac{1}{r}\,|cd\rangle =
\nonumber \\
&& \qquad 2\sum_i^{N/2} \varepsilon_i - \sum_{a,b,c,d}
 D_{ac}[c]D_{bd}[c]\langle ab|\,\frac{1}{r}\,|cd\rangle\ ,
\end{eqnarray}

where a convenient rearrangement of the indices in the two sums
has been performed from the first to the second line.

\section[Gaussian basis sets]{Introduction to Gaussian basis sets}
\label{sec:QC_basis_sets}

In principle, arbitrary functions may be chosen as the $\chi_a$ to
solve the Roothaan-Hall equations in the previous section, however, in
eq.~(\ref{eq:chQC_matrices_2}), we see that one of the main numerical
bottlenecks in SCF calculations arises from the necessity of
calculating the $O(M^4)$ four-centre
integrals\footnote{\label{foot:exact_ERIs} If the symmetry
    properties of the integrals are used, the precise number of ERIs
    is found to be $\frac{1}{8}M(M+1)(M^2+M+2)$ \cite{Bra1971IJQC}.}
$\langle ab|\,\frac{1}{r}\,|cd\rangle$, since the solution of the
generalized eigenvalue problem in eq.~(\ref{eq:chQC_RH_2}) typically
scales only like $O(M^3)$, and there are \mbox{$O(M^2)$} two-centre
$\langle a|\,\hat{h}\,|b\rangle$ integrals (see however the next
section). Either if these integrals
are calculated at each iterative step and directly taken from RAM
memory (\emph{direct SCF}) or if they are calculated at the first
step, written to disk, and then read from there when needed
(\emph{conventional SCF}), an appropriate choice of the
functions~$\chi_a$ in the finite basis set is essential if accurate
results are sought, $M$ is intended to be kept as small as possible
and the integrals are wanted to be computed rapidly. When one moves
into higher-level theoretical descriptions and the numerical
complexity scales with $M$ even more unpleasantly, the importance of
this choice greatly increases.

In this section, in order to support that study, we shall introduce
some of the concepts involved in the interesting field of basis-set
design. For further details not covered here, the reader may want to
check
refs.~\cite{Gar2003BOOK,Jen1998BOOK,Sza1996BOOK,Hel1995BOOK}.

The only analytically solvable molecular problem in non-relativistic
quantum mechanics is the \emph{hydrogen-like atom}, i.e., the system
formed by a nucleus of charge $Z$ and only one electron (H, He$^+$,
Li$^{2+}$, etc.). Therefore, it is not strange that all the thinking
about atomic-centred basis sets in quantum chemistry is much
influenced by the particular solution to this problem.

The spatial eigenfunctions of the Hamiltonian operator of an
hydrogen-like atom, in atomic units and spherical coordinates,
read\footnote{\label{foot:BO_hydrogen-like} For consistency with the
rest of the text, the Born-Oppenheimer approximation has been also
assumed here. So that the \emph{reduced mass} $\mu:=m_e M_N / (m_e +
M_N)$ that should enter the expression is considered to be the mass of
the electron $\mu \simeq m_e$ (recall that, in atomic units, $m_e=1$
and $M_N \gtrsim 2000$).}

\begin{equation}
\label{eq:chQC_hydrogen-like}
\phi_{nlm}(r,\theta,\varphi)=\sqrt{\left(\frac{2Z}{n}\right)^3 
                                   \frac{(n-l-1)!}{2n[(n+l)!]^3}}
 \left( \frac{2Z}{n} r \right)^l
 L^{2l+1}_{n-l-1}\left( \frac{2Z}{n} r \right)
 e^{-Zr/n} Y_{lm}(\theta,\varphi) \ ,
\end{equation}

where $n$, $l$ and $m$ are the energy, total angular momentum and
$z$-angular momentum quantum numbers, respectively. Their ranges of
variation are coupled: all being integers, $n$ runs from 1 to
$\infty$, $l$ from 0 to $n-1$ and $m$ from $-l$ to $l$. The function
$L^{2l+1}_{n-l-1}$ is a \emph{generalized Laguerre polynomial}
\cite{Abr1964BOOK}, for which it suffices to say here that it is of
order $n-l-1$ (thus having, in general, $n-l-1$ zeros), and the
function $Y_{lm}(\theta,\varphi)$ is a \emph{spherical harmonic},
which is a simultaneous eigenfunction of the total angular momentum
operator $\hat{l}^{\,2}$ (with eigenvalue $l(l+1)$) and of its
$z$-component $\hat{l}_z$ (with eigenvalue $m$).

The hope that the one-electron orbitals that are the solutions of the
Hartree-Fock problem in many-electron atoms could not be very
different from the $\phi_{nlm}$ \label{page:foot:orbitals_only_in_HF}
above\footnote{\label{foot:orbitals_only_in_HF} Note that the
$N$-electron wavefunction of the exact ground-state of a
non-hydrogen-like atom depends on 3$N$ spatial variables in a way that
cannot be written, in general, as a Slater determinant of one-electron
functions. The image of single electrons occupying definite orbitals,
together with the possibility of comparing them with the one-particle
eigenfunctions of the Hamiltonian of hydrogen-like atoms, vanishes
completely outside the Hartree-Fock formalism.}, together with the
powerful chemical intuition that states that `atoms-in-molecules are
not very different from atoms-alone', is what mainly drives the choice
of the functions $\chi_a$ in the basis set, and, in the end, the
variational procedure that will be followed is expected to fix the
largest failures coming from these too-simplistic assumptions.

Hence, it is customary to choose functions that are centred at atomic
nuclei and that partially resemble the exact solutions for
hydrogen-like atoms. In this spirit, the first type of AOs to be tried
\cite{Pop1999RMP} were the
\emph{\underline{S}later-\underline{t}ype \underline{o}rbitals}
(STOs), proposed by Slater \cite{Sla1930PRb} and Zener
\cite{Zen1930PR} in 1930:

\begin{equation}
\label{eq:chQC_STO}
\chi^{\mathrm{STO}}_a({\bm r}\,;{\bm R}_{\alpha_a}):=
 \mathcal{N}^{\mathrm{STO}}_a \widetilde{Y}^{c,s}_{l_a
 m_a}(\theta_{\alpha_a},\varphi_{\alpha_a})\,
 \big|{\bm r}-{\bm R}_{\alpha_a}\big|^{\,n_a-1} \exp \Big( - \zeta_a
 \,|{\bm r}-{\bm R}_{\alpha_a}| \Big) \ ,
\end{equation}

where $\mathcal{N}^{\mathrm{STO}}_a$ is a normalization constant and
$\zeta_a$ is an adjustable parameter. The index~$\alpha_a$ is that of
the nucleus at which the function is centred, and, of course, in the
majority of cases, there will be several $\chi^{\mathrm{STO}}_a$
corresponding to different values of $a$ centred at the same
nucleus. The integers $l_a$ and $m_a$ can be considered quantum
numbers, since, due to the fact that the only angular dependence is in
$\widetilde{Y}^{c,s}_{l_a m_a}$ (see below for a definition), the STO
defined above is still a simultaneous eigenstate of the one-electron
angular momentum operators $\hat{l}^{\,2}$ and~$\hat{l}_z$ (with the
origin placed at ${\bm R}_{\alpha_a}$). The parameter $n_a$, however,
should be regarded as a `principal (or energy) quantum number' only by
analogy, since, on the one hand, it does not exist a `one-atom
Hamiltonian' whose exact eigenfunctions it could label and, on the
other hand, only the leading term of the Laguerre polynomial in
eq.~(\ref{eq:chQC_hydrogen-like}) has been kept in the
STO\footnote{\label{foot:basis_STO} If we notice that, within the set
of all possible STOs (as defined in eq.~(\ref{eq:chQC_STO})), every
hydrogen-like energy eigenfunction (see
eq.~(\ref{eq:chQC_hydrogen-like})) can be formed as a linear
combination, we easily see that the STOs constitute a complete basis
set. This is important to ensure that the Hartree-Fock limit could be
actually approached by increasing $M$.}.

Additionally, in the above notation, the fact that
$\chi^{\mathrm{STO}}_a$ parametrically depends on the position of a
certain $\alpha_a$-th nucleus has been stressed, and the functions
$\widetilde{Y}^{c,s}_{l_a m_a}$, which are called \emph{real spherical
harmonics} \cite{Mat2002IJQC} (remember that we want to do real
RHF), are defined in terms of the classical \emph{spherical harmonics}
$Y_{l_a m_a}$ by

\begin{subequations}
\label{eq:chQC_real_spherical_harmonics}
\begin{align}
& \widetilde{Y}^{c}_{l_a m_a}(\theta_{\alpha_a},\varphi_{\alpha_a}):=
  \frac{Y_{l_a m_a} + 
  Y^{*}_{l_a m_a}}{\sqrt{2}} \varpropto
  P_{l_a}^{m_a}(\cos \theta_{\alpha_a})\cos (m_a \varphi_{\alpha_a}) \ ,
  \label{eq:chQC_real_spherical_harmonics_a} \\
& \widetilde{Y}^{s}_{l_a m_a}(\theta_{\alpha_a},\varphi_{\alpha_a}):=
  -i \frac{Y_{l_a m_a} - 
  Y^{*}_{l_a m_a}}{\sqrt{2}} \varpropto
  P_{l_a}^{m_a}(\cos \theta_{\alpha_a})\sin (m_a \varphi_{\alpha_a}) \ ,
  \label{eq:chQC_real_spherical_harmonics_b}
\end{align}
\end{subequations}

where $c$ stands for \emph{cosine}, $s$ for \emph{sine}, the functions
$P_{l_a}^{m_a}$ are the \emph{associated Legendre polynomials}
\cite{Abr1964BOOK}, and the spherical coordinates
$\theta_{\alpha_a}$ and $\varphi_{\alpha_a}$ also carry the
$\alpha_a$-label to remind that the origin of coordinates in terms of
which they are defined is located at ${\bm R}_{\alpha_a}$. Also note
that, using that $\widetilde{Y}^{c}_{l_a 0}=\widetilde{Y}^{s}_{l_a
0}$, there is the same number of real spherical harmonics as of
classical ones.

\begin{figure}
\centerline{
\epsfxsize=11cm
\epsfbox{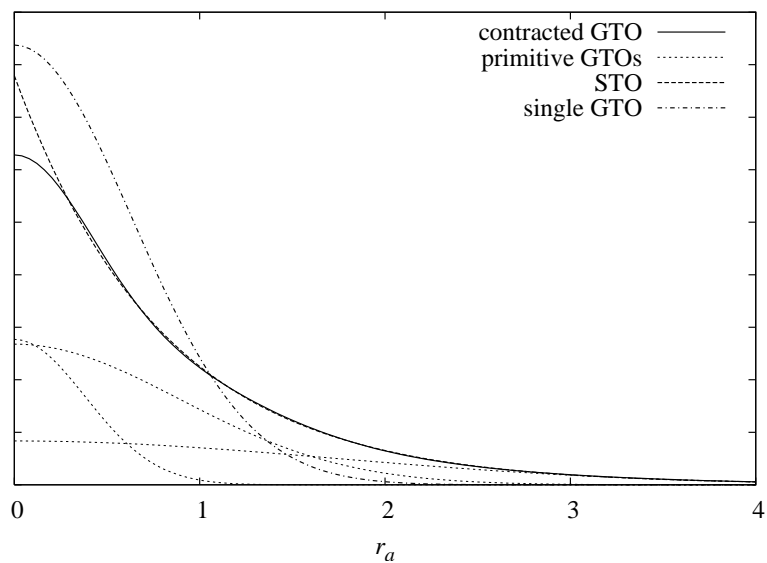}
}
\caption{\label{fig:contraction} Radial behaviour of the
1s-contracted GTO of the hydrogen atom in the STO-3G basis set
\cite{Heh1969JCP}, the three primitive GTOs that form it, the STO
that is meant to be approximated and a single GTO with the same norm
and exponent as the STO. The notation $r_a$ is shorthand for the
distance to the $\alpha_a$-th nucleus
$|{\bm r}-{\bm R}_{\alpha_a}|$.}
\end{figure}

These $\chi^{\mathrm{STO}}_a$ have some good physical
properties. Among them, we shall mention that, for
$|{\bm r}-{\bm R}_{\alpha_a}| \to 0$, they present a \emph{cusp} (a
discontinuity in the radial derivative), as required by Kato's theorem
\cite{Kat1957CPAM}; and also that they decay at an exponential rate
when $|{\bm r}-{\bm R}_{\alpha_a}| \to \infty$, which is consistent
with the image that, an electron that is taken apart from the vicinity
of the nucleus must `see', at large distances, an unstructured
point-like charge (see, for example, the STO in
fig.~\ref{fig:contraction}). Finally, the fact that they do not
present radial nodes (due to the aforementioned absence of the
non-leading terms of the Laguerre polynomial in
eq.~(\ref{eq:chQC_hydrogen-like})) can be solved by making linear
combinations of functions with different values
\label{page:foot:r_exponent_Idontcare} of
$\zeta_a$\footnote{\label{foot:r_exponent_Idontcare} This way of
proceeding renders the choice of the exponent carried by the
$|{\bm r}-{\bm R}_{\alpha_a}|$ part ($n_a-1$ in the case of the STO in
eq.~(\ref{eq:chQC_STO})) a rather arbitrary one. As a consequence,
different definitions may be found in the literature and the
particular exponent chosen in actual calculations turns out to be
mostly a matter of computational convenience.}.

Now, despite their being good theoretical candidates to expand the MO
$\phi_I$ that make up the $N$-particle solution of the Hartree-Fock
problem, these STOs have serious computational drawbacks: Whereas the
two-centre integrals (such as $\langle a|\,\hat{h}\,|b\rangle$ in
eq.~(\ref{eq:chQC_matrices_2})) can be calculated analytically, the
four-centre ERIs $\langle ac|\,1/r\,|bd\rangle$ can not
\cite{Pop1999RMP,Jen1998BOOK} if functions like the ones in
eq.~(\ref{eq:chQC_STO}) are used. This fact, which was known as ``the
nightmare of the integrals'' in the first days of computational
quantum chemistry \cite{Pop1999RMP}, precludes the use of STOs in
practical ab initio calculations of large molecules.

A major step to overcome these difficulties that has revolutioned the
whole field of quantum chemistry \cite{Can2003BOOK,Pop1999RMP}
was the introduction of \emph{\underline{C}artesian
\underline{G}aussian-\underline{t}ype \underline{o}rbitals} (cGTO):

\begin{eqnarray}
\label{eq:chQC_GTO}
\lefteqn{\chi^{\mathrm{cGTO}}_a({\bm r}\,;{\bm R}_{\alpha_a}):=}\nonumber\\
&& \!\!\!\! \mathcal{N}^{\mathrm{cGTO}}_a
 \Big(r^1-R^1_{\alpha_a}\Big)^{l^{\,x}_a}
 \Big(r^2-R^2_{\alpha_a}\Big)^{l^{\,y}_a}
 \Big(r^3-R^3_{\alpha_a}\Big)^{l^{\,z}_a}\,
 \exp \Big( - \zeta_a \,|{\bm r}-{\bm R}_{\alpha_a}|^2 \Big) \ ,
\end{eqnarray}

where the $r^p$ and the $R^p_{\alpha_a}$, with $p=1,2,3$, are the
Euclidean coordinates of the electron and the $\alpha_a$-th nucleus
respectively, and the integers $l^{\,x}_a$, $l^{\,y}_a$ and
$l^{\,z}_a$, which take values from 0 to $\infty$, are called
\emph{orbital quantum numbers}\footnote{\label{foot:basis_GTO} Since
the harmonic-oscillator energy eigenfunctions can be constructed as
linear combinations of Cartesian GTOs, we have that the latter
constitute a complete basis set and, like in the case of the STOs, we
may expect that the Hartree-Fock limit is approached as $M$ is
increased.}.

Although these GTOs do not have the good physical properties of the
STOs (compare, for example, the STO and the GTO in
fig.~\ref{fig:contraction}), in 1950, Boys \cite{Boy1950PRSLA}
showed that all the integrals appearing in SCF theory could be
calculated analytically if the $\chi_a$ had the form in
eq.~(\ref{eq:chQC_GTO}). The enormous computational advantage that
this entails makes possible to use a much larger number of functions
to expand the one-electron orbitals $\phi_i$ if GTOs are used,
partially overcoming their bad short- and long-range behaviour and
making the Gaussian-type orbitals the universally preferred choice in
SCF calculations \cite{Jen1998BOOK}.

To remedy the fact that the angular behaviour of the Cartesian GTOs in
eq.~(\ref{eq:chQC_GTO}) is somewhat hidden, they may be linearly
combined to form \emph{\underline{S}pherical
\underline{G}aussian-\underline{t}ype \underline{o}rbitals} (sGTO):

\begin{equation}
\label{eq:chQC_sGTO}
\chi^{\mathrm{sGTO}}_a({\bm r}\,;{\bm R}_{\alpha_a}):=
 \mathcal{N}^{\mathrm{sGTO}}_a
 \widetilde{Y}^{c,s}_{l_a m_a}(\theta_{\alpha_a},\varphi_{\alpha_a})\,
 \big|{\bm r}-{\bm R}_{\alpha_a}\big|^{\,l_a}
 \exp \Big( - \zeta_a \,|{\bm r}-{\bm R}_{\alpha_a}|^2 \Big) \ ,
\end{equation}

which are proportional to the real spherical harmonic
$\widetilde{Y}^{c,s}_{l_a m_a}(\theta_{\alpha_a},\varphi_{\alpha_a})$, and to
which the same remarks made in footnote~\ref{foot:r_exponent_Idontcare} in
page~\pageref{page:foot:r_exponent_Idontcare} for the STOs, regarding the
exponent in the $|{\bm r}-{\bm R}_{\alpha_a}|$ part, may be applied.

The fine mathematical details about the linear combination that
relates the Cartesian GTOs to the spherical ones are beyond the scope
of this introduction. We refer the reader to
refs.~\cite{Sch1995IJQC} and~\cite{Mat2002IJQC} for
further information and remark here some points that will have
interest in the subsequent discussion.

First, the cGTOs that are combined to make up a sGTO must have all the
same value of $l_a:=l^{\,x}_a+l^{\,y}_a+l^{\,z}_a$ and, consequently,
this sum of the three orbital quantum numbers $l^{\,x}_a$, $l^{\,y}_a$
and $l^{\,z}_a$ in a particular Cartesian GTO is typically (albeit
dangerously) referred to as the \emph{angular momentum} of the
function. In addition, apart from the numerical value of $l_a$, the
spectroscopic notation is commonly used in the literature, so that
cGTOs with $l_a=0,1,2,3,4,5,\ldots$ are called \emph{s}, \emph{p},
\emph{d}, \emph{f}, \emph{g}, \emph{h}, \dots, respectively. Where the
first four come from the archaic words \emph{\underline{s}harp},
\emph{\underline{p}rincipal}, \emph{\underline{d}iffuse} and
\emph{\underline{f}undamental}, while the subsequent ones proceed in
alphabetical order.

Second, for a given $l_a>1$, there are more Cartesian GTOs
($(l_a+1)(l_a+2)/2$) than spherical ones ($2l_a+1$), in such a way
that, from the $(l_a+1)(l_a+2)/2$ functionally independent linear
combinations that can be formed using the cGTOs of angular
momentum~$l_a$, only the angular part of $2l_a+1$ of them turns out to
be proportional to a real spherical harmonic $\widetilde{Y}^{c,s}_{l_a
m_a}(\theta_{\alpha_a},\varphi_{\alpha_a})$; the rest of them are
proportional to real spherical harmonic functions with a different
value of the angular momentum quantum number. For example, from the
six different d-Cartesian GTOs, whose polynomial parts are $x^2$,
$y^2$, $z^2$, $xy$, $xz$ and $yz$ (using an evident, compact
notation), only five different spherical GTOs can be constructed: the
ones with polynomial parts proportional to $2z^2-x^2-y^2$, $xz$, $yz$,
$x^2-y^2$ and $xy$ \cite{Sch1995IJQC}. Among these new sGTOs,
which, in turn, are proportional (neglecting also powers of $r$, see
footnote~\ref{foot:r_exponent_Idontcare} in
page~\pageref{page:foot:r_exponent_Idontcare}) to the real spherical
harmonics $\widetilde{Y}_{2 0}$, $\widetilde{Y}^{c}_{2 1}$,
$\widetilde{Y}^{s}_{2 1}$, $\widetilde{Y}^{c}_{2 2}$ and
$\widetilde{Y}^{s}_{2 2}$, the linear combination $x^2+y^2+z^2$ is
missing, since it presents the angular behaviour of an s-orbital
(proportional to $\widetilde{Y}_{0 0}$).

Finally, let us remark that, whereas Cartesian GTOs in
eq.~(\ref{eq:chQC_GTO}) are easier to be coded in computer
applications than sGTOs\cite{Sch1995IJQC}, it is commonly accepted
that these spurious spherical orbitals of lower angular momentum that
appear when cGTOs are used do not constitute efficient choices to be
included in a basis set \cite{Mat2002IJQC} (after all, if we wanted
an additional s-function, why include it in such an indirect and
clumsy way instead of just designing a specific one that suits our
particular needs?). Consequently, the most common practice in the
field is to use Cartesian GTOs removing from the basis sets the linear
combinations such as the $x^2+y^2+z^2$ above.

Now, even if the integrals involving cGTOs can be computed
analytically, there are still $O(M^4)$ of them in a SCF
calculation. For example, in the model dipeptide HCO-{\small
L}-Ala-NH$_2$, which is commonly used to mimic an alanine residue in a
protein
\cite{Ech2006JCCb,Ech2006JCCa,Rod1998JMS,Yu2001JMS}, there
are 62 electrons and henceforth 31 RHF spatial orbitals $\phi_I$. If a
basis set with only 31 functions is used (this is a lower bound that
will be rarely reached in practical calculations due to symmetry
issues, see below), near a million of four-centre $\langle
ac|\,1/r\,|bd\rangle$ integrals must be computed. This is why, one
must use the freedom that remains once the decision of sticking to
cGTOs has been taken (namely, the choice of the \emph{exponents}
$\zeta_a$ and the \emph{angular momentum} $l_a$) to design basis sets
that account for the relevant behaviour of the systems studied while
keeping $M$ below the `pain threshold'.

The work by Nobel Prize John Pople's group has been a major reference
in this discipline, and their STO-$n$G family \cite{Heh1969JCP},
together with the split-valence Gaussian basis sets, 3-21G, 4-31G,
6-31G,
etc. \cite{Dit1971JCP,Heh1972JCP,Har1973TCHA,Fri1984JCP,Kri1980JCP,Bin1980JACS,Spi1982JCC,Cla1983JCC},
shall be used here to exemplify some relevant issues. However,
note that most of the concepts introduced are also applicable
to more modern basis sets, such as those by Dunning
\cite{Dun1989JCP}.

To begin with, let us recall that the short- and long-range behaviour
of the Slater-type orbitals in eq.~(\ref{eq:chQC_STO}) is better than
that of the more computationally efficient GTOs. In order to improve
the physical properties of the latter, it is customary to linearly
combine $M_a$ Cartesian GTOs, denoted now by $\xi^{\,\mu}_a$
($\mu=1,\ldots,M_a$), and termed \emph{\underline{p}rimitive
\underline{G}aussian-\underline{t}ype \underline{o}rbitals} (PGTO),
having the same atomic centre ${\bm R}_{\alpha_a}$, the same set of
orbital quantum numbers, $l^{\,x}_a$, $l^{\,y}_a$ and $l^{\,z}_a$, but
different exponents $\zeta_a^{\,\mu}$, to make up a
\emph{\underline{c}ontracted \underline{G}aussian-\underline{t}ype
\underline{o}rbitals} (CGTO), defined by

\begin{eqnarray}
\label{eq:chQC_contracted_GTO}
\lefteqn{\chi_a({\bm r}\,;{\bm R}_{\alpha_a}):= \sum_\mu^{M_a} g^{\,\mu}_a 
          \xi^{\,\mu}_a({\bm r}\,;{\bm R}_{\alpha_a}) = }\nonumber \\
&& \!\!\!\!\!\!\!\!\!\!\!\!\Big(r^1-R^1_{\alpha_a}\Big)^{l^{\,x}_a}
   \Big(r^2-R^2_{\alpha_a}\Big)^{l^{\,y}_a}
   \Big(r^3-R^3_{\alpha_a}\Big)^{l^{\,z}_a}
 \sum_\mu^{M_a} g^{\,\mu}_a \mathcal{N}_a^{\,\mu}
 \exp \Big( - \zeta^{\,\mu}_a \,|{\bm r}-{\bm R}_{\alpha_a}|^2 \Big) \ ,
\end{eqnarray}

where the normalization constants $\mathcal{N}_a^{\,\mu}$ have been
kept inside the sum because they typically depend on
$\zeta^{\,\mu}_a$. Also, we denote now by $M_C$ the number of
contracted GTOs and, by $M_P:=\sum_a M_a$, the number of primitive
ones.

In the STO-$n$G family of basis sets, for example, $n$ primitive GTOs
are used for each contracted one, fitting the coefficients
$g^{\,\mu}_a$ and the exponents $\zeta_a^{\,\mu}$ to resemble the
radial behavior of Slater-type orbitals \cite{Heh1969JCP}. In
fig.~\ref{fig:contraction}, the 1s-contracted GTO (see the discussion
below) of the hydrogen atom in the STO-3G basis set is depicted,
together with the three primitive GTOs that form it and the STO that
is meant to \label{page:foot:basisform} be
approximated\footnote{\label{foot:basisform}
Basis sets were obtained from the Extensible Computational Chemistry
Environment Basis Set Database at
{\texttt{http://www.emsl.pnl.gov/forms/basisform.html}}, Version
02/25/04, as developed and distributed by the Molecular Science
Computing Facility, Environmental and Molecular Sciences Laboratory
which is part of the Pacific Northwest Laboratory, P.O. Box 999,
Richland, Washington 99352, USA, and funded by the U.S. Department of
Energy. The Pacific Northwest Laboratory is a multi-program laboratory
operated by Battelle Memorial Institute for the U.S. Department of
Energy under contract DE-AC06-76RLO 1830. Contact Karen Schuchardt for
further information.}. We can see that the contracted GTO has a very
similar behavior to the STO in a wide range of distances, while the
single GTO that is also shown in the figure (with the same norm and
exponent as the STO) has not.

Typically, the fitting procedure that leads to contracted GTOs is
performed on isolated atoms and, then, the already mentioned chemical
intuition that states that `atoms-in-molecules are not very different
from atoms-alone' is invoked to keep the linear combinations fixed
from there on. Obviously, better results would be obtained if the
contraction coefficients were allowed to vary. Moreover, the number of
four-centre integrals that need to be calculated depends on the number
of \emph{primitive} GTOs (like $O(M_P^4)$), so that we have not
gained anything on this point by contracting. However, the size of the
variational space is $M_C$ (i.e., the number of \emph{contracted}
GTOs), in such a way that, once the integrals $\langle
ac|\,1/r\,|bd\rangle$ are calculated (for non-direct SCF), all
subsequent steps in the iterative self-consistent procedure scale as
powers of $M_C$. Also, the disk storage (again, for non-direct
schemes) depends on the number of \emph{contracted} GTOs and, frequently,
it is the disk storage and not the CPU time the limiting factor of a
calculation.

An additional chemical concept that is usually defined in this
context and that is needed to continue with the discussion is that of
\emph{shell}: Atomic shells in quantum chemistry are defined
analogously to those of the hydrogen atom, so that each electron is
regarded as `filling' the multi-electron atom `orbitals' according to
Hund's rules \cite{Woo1983BOOK}. Hence, the \emph{occupied shells}
of carbon, for example, are defined to be 1s, 2s and 2p, whereas those
of, say, silicon, would be 1s, 2s, 2p, 3s an 3p. Each shell may
contain $2(2l+1)$ electrons if complete, where $2l+1$ accounts for the
orbital angular momentum multiplicity and the 2 factor for that of
electron spin.

\begin{table}
\tbl{Exponents $\zeta^{\,\mu}_a$
and contraction coefficients $g^{\,\mu}_a$ of the primitive Gaussian
shells that make up the three different contracted ones in the STO-3G
basis set for carbon (see ref.~\cite{Heh1969JCP} and
footnote~\ref{foot:basisform} in page~\pageref{page:foot:basisform}). The exponents of the 2s- and 2p-shells
are constrained to be the same.}
{\begin{tabular}{r@{\hspace{20pt}}r@{\hspace{40pt}}r@{\hspace{20pt}}r@{\hspace{20pt}}r}
\toprule
 \multicolumn{2}{c}{1s-shell \rule{34pt}{0pt}} & 
 \multicolumn{3}{c}{2sp-shell \rule{10pt}{0pt}} \\
 \multicolumn{1}{c}{$\zeta_a^{\,\mu}$ \rule{7pt}{0pt}} &
 \multicolumn{1}{c}{$g_a^{\,\mu}$     \rule{34pt}{0pt}} &
 \multicolumn{1}{c}{$\zeta_a^{\,\mu}$ \rule{17pt}{0pt}} &
 \multicolumn{1}{c}{$g_a^{\,\mu}$ (s) \rule{13pt}{0pt}} &
 \multicolumn{1}{c}{$g_a^{\,\mu}$ (p) \rule{4pt}{0pt}} \\
\colrule
71.6168370 &  0.15432897 &
 2.9412494 & 0.15591627 & -0.09996723 \\
13.0450960 &  0.53532814 &
 0.6834831 & 0.60768372 &  0.39951283 \\
 3.5305122 &  0.44463454 &
 0.2222899 & 0.39195739 &  0.70115470 \\
\botrule
\end{tabular}}
\label{tab:contraction}
\end{table}

Thus, using these definitions, all the basis sets in the
aforementioned STO-$n$G family are \emph{minimal}; in the sense that
they are made up of only $2l+1$ contracted GTOs for each completely or
partially occupied shell, so that the STO-$n$G basis sets for carbon,
for example, contain two s-type contracted GTOs (one for the 1s- and
the other for the 2s-shell) and three p-type ones (belonging to the
2p-shell). Moreover, due to rotational-symmetry arguments in the
isolated atoms, all the $2l+1$ functions in a given shell are chosen
to have the same exponents and the same contraction coefficients,
differing only on the polynomial that multiplies the Gaussian part.
Such $2l+1$ CGTOs shall be said to constitute a
\emph{\underline{G}aussian \underline{s}hell} (GS), and we shall also
distinguish between the \emph{\underline{p}rimitive} (PGS) and
\emph{\underline{c}ontracted} (CGS) versions.

In table~\ref{tab:contraction}, the exponents $\zeta^{\,\mu}_a$ and the
contraction coefficients $g^{\,\mu}_a$ of the primitive GTOs that make up the
three different shells in the STO-3G basis set for carbon are presented (see
ref.~\cite{Heh1969JCP} and footnote~\ref{foot:basisform} in
page~\pageref{page:foot:basisform}). The fact that the exponents
$\zeta^{\,\mu}_a$ in the 2s- and 2p-shells are constrained to be the same is a
particularity of some basis sets (like this one) which saves some
computational effort and deserves no further attention.

Next, let us introduce a common notation that is used to describe the
contraction scheme: It reads (\emph{primitive shells}) /
[\emph{contracted shells}], or alternatively (\emph{primitive shells})
$\to$ [\emph{contracted shells}]. According to it, the STO-3G basis
set for carbon, for example, is denoted as (6s,3p) $\to$ [2s,1p], or
(6,3) $\to$ [2,1]. Moreover, since for organic molecules it is
frequent to have only hydrogens and the 1st-row atoms C, N and O
(whose occupied shells are identical)\footnote{\label{foot:QC_sulphur}
In proteins, one may also have sulphur in cysteine and methionine
residues.}, the notation is typically extended and the two groups of
shells are separated by a slash; as in (6s,3p/3s) $\to$ [2s,1p/1s] for
STO-3G.

\begin{table}
\tbl{Exponents $\zeta^{\,\mu}_a$
and contraction coefficients $g^{\,\mu}_a$ of the primitive Gaussian
shells that make up the three different constrained ones in the 6-31G
basis set for carbon (see ref.~\cite{Heh1972JCP} and
footnote~\ref{foot:basisform} in page~\pageref{page:foot:basisform}). In the 1s-shell, there is only one
contracted Gaussian shell made by six primitive ones, whereas, in the
2s- and 2p-valence shells, there are two CGSs, one of them made by
three PGSs and the other one only by a single PGS. The exponents of
the 2s- and 2p-shells are constrained to be the same.}
{\begin{tabular}{r@{\hspace{20pt}}r@{\hspace{40pt}}r@{\hspace{20pt}}r@{\hspace{20pt}}r}
\toprule
 \multicolumn{2}{c}{1s-shell \rule{34pt}{0pt}} & 
 \multicolumn{3}{c}{2sp-shell \rule{10pt}{0pt}} \\
 \multicolumn{1}{c}{$\zeta_a^{\,\mu}$ \rule{7pt}{0pt}} &
 \multicolumn{1}{c}{$g_a^{\,\mu}$     \rule{34pt}{0pt}} &
 \multicolumn{1}{c}{$\zeta_a^{\,\mu}$ \rule{17pt}{0pt}} &
 \multicolumn{1}{c}{$g_a^{\,\mu}$ (p) \rule{13pt}{0pt}} &
 \multicolumn{1}{c}{$g_a^{\,\mu}$ (s) \rule{4pt}{0pt}} \\
\colrule
3047.52490 &  0.0018347 &
 7.8682724 & -0.1193324 & 0.0689991 \\
 457.36951 &  0.0140373 &
 1.8812885 & -0.1608542 & 0.3164240 \\
 103.94869 &  0.0688426 &
 0.5442493 &  1.1434564 & 0.7443083 \\
  29.21015 &  0.2321844 &
  &  & \\
  9.286663 &  0.4679413 &
 0.1687144 &  1.0000000 & 1.0000000 \\
  3.163927 &  0.3623120 &
  &  & \\
\botrule
\end{tabular}}
\label{tab:contraction2}
\end{table}

The first improvement that can be implemented on a minimal basis set
such as the ones in the STO-$n$G is the \emph{splitting}, which
consists in including more than one Gaussian shell for each occupied
one. If the splitting is evenly performed, i.e., each shell has the
same number of GSs, then the basis set is called
\emph{\underline{d}ouble \underline{z}eta} (DZ),
\emph{\underline{t}riple \underline{z}eta} (TZ),
\emph{\underline{q}uadruple \underline{z}eta} (QZ),
\emph{\underline{q}uintuple \underline{z}eta} (5Z),
\emph{\underline{s}extuple \underline{z}eta} (6Z), and so on; where
the word \emph{zeta} comes from the Greek letter $\zeta$ used for the
exponents. A hypothetical TZ basis set in which each CGTO is made by,
say, four primitive GTOs, would read (24s,12p/12s) $\to$ [6s,3p/3s] in
the aforementioned notation.

At this point, the already familiar intuition that says that
`atoms-in-molecules are not very different from atoms-alone' must be refined
with another bit of chemical experience and qualified by noticing that `core
electrons are less affected by the molecular environment and the formation of
bonds than valence electrons'\footnote{\label{foot:orbitals_only_in_HF2}
  Recall that, for the very concept of `core' or `valence electrons' (actually
  for any label applied to a single electron) to have any sense, we must be in
  the Hartree-Fock formalism (see footnote~\ref{foot:orbitals_only_in_HF} in
  page~\pageref{page:foot:orbitals_only_in_HF}).}. In this spirit, the above
evenness among different shells is typically broken, and distinct basis
elements are used for the energetically lowest lying (\emph{core}) shells than
for the highest lying (\emph{valence}) ones.

On one side, the contraction scheme may be different. In which case,
the notation used up to now becomes ambiguous, since, for example, the
designation (6s,3p/3s) $\to$ [2s,1p/1s], that was said to correspond
to STO-3G, would be identical for a \emph{different} basis set in
which the 1s-Gaussian shell of heavy atoms be formed by 4 PGSs and the
2s-Gaussian shell by 2~PGSs (in 1st-row atoms, the 2s- and 2p-shells
are defined as valence and the 1s-one as core, while in hydrogen
atoms, the 1s-shell is a valence one). This problem can be solved by
explicitly indicating how many primitive GSs form each contracted one,
so that, for example, the STO-3G basis set is denoted by (33,3/3)
$\to$ [2,1/1], while the other one mentioned would be (42,3/3) $\to$
[2,1/1] (we have chosen to omit the angular momentum labels this
time).

The other point at which the core and valence Gaussian shells may
differ is in their respective `zeta quality', i.e., the basis set may
contain a different number of contracted Gaussian shells in each
case. For example, it is very common to use a single CGS for the core
shells and a multiple splitting for the valence ones. These type of
basis sets are called \emph{split-valence} and the way of naming their
quality is the same as before, except for the fact that a capital V,
standing for \emph{\underline{v}alence}, is added either at the
beginning or at the end of the acronyms DZ, TZ, QZ, etc., thus
becoming VDZ, VTZ, VQZ, etc. or DZV, TZV, QZV, etc.

Pople's 3-21G \cite{Bin1980JACS}, 4-31G \cite{Dit1971JCP}, 6-31G
\cite{Heh1972JCP} and 6-311G \cite{Kri1980JCP} are well-known
examples of split-valence basis sets that are commonly used for SCF
calculations in organic molecules and that present the two
characteristics discussed above. Their names indicate the contraction
scheme, in such a way that the number before the dash represents how
many primitive GSs form the single contracted GSs that is used for
core shells, and the numbers after the dash how the valence shells are
contracted, in much the same way as the notation in the previous
paragraphs. For example, the 6-31G basis set (see
table~\ref{tab:contraction2}), contains one CGS made up of six
primitive GSs in the 1s-core shell of heavy atoms (the 6 before the
dash) and two CGS, formed by three and one PGSs respectively, in the
2s- and 2p-valence shells of heavy atoms and in the 1s-shell of
hydrogens (the 31 after the dash). The 6-311G basis set, in turn, is
just the same but with an additional single-primitive Gaussian shell
of functions in the valence region. Finally, to fix the concepts
discussed, let us mention that, using the notation introduced above,
these two basis sets may be written as (631,31/31) $\to$ [3,2/2] and
(6311,311/311) $\to$ [4,3/3], respectively.

Two further improvements that are typically used and that may also be
incorporated to Pople's split-valence basis sets are the addition of
\emph{polarization} \cite{Har1973TCHA,Fri1984JCP} or
\emph{diffuse functions}
\cite{Spi1982JCC,Cla1983JCC,Fri1984JCP}. We shall discuss them
both to close both this section and the work.

Up to now, neither the contraction nor the splitting involved GTOs of
larger angular momentum than the largest one among the occupied
shells. However, the molecular environment is highly anisotropic and,
for most practical applications, it turns out to be convenient to add
these \emph{polarization} (large angular momentum) Gaussian shells to
the basis set, since they present lower symmetry than the GSs
discussed in the preceding paragraphs. Typically, the polarization
shells are single-primitive GSs and they are denoted by adding a
capital P to the end of the previous acronyms, resulting into, for
example, DZP, TZP, VQZP, etc., or, say, DZ2P, TZ3P, VQZ4P if more than
one polarization shell is added. In the case of Pople's basis sets
\cite{Har1973TCHA,Fri1984JCP}, these improvements are denoted by
specifying, in brackets and after the letter G, the number and type of
the polarization shells separating heavy atoms and hydrogens by a
comma\footnote{\label{foot:old_polarization} There also exists an old
notation for the addition of a single polarization shell per atom that
reads 6-31G** and that is equivalent to 6-31G(d,p).}. For example, the
basis set 6-31G(2df,p) contains the same Gaussian shells as the
original 6-31G plus two d-type shells and one f-type shell centred at
the heavy atoms, as well as one p-type shell centred at the
hydrogens.

Finally, for calculations in charged species (specially anions), where
the charge density extends further in space and the tails of the
distribution are more important to account for the relevant behaviour
of the system, it is common to augment the basis sets with
\emph{diffuse functions}, i.e., single-primitive Gaussian shells of
the same angular momentum as some preexisting one but with a smaller
exponent $\zeta$ than the smallest one in the shell. In general, this
improvement is commonly denoted by adding the prefix \emph{aug}- to
the name of the basis set. In the case of Pople's basis sets, on the
other hand, the insertion of a plus sign `+' between the contraction
scheme and the letter G denotes that the set contains one diffuse
function in the 2s- and 2p-valence shells of heavy atoms. A second +
indicates that there is another one in the 1s-shell of hydrogens. For
example, one may have the doubly augmented (and doubly polarized)
6-31++G(2d,2p) basis set.

\section[Modern developments]
        {Modern developments: An introduction to linear-scaling methods}
\label{sec:QC_modern_developments}

The ground-breaking advances reviewed in the previous sections allow
to routinely calculate, using Hartree-Fock SCF methods, physical
properties of molecules of tens of atoms in present day computers.
However, the simplest algorithms that can be devised to perform the
limiting steps in such calculations are far from optimal. This large
room for improvement, which may be enough to accommodate the exciting
possibility of linearly scaling approaches that could open the
door to thousands atoms computations, has been steering many
innovative lines of research in the last years

To close this review, we shall briefly outline here some of the
hottest areas of modern development related to the topics discussed,
specially those aimed to the reduction of the in principle $O(M^4)$
complexity associated to the calculation of the $\langle
ac|\,1/r\,|bd\rangle$ ERIs in eq.~(\ref{eq:chQC_twoe_integrals}), as
well as the $O(M^3)$ cost of the diagonalization of the Fock operator
in eq.~(\ref{eq:chQC_RH_2})\footnote{\label{foot:M} In principle,
careful distinction must be made between the number of primitive GTOs
$M_P$ and the number of contracted ones $M_C$ (see the previous
section). However, since in this section we will be dealing only with
approximate scalings without worrying much about the prefactor, the
`neutral' notation $M$ has been chosen to denote a quantity which is
linear on both $M_P$ and $M_C$.}. For wider reviews on these topics,
we recommend to the interested reader the accounts in
refs.~\cite{Sha2006PCCP,Cha1996BOOK,Och2007RCC}.

The first class of attempts to reduce the cost associated to the
calculation of the $O(M^4)$ ERIs $\langle ac|\,1/r\,|bd\rangle$ are
those aimed to the improvement of the algorithms for analytically
calculating them \emph{without approximations}. There are basically
two issues that render the construction of these algorithms a non
trivial task: First, the fact that the only four-center ERIs that
can be straightforwardly computed are the ones corresponding to a
product of four s-type GTOs, while higher angular momentum ERIs
may obtained from them in a non unique way.

After using the \emph{Gaussian product rule} \cite{Sza1996BOOK} (see
the previous section for the notation used)

\begin{eqnarray}
\label{eq:chQC_GPR}
\lefteqn{\exp \Big( - \zeta_a \,|{\bm r}-{\bm R}_{\alpha_a}|^2 \Big)
\exp \Big( - \zeta_b \,|{\bm r}-{\bm R}_{\alpha_b}|^2 \Big) = }\nonumber \\
&& \underbrace{\exp \left( - \frac{\zeta_a \zeta_b}{\zeta_a + \zeta_b}
             |{\bm R}_{\alpha_a}-{\bm R}_{\alpha_b}|^2 \right)}_{
   \displaystyle \mathcal{E}_{ab}}
   \exp \left( - ( \zeta_a + \zeta_b )
        \left|{\bm r} -
        \frac{\zeta_a {\bm R}_{\alpha_a}+
              \zeta_b {\bm R}_{\alpha_b}}{\zeta_a + \zeta_b}
        \right|^2 \right) \ ,
\end{eqnarray}

which allows the four-centre integral to be turned into a two-center
one, and whose absence for the case of STOs is the essential reason
for their being non practical, we can turn the 6-dimensional ERI into
a simple one-dimensional integral that can be readily calculated by
different means \cite{Gil1991IJQC,Gil1994AQC}:

\begin{eqnarray}
\label{eq:chQC_fund}
&& \int\!\!\!\!\int
 \frac{e^{- \zeta_a |{\bm r}-{\bm R}_{\alpha_a}|^2}
       e^{- \zeta_b |{\bm r}^{\prime}-{\bm R}_{\alpha_b}|^2}
       e^{- \zeta_c |{\bm r}-{\bm R}_{\alpha_c}|^2}
       e^{- \zeta_d |{\bm r}^{\prime}-{\bm R}_{\alpha_d}|^2}}
   {|{\bm r}-{\bm r}^{\prime}|}\,\mathrm{d}{\bm r}
   \,\mathrm{d}{\bm r}^{\prime} = \nonumber \\
&&   A_{abcd} \int_0^1 e^{-B_{abcd}x^2} \mathrm{d}x \ ,
\end{eqnarray}

with

\begin{subequations}
\label{eq:chQC_fund_expl}
\begin{align}
& A_{abcd} := \mathcal{E}_{ac}\mathcal{E}_{bd}
   \frac{2\pi^{5/2}}{(\zeta_a + \zeta_c)(\zeta_b + \zeta_d)
   (\zeta_a + \zeta_c + \zeta_b + \zeta_d)^{1/2}} \ ,
  \label{eq:chQC_fund_expl_a} \\
& B_{abcd} := \left(\frac{(\zeta_a + \zeta_c)(\zeta_b + \zeta_d)}{
   \zeta_a + \zeta_c + \zeta_b + \zeta_d}\right)^2 
   \left(\frac{\zeta_a {\bm R}_{\alpha_a}+
              \zeta_c {\bm R}_{\alpha_c}}{\zeta_a + \zeta_c} -
   \frac{\zeta_b {\bm R}_{\alpha_b}+
              \zeta_d {\bm R}_{\alpha_d}}{\zeta_b + \zeta_d}
  \right)^2 \ .
  \label{eq:chQC_fund_expl_b}
\end{align}
\end{subequations}

Such an integral is called a \emph{fundamental} ERI and, as we said,
ERIs involving GTOs with higher angular momentum than $l=0$ are
obtained from this fundamental one through iterative differentiations
of eq.~(\ref{eq:chQC_fund}) with respect to the nuclear positions,
leading to recurrence relations expressing ERIs of a given angular
momentum as a function of the lower angular momentum ones. The
particular flavour of these recurrence relations that is used is one
of the matters in which the algorithms for calculating ERIs differ.

The second issue that renders the construction of algorithms for
computing ERIs non-trivial is related to how the contraction of
primitive GTOs is handled.  In the geminal paper by Boys
\cite{Boy1950PRSLA}, the most naive procedure was suggested, namely,
the conversion of each ERI of contracted GTOs into a quadruple sum of
ERIs of primitive GTOs. However, this does not take profit, for
example, from the fact, mentioned in the previous section, that all
CGTOs in the same contracted Gaussian shell are formed by PGTOs with
the same set of exponents $\zeta_a^\mu$. Much profit can be taken from
this and other constraints and, in fact, the cost-scaling profile of
each algorithm with the contraction degree $M_a$ of the CGTOs is
strongly correlated to the moment at which the transformation between
CGTOs and PGTOs is performed \cite{Gil1994AQC}.

Among the most used of these analytical algorithms, we can mention the
Pople-Hehre (PH) one \cite{Pop1978JCOP}, which additionally exploits
the fact that for each four-center ERI of low angular momentum there
is a privileged Cartesian axis system in which many primitive
integrals vanish by symmetry; the
\emph{\underline{M}cMurchie-\underline{D}avidson} (MD) approach
\cite{McM1978JCOP}, which avoids the rotation in PH thus being more
efficient for high angular momentum ERIs; the
\emph{\underline{O}bara-\underline{S}aika-\underline{S}chlegel} (OSS)
algorithm \cite{Sch1989JCP,Oba1988JCP}; and a better defined and
improved version of it: the
\emph{\underline{H}ead-\underline{G}ordon-\underline{P}ople} (HGP)
algorithm \cite{Hea1988JCP}. Finally, if the moment at which the
contraction is handled is chosen dynamically depending on the type of
GTO appearing in the ERI, we have the PRISM modifications of MD and
HGP: the MD-PRISM \cite{Gil1989IJQC,Gil1990JCP,Gil1991IJQC} and
HGP-PRISM \cite{Gil1994AQC} algorithms, as well as a generalization of
all the previous methods, called COLD PRISM \cite{Ada1997JCP}.

Now, even if we implement any of these efficient methods for
calculating the ERIs, there is still $O(M^4)$ of them. This scaling is
simply too harsh for a too large class of applications. Therefore, the
next natural step is to try to devise approximate methods that
minimize the necessary decrease in accuracy at the same time that
maximize the savings in computer time. Thanks to the particular
characteristics of the ERIs, the Gaussian basis functions and the
concrete physical problem intended to solve\footnote{\label{locality}
The locality of many-electron Quantum Mechanics \cite{Goe1999RMP},
which is related to the \emph{nearsightedness} concept introduced by
Kohn \cite{Koh1996PRL}, is the main property that allows to finally
achieve linearity.}, this endeavour has been successfully pursued by
many researchers and the `holy grail' \cite{Jen1998BOOK} of linear
scaling with $M$ is asymptotically approaching current calculations in
large systems. Here, we shall discuss the basic issues that make this
possible. For more in depth reviews, we point the readers to the
accounts in references \cite{Cha1996BOOK,Och2007RCC,Goe1999RMP}.

The first point to consider in order to reduce the $O(M^4)$ scaling
attains the so-called \emph{radial overlap}. If we take a look at
eq.~(\ref{eq:chQC_contracted_GTO}), we can see that, irrespective of
the polynomial prefactor which contains all the angular dependence of
the GTO, every function $\chi_a({\bm r})$ contains a radial
exponential part (actually, a sum of exponentials). Therefore, if we
consider any product $\chi_a ({\bm r})\chi_b({\bm r})$ of two GTOs, we
will always find a multiplying sum of terms such as those depicted in
the expression for the Gaussian product rule in~(\ref{eq:chQC_GPR}).
The exponential decay of all quantities $\mathcal{E}_{ab}$ in those
terms with the distance $|{\bm R}_{\alpha_a}-{\bm R}_{\alpha_b}|$
between the nuclei on which the GTOs are centred indicates that, among
the $M(M+1)/2$ possible pair products $\chi_a ({\bm r})\chi_b({\bm
r})$, only $O(M)$ of them will be non-negligible. To see this, note
that if we fix (say) $a$, the only values of $b$ that will yield a
non-negligible product $\chi_a ({\bm r})\chi_b({\bm r})$ are those for
which $|{\bm R}_{\alpha_a}-{\bm R}_{\alpha_b}|$ is `small'. Since
atoms do not interpenetrate in most of the conformations that shall be
studied, this can only happen for a number of different $b$'s which is
not $O(M)$ but a constant independent of the size of the
molecule. There are $M$ different possible values of $a$ for which the
above reasoning can be repeated, and the result follows.

As a consequence, if only $O(M)$ pairs $\chi_a ({\bm r})\chi_b({\bm
  r})$ are non-negligible, then only $O(M^2)$ ERIs $\langle
ac|\,1/r\,|bd\rangle$ may in principle contribute in a significant way
to the Fock operator in eq.~(\ref{eq:chQC_matrices_2}) and not
$O(M^4)$. (For similar estimations based on slightly different
hypotheses, see \cite{Dyc1973TCA,Ahl1974TCA}.) One must also note that
the discussion is complicated by the fact that the ERIs do not appear
alone, but contracted with the density matrix elements
$D_{cd}$. This allows for further improvements of the scaling beyond
$O(M^2)$ which are discussed later in this section.

Now, knowing that most of the ERIs are too small to be relevant,
\emph{we do not know which ones} and, if we calculated the $O(M^4)$ of
them in order to spot the little ones, then we would have not gained
anything. This simple argument shows the necessity of finding a set of
\emph{estimators} that allow us to selectively drop ERIs \emph{without
calculating them}. Of course, the number of estimators must also scale
at worst like $O(M^2)$ in order for the scheme to be useful.

One of the first and simplest such estimators, was introduced by
Alml\"of et al. at the same time that they proposed the Direct SCF
method \cite{Alm1982JCC}. In their scheme, each ERI $\langle
ac|\,1/r\,|bd\rangle$ was approximated by the corresponding radial
overlap factor $\mathcal{E}_{ab}\mathcal{E}_{cd}$ (note that there are
only $O(M^2)$ numbers $\mathcal{E}_{ab}$). Although this estimator was
relatively successful, it presented the important drawback of not
being an upper-bound for the ERIs, thus rendering the control of
errors an a priori impossible task. In order to overcome this problem,
H\"aser and Ahlrichs \cite{Has1989JCC} later proposed a different
estimator which can be assured to be always greater than the
associated ERI. Following them, after the proof of positive
definiteness by Roothaan \cite{Roo1951RMP}, the electrostatic
interaction energy between two continuous charge distributions,

$$
\int \frac{\rho_1({\bm r})\rho_2({\bm r^\prime})}{|{\bm r}-{\bm r^\prime}|}
\mathrm{d}{\bm r}\mathrm{d}{\bm r^\prime}\ ,
$$

can be easily shown to satisfy the properties of an inner
product. Hence, if we choose $\rho_1:=\chi_a\chi_b$, and
$\rho_2:=\chi_c\chi_d$, we can use the well-known Schwarz inequality
to show that

\begin{eqnarray}
\label{eq:chQC_schwarz}
\langle ac | \frac{1}{r} | bd \rangle & = &
\int \frac{\rho_1({\bm r})\rho_2({\bm r^\prime})}{|{\bm r}-{\bm r^\prime}|}
\mathrm{d}{\bm r}\mathrm{d}{\bm r^\prime} \leq \nonumber \\
&& \mbox{} \leq
\left(
\int \frac{\rho_1({\bm r})\rho_1({\bm r^\prime})}{|{\bm r}-{\bm r^\prime}|}
\mathrm{d}{\bm r}\mathrm{d}{\bm r^\prime}\right)^{1/2}
\left(
\int \frac{\rho_2({\bm r})\rho_2({\bm r^\prime})}{|{\bm r}-{\bm r^\prime}|}
\mathrm{d}{\bm r}\mathrm{d}{\bm r^\prime}\right)^{1/2} =  \nonumber \\
&& \mbox{} \qquad \langle aa | \frac{1}{r} | bb \rangle^{1/2}
   \langle cc | \frac{1}{r} | dd \rangle^{1/2} \ .
\end{eqnarray}

In such a way that, by calculating only the $M(M+1)/2$ different
two-index ERIs in the last term, we can safely bound from above the
whole set containing $O(M^4)$ of them.

After these seminal works, more sophisticated and tighter bounds have
been developed through the years
\cite{Lam2005JCP,Jia1991IJQC,Gil1994CPL}, their application being
nowadays routine in Quantum Chemistry packages.

In a second generation of methods, the $O(M^2)$ formal scaling
achieved in practical calculations \cite{Str1995JCP} by using the
above ideas has been recently attacked. To this end, more specifically
physical properties of the problem are exploited (see
footnote~\ref{locality} in page~\pageref{locality}), and different
strategies are used to deal with the Coulomb and exchange parts of the
Fock operator in~(\ref{eq:chQC_matrices_2}).  The main difference
between the behaviours of these two contributions lies in the way in
which the ERIs are contracted with the density matrix $D_{cd}$: In the
`classical' Coulomb part, the relative sizes of the ERIs $\langle
ac|\,1/r\,|bd\rangle$ are largely correlated with the relative sizes
of the associated elements $D_{cd}$ \cite{Pan1991IJQC}, so that no
decrease in the $\sim M^2$ scaling is expected from density matrix
considerations. Differently, in the exchange terms, the fact that the
elements $D_{cd}$ couple `exchanged' indices in the $\langle
ac|\,1/r\,|db\rangle$ ERIs produces cancellations which make this type
of `quantum' contributions rather short-range (for non-metallic
species) \cite{Pan1991IJQC,Ter1994CPL,Bur1996JCP}. Hence, as for every
short-range interaction, a number of terms scaling \emph{linearly}
with the system size is expected. Apart from the different treatment
that this difference suggests, note that the separation of the Coulomb
and exchange tasks allows for improved parallelization the computer
codes \cite{Cha1996BOOK}.

For the exchange part, the aforementioned short-range behaviour allows
to devise $O(M)$ algorithms just by intelligently ordering the loops
in which the ERIs are calculated. The fine details of these methods
are rather technical and they are beyond the scope of this review. We
point the reader to \cite{Och2000CPL,Och1998JCP,Sch1997JCP,Bur1996JCP}
and references therein for further information.

In the Coulomb case, on the other hand, $O(M^2)$ terms still enter the
sum in~(\ref{eq:chQC_matrices_2}), and more physically-based
approximations must be used. The \emph{\underline{c}ontinuous
  \underline{f}ast \underline{m}ultipole \underline{m}ethod} (CFMM) by
White et al. \cite{Whi1994CPL}, for example, is probably one of the
most celebrated algorithms for calculating the Coulomb part of the
Fock operator. It is a generalization for continuous charge
distributions of the \emph{\underline{f}ast \underline{m}ultipole
  \underline{m}ethod} (FMM), introduced by Greengard and Rokhlin
\cite{Gre1987JCOP} in a ground-breaking paper and aimed for point-like
charges. In both~FMM and~CFMM, a clever hierarchical tree-like
division of the space into cells is performed\footnote{ Basically, a
  truncation of a \emph{\underline{B}arnes and \underline{H}ut} (BH)
  tree \cite{Bar1986NAT}.}, and the far away regions are approximated
via truncated multipole expansions.

These two ingredients, which allow to calculate the Coulomb
contribution in $O(M)$ steps for large systems, are common to most of
the fast Coulomb algorithms\footnote{ For a review of different
approaches, see~\cite{Goe1999RMP}.}. Despite this similarity, the room
for improvement seems still large enough to accommodate a vigorous
field with many publications appearing each year. Let us mention here,
for example, the \emph{\underline{g}eneralized \underline{c}ell
\underline{m}ultipole \underline{m}ethod} (GCMM) by Kutteh et al.,
which can use moments higher than monopole \cite{Kut1995CPL}; the
\emph{\underline{q}uantum \underline{c}hemical \underline{t}ree
\underline{c}ode} (QCTC) by Challacombe et al.  \cite{Cha1995JCP},
which independently thresholds `bra' and `ket' distributions; and the
\emph{\underline{G}aussian \underline{v}ery \underline{f}ast
\underline{m}ultipole \underline{m}ethod} (GvFMM) by Strain et
al. \cite{Str1996SCI}, which benefits from the idea, introduced
in~\cite{Pet1994JCP} for point-like charges, of using a dynamical
maximum angular momentum to further speed up the calculations

Note however, that the near-field contributions in these methods are
still calculated without approximations and represent a great portion
of the computer time. In this line, some modern algorithms are
appearing to alleviate this part of the work, such as, for example,
the \emph{J matrix engine} by White and Head-Gordon \cite{Whi1996JCP},
which uses and improves the ideas discussed in the first part of this
section about analytically calculating the ERIs; or the method by
Izmaylov et al. \cite{Izm2006JCP}, which implements a hierarchy of
screening levels to eliminate negligible integrals. According to
recent reports \cite{Fus2005JCP}, the combination of
CFMM, with the J matrix engine technique and with the
\emph{\underline{F}ourier \underline{t}ransform \underline{C}oulomb}
(FTC) method by F\"usti-Moln\'ar and Pulay \cite{Fus2002JCP} is
nowadays probably the fastest way for assembling the Coulomb matrix.

Now, once the construction process of the whole Fock matrix $F_{ab}$
(via its Coulomb and exchange parts) has been cast into the form of an
$O(M)$ algorithm, the importance of the second rate-limiting step in
SCF procedures, the diagonalization of $F_{ab}$, comes into
focus. Although the prefactor of the (in principle) $O(M^3)$
diagonalization step is very small and, for systems of less than a few
thousands of atoms, the absolute time spent on it is smaller than the
one needed for the formation of the Fock matrix
\cite{Fus2005JCP,Goe1999RMP,Scu1999JPCA,Str1995JCP}, it is clear that,
in the long run, it will dominate the calculations in larger systems
and will become the relevant bottleneck \cite{Scu1999JPCA}.

We shall point out the essentials regarding the methods aimed to
reduce the scaling the diagonalization step. For more in depth reviews
on the topic, we suggest to the reader the accounts
in~\cite{Bow2002JPCM,Och2000CPL,Goe1999RMP,Scu1999JPCA}.

The $O(M^3)$ complexity of classical diagonalization methods (such as
the Givens-Householder one \cite{Ort1957BOOK}) can be easily
understood if we think that the core of the algorithm is just the
multiplication of $M \times M$ matrices. Nevertheless, if the matrices
multiplied are \emph{sparse}, i.e., they have a number of
non-negligible elements that scale not as $O(M^2)$ but as $O(M)$, then
the product can be obtained in $O(M)$ steps. As a result of the
already mentioned locality properties of many-body Quantum Mechanics
(see footnote~\ref{locality} in page~\pageref{locality}), some
matrices appearing in the SCF methods discussed in this review,
namely, the density matrix $D_{ab}$ and the Fock operator $F_{ab}$,
are indeed sparse if the system is non-metallic, i.e., if it has a
non-vanishing HOMO-LUMO gap \cite{Scu1999JPCA,Mas1998JPCA}. The idea
behind most of the modern algorithms for achieving (or avoiding)
diagonalization with $O(M)$ effort consists essentially in performing
all operations using only these (local) sparse matrices, and avoiding
(non-local) dense ones, such as the MO coefficients matrix $c_{ab}$.

The attempts to improve the scaling of the diagonalization steps fall
basically in two groups \cite{Sha2003JCP}. In the first one, profit is
taken from the use of MOs which, instead of being extended over the
whole molecule, such as the canonical orbitals used in the previous
sections, are \emph{localized} in a small region of space. In these
class of methods, diagonalization \cite{Mau1993PRB} or
pseudo-diagonalization (annihilation of the occupied-virtual blocks of
the Fock matrix) \cite{Ste1996IJQC,Ste1982JCC} is still performed, and
the $O(M)$ scaling is achieved because the representation of all
operators in the basis of localized MOs is sparse. The second group of
algorithms do not use the MOs as variables but the density matrix
itself. Among them, two subfamilies of methods may be found: In the
first one, the search for the optimal density matrix is simply treated
as a standard optimization problem, being the score function the HF
energy, and the variables the density matrix elements or a set of
parameters of some suitable truncated expansion of it
\cite{Lia2003JCP}. See, for example, the approaches by Ochsenfeld and
Head-Gordon \cite{Och1997CPL}, by Salek et al. \cite{Sal2007JCP}, by
Millam and Scuseria \cite{Mil1997JCP}, or by Ordej\'on et
al. \cite{Ord1993PRB}. The other subfamily of density matrix-based
methods use iterative procedures, in such a way that, at each step,
the Fock operator is considered fixed and the equations are solved for
the density matrix (much in the spirit of the MOs-based algorithms
discussed in the previous sections). In this group, we can find, for
example, the approach in \cite{Lia2003JCP} using the Lanczos algorithm
\cite{Cul1985BOOK,Eri1980MC}, or the method by Helgaker et
al. \cite{Hel2000CPL}.

These algorithms, combined with new strategies that also avoid
diagonalization and improve SCF convergence properties, such as the
one described in~\cite{Tho2004JCP}, represent the final step towards
linear Hartree-Fock methods in Quantum Chemistry.

To close this section, although we have been concerned, up to now,
with the calculation of the electronic ground-state given a fixed
position of the nuclei, let us stress that it is also very common to
use quantum chemical methods for finding the local energy minima of
molecules. To this end, geometry optimizations must be performed and
not only must we be able to compute the energy of the molecule, but
also their derivatives with respect to the nuclear
coordinates\footnote{ Monte Carlo methods, in which the derivatives of
the energy function are not needed, could also be used. However,
although they are efficient (and often the only choice) for global
optimization problems, most of the minimizations performed in Quantum
Chemistry aim only for the closest local minimum. In such a case,
methods which do need the derivatives, such as Newton-Raphson,
steepest-descent or conjugate-gradient, usually perform
better.}. Additionally, these derivatives are also needed to do ab
initio molecular dynamics in the ground-state Born-Oppenheimer PES
\cite{Mar2000TR}.

The most naive approach, namely, the computation of the gradient of
the energy $E(\underbar{R})$ (using a simpler notation for it than
$V_N^\mathrm{eff}(\underbar{R})$ in~(\ref{eq:chQC_V_eff})) by finite
differences, is very inefficient for anything but the smallest
molecules. To see this, one just need to notice that the gradient has
as many components as the system degrees of freedom $n$. Hence, in
order to obtain it, say, at a point $\underbar{R}_0$, we would have to
compute $n+1$ times a single point energy, in order to know
$E(\underbar{R}_0)$ and $E(\underbar{R}_0+\Delta \underbar{R}_i)$,
being $\Delta \underbar{R}_i$, with $i=1,\ldots,n$, a small
displacement in each of the nuclear degrees of freedom.

This drawback was overcome in the late 60s by Pulay and others (see
\cite{Pul1987ACP,Sch1987ACP} and references therein) with the
introduction of the so-called \emph{analytical derivatives}, in which
the gradient (and higher-order derivatives) are expressed, like the
energy itself, just as a function of ERIs involving the wavefunction
at the point $\underbar{R}_0$. This marked an inflexion point in the
development of optimization and molecular dynamics algorithms that
continues nowadays, as analytical derivatives are routinely introduced
together with almost any new method for calculating the energy. In
relation to the improvements reviewed in this work, for example, let
us note that, in \cite{Sha2001JCP}, the extension of the J matrix
engine method to calculate the derivatives of the Coulomb part with
respect of the nuclei coordinates is introduced; in \cite{Och2000CPL},
linear scaling exchange gradients are developed; analytic derivatives
for the GvFMM are provided in \cite{Bur1996CPL}; the HGP algorithm for
calculating the ERIs is extended to the computation of derivatives of
the ERIs as well in \cite{Hea1988JCP}; and we may find similar
developments for the FTC method \cite{Fus2005JCP} or for algorithms
that achieve diagonalization with linear effort
\cite{Och1997CPL,Ord1993PRB}.

\section*{Acknowledgments}

\hspace{0.5cm} This work has been supported by the research projects
E24/3 and PM048 (Arag\'on Government), MEC (Spain)
\mbox{FIS2006-12781-C02-01} and MCyT (Spain)
\mbox{FIS2004-05073-C04-01}. P. Echenique has been supported by a BIFI
research contract and by a MEC (Spain) postdoctoral grant.

\appendices

\section[Functional derivatives]{Functional derivatives}

A {\it functional} $\mathcal{F}[\Psi]$ is a mapping that takes
functions to numbers (in this work, only functionals in the real
numbers are considered):

\begin{displaymath}
\begin{array}{cccc}
\mathcal{F}: & \mathcal{G} & \longrightarrow & \mathbb{R} \\
 & \Psi & \longmapsto & \mathcal{F}[\Psi]
\end{array}
\end{displaymath}

For example, if the function space $\mathcal{G}$ is the Hilbert space
of square-integrable functions $L^{2}$ (the space of states of quantum
mechanics), the objects in the domain of $\mathcal{F}$ (i.e., the
functions in $L^{2}$) can be described by infinite-tuples
$(c_{1},c_{2},\ldots )$ of complex numbers and~$\mathcal{F}$ may be
pictured as a function of infinite variables.

When dealing with function spaces $\mathcal{G}$ that meet certain
requirements\footnote{\label{foot:F_requirements} We will not discuss
the issue further but let it suffice to say that $L^{2}$ does satisfy
these requirements.}, the limit on the left-hand side of the
following equation can be written as the integral on the right-hand
side:

\begin{equation}
\label{eq:def_funct_der}
\lim_{\varepsilon \rightarrow 0}
 \frac{\mathcal{F}[\Psi_{0} + \varepsilon\delta\Psi] - \mathcal{F}[\Psi_{0}]}{\varepsilon} :=
 \int \frac{\delta \mathcal{F}[\Psi_{0}]}{\delta\Psi}(x) \delta\Psi(x)
 \mathrm{d}x\  ,
\end{equation}

where $x$ denotes a point in the domain of the
functions in $\mathcal{G}$, and the the object $(\delta
\mathcal{F}[\Psi_{0}] / \delta\Psi)(x)$ (which is a function of $x$ not
necessarily belonging to $\mathcal{G}$) is called the {\it functional
derivative} of $\mathcal{F}[\Psi]$ in the in the {\it point}
$\Psi_{0}$.

One common use of this functional derivative is to find
stationary points of functionals. A function $\Psi_{0}$ is said to
be an {\it stationary point} of $\mathcal{F}[\Psi]$ if:

\begin{equation}
\label{eq:def_funct_extr}
\frac{\delta \mathcal{F} [\Psi_{0}]}{\delta\Psi}(x) = 0 \  .
\end{equation}

In order to render this definition operative, one must have a method
for computing $(\delta \mathcal{F}[\Psi_{0}] / \delta\Psi)(x)$. Interestingly,
it is possible, in many useful cases (and in all the applications
of the formalism in this work), to calculate the
sought derivative directly from the
left-hand side of eq.~(\ref{eq:def_funct_der}). The procedure, in
such a situation, begins by writing out
$\mathcal{F}[\Psi_{0} + \varepsilon\delta\Psi]$ and clearly separating the
different orders in $\varepsilon$. Secondly, one drops the
terms of zero order (by virtue of the subtraction of the quantity
$\mathcal{F}[\Psi_{0}]$) and those of second order or higher (because they
vanish when divided by $\varepsilon$ and the limit
$\varepsilon \rightarrow 0$ is taken). The remaining terms,
all of order one, are divided by $\varepsilon$ and, finally,
$(\delta \mathcal{F}[\Psi_{0}] / \delta\Psi)(x)$ is
identified out of the resulting expression (which must written in
the form of the right-hand side of eq.~(\ref{eq:def_funct_der}).
For a practical example of this process, see secs.~\ref{sec:QC_Hartree}
and \ref{sec:QC_Hartree-Fock}.

\section[Lagrange multipliers]
        {Lagrange multipliers and constrained stationary points}

Very often, when looking for the stationary points of a function (or
a functional), the search space is not the whole one, in which the
derivatives are taken, but a certain subset of it defined by a number
of constraints. An elegant and useful method for solving the
constrained problem is that of the {\it Lagrange multipliers}.

Although it can be formally generalized to infinite dimensions (i.e.,
to functionals, see appendix~A),
here we will introduce the method in $\mathbb{R}^{\mathrm{N}}$ in
order to gain some geometrical insight and intuition.

The general framework may be described as follows: we have a
differentiable function $f({\bm x})$ that takes points in
$\mathbb{R}^{\mathrm{N}}$ to real numbers and we want to find the
stationary points of $f$ restricted to a certain subspace $\Sigma$ of
$\mathbb{R}^{\mathrm{N}}$, which is defined by $K$ {\it
constraints}\footnote{\label{foot:K_lt_N} If the constraints are
functionally independent, one must also ask that $K<N$. If not,
$\Sigma$ will be either a point (if $K=N$) or empty (if $K>N$).}:

\begin{equation}
\label{eq:constraints}
L_{i}({\bm x}) = 0 \quad i=1,\ldots,K \  .
\end{equation}

The points that are the solution of the constrained problem
are those ${\bm x}$ belonging to~$\Sigma$ where
the first order variation of $f$ would be zero if the derivatives
were taken `along'~$\Sigma$. In other words, the
points ${\bm x}$ where the gradient ${\bm \nabla}f$ has
only components (if any) in directions that `leave' $\Sigma$
(see below for a rigorous formalization of these intuitive ideas).
Thus, when comparing the solutions of the unconstrained problem
to the ones of the constrained problem, three distinct situations
arise (see fig.~\ref{fig:lagrange}):

\begin{figure}
\centerline{
\epsfxsize=8cm
\epsfbox{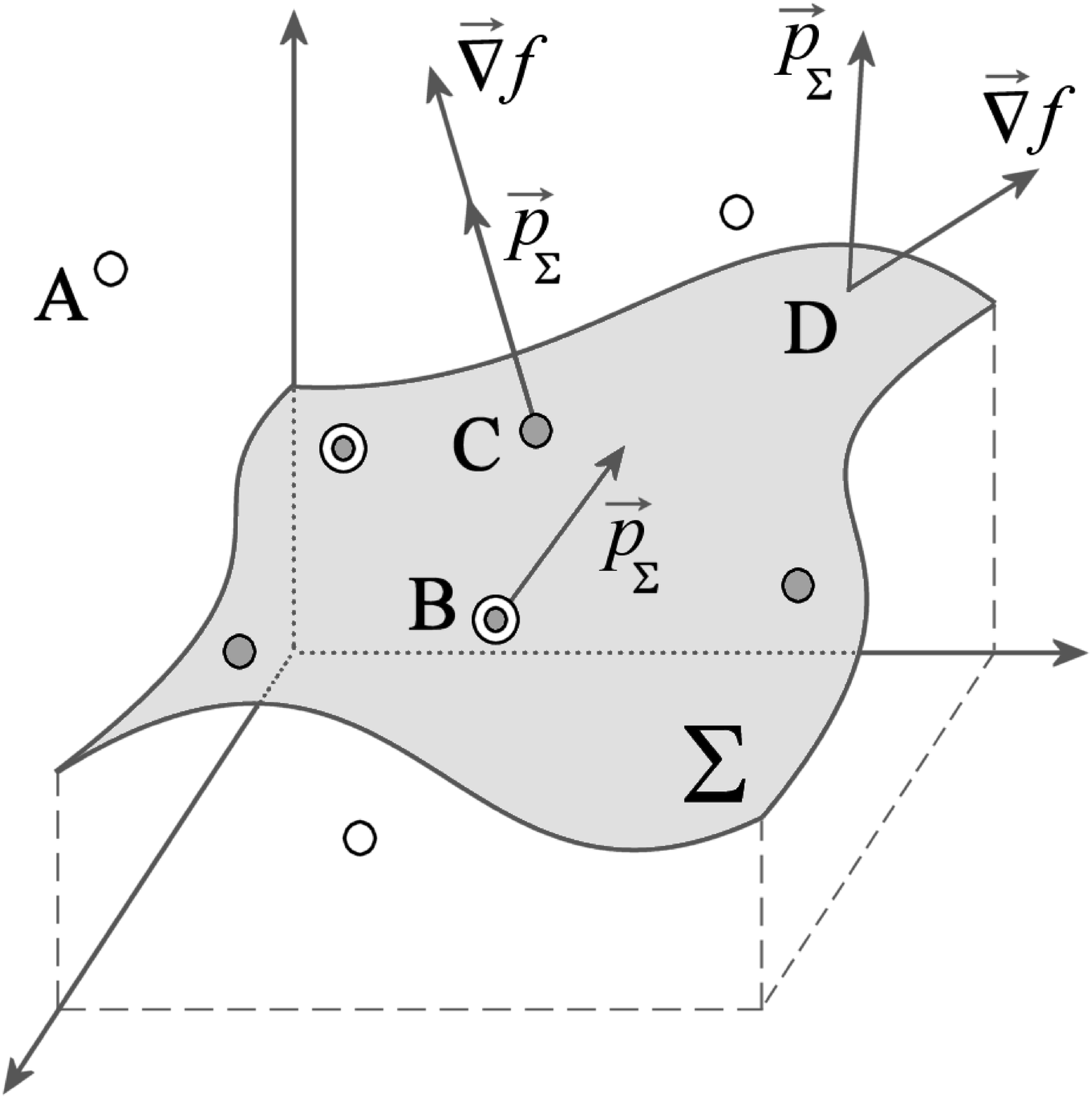}
}\vspace{0.2cm}
\caption{\label{fig:lagrange} Schematic depiction of a 
constrained stationary points problem. $\Sigma$ is the
\mbox{2-dimensional} search space, which is embedded in
$\mathbb{R}^{3}$.  The \emph{white-filled circles} are solutions of
the unconstrained problem only, the \emph{gray-filled circles} are
solutions of only the constrained one and the \emph{gray-filled
circles inside white-filled circles} are solutions of both. A, B, C
and D are examples of different situations discussed in the text.}
\end{figure}

\begin{enumerate}
\item A point ${\bm x}$ is a solution of the unconstrained problem
 (i.e. it satisfies ${\bm \nabla}f({\bm x})=0$) but it does not belong
 to $\Sigma$. Hence, it is not a solution of the constrained
 problem. This type of point is depicted as a white-filled circle in
 fig.~\ref{fig:lagrange}.
\item A point ${\bm x}$ is a solution of the unconstrained
 problem (i.e. it satisfies ${\bm \nabla}f({\bm x})=0$) and it belongs
 to $\Sigma$. Hence, it is also a solution of the constrained problem,
 since, in particular, the components of the gradient in directions
 that do not leave $\Sigma$ are zero.  This type of point is depicted
 as a gray-filled circle inside a white-filled circle in
 fig.~\ref{fig:lagrange}.
\item A point ${\bm x}$ is not a solution of the unconstrained
 problem (i.e., one has \mbox{${\bm \nabla}f({\bm x}) \ne 0$}) but it
 belongs to $\Sigma$ and the only non-zero components of the gradient
 are in directions that leave $\Sigma$. Hence, it is a solution of the
 constrained problem. This type of point is depicted as a gray-filled
 circle in fig.~\ref{fig:lagrange}.
\end{enumerate}

From this discussion, it can be seen that, in principle, no
conclusions about the number (or existence) of solutions of the
constrained problem may be drawn only from the number of solutions
of the unconstrained one. This must be investigated for
each particular situation.

In fig.~\ref{fig:lagrange}, an schematic example in $\mathbb{R}^{3}$
is depicted. The constrained search space $\Sigma$ is a 2-dimensional
surface and the direction\footnote{\label{foot:vector_in_2D} Note
that, only if $K=1$, i.e., if the dimension of $\Sigma$ is $N-1$,
there will be a vector perpendicular to the constrained space. For
$K>1$, the dimensionality of the vector space of the directions in
which one `leaves'~$\Sigma$ will be also larger than 1.} in which one
leaves $\Sigma$ is shown at several points as a perpendicular vector
${\bm p}_{\Sigma}$. In such a case, the criterium that ${\bm \nabla}f$
has only components in the direction of leaving $\Sigma$ may be
rephrased by asking ${\bm \nabla}f$ to be parallel to
${\bm p}_{\Sigma}$, i.e., by requiring that there exists a number
$\lambda$ such that ${\bm \nabla}f=-\lambda {\bm p}_{\Sigma}$.  The
case $\lambda=0$ is also admitted and the explanation of the minus
sign will be given in the following.

In this case, $K=1$, and one may note that the perpendicular
vector ${\bm p}_{\Sigma}$ is precisely ${\bm p}_{\Sigma}={\bm \nabla}L_{1}$.
Let us define $\widetilde{f}$ as

\begin{equation}
\label{eq:aux_f_2D}
\widetilde{f}({\bm x}) := f({\bm x}) + \lambda L_{1}({\bm x}) \  .
\end{equation}

It is clear that, requiring the gradient of $\widetilde{f}$ to be
zero, one recovers the condition ${\bm \nabla}f=-\lambda
{\bm p}_{\Sigma}$, which is satisfied by the points solution of the
constrained stationary points problem. If one also asks that the
derivative of $\widetilde{f}$ with respect to $\lambda$ be zero, the
constraint $L_{1}({\bm x})=0$ that defines $\Sigma$ is obtained as
well.

This process illustrates the {\it Lagrange multipliers} method in this
particular example. In the general case, described by
eq.~(\ref{eq:constraints}) and the paragraph above it, it can be proved
that the points ${\bm x}$ which are stationary subject to the constraints
imposed satisfy

\begin{equation}
\label{eq:lag_mult}
{\bm \nabla}\widetilde{f}({\bm x}) = 0 \quad \mbox{and} \quad
 \frac{\partial \widetilde{f}({\bm x})}{\partial \lambda_{i}}=0
 \quad i=1,\ldots,K \  ,
\end{equation}

where

\begin{equation}
\label{eq:aux_f}
\widetilde{f}({\bm x}) := f({\bm x}) +
 \sum_{i=1}^{K}\lambda_{i}L_{i}({\bm x}) \  .
\end{equation}

Of course, if one follows this method, the parameters $\lambda_{i}$
(which are, in fact, the {\it Lagrange multipliers}) must also
be determined and may be considered as part of the solution.

Also, it is worth remarking here that any two pair of functions,
$f_{1}$ and $f_{2}$, of $\mathbb{R}^{\mathrm{N}}$ whose restrictions
to $\Sigma$ are equal (i.e., that satisfy
$f_{1}|_{\Sigma}=f_{2}|_{\Sigma}$) obviously represent the same
constrained problem and they may be used indistinctly to construct the
auxiliary function $\widetilde{f}$. This fact allows us, after having
constructed $\widetilde{f}$ from a particular $f$, to use the
equations of the constraints to change $f$ by another simpler function
which is equal to $f$ when restricted to~$\Sigma$. This freedom is
used to derive the Hartree and Hartree-Fock equations, in
secs.~\ref{sec:QC_Hartree} and~\ref{sec:QC_Hartree-Fock},
respectively.

The formal generalization of these ideas to functionals (see
appendix~A) is straightforward if
the space $\mathbb{R}^{\mathrm{N}}$ is substituted by a functions
space $\mathcal{F}$, the points ${\bm x}$ by functions, the functions
$f$, $L_{i}$ and $\widetilde{f}$ by functionals and the requirement
that the gradient of a function be zero by the requirement that the
functional derivative of the analogous functional be zero.

Finally, let us stress something that is rarely mentioned in the
literature: \emph{There is another (older) method, apart from the Lagrange
multipliers one, for solving a constrained optimization problem:
simple substitution.} I.e., if we can find a set of $N-K$ independent
\emph{adapted coordinates} that parameterize $\Sigma$ and we can write
the score function $f$ in terms of them, we would be automatically
satisfying the constraints. Actually, in practical cases, the method
chosen is a suitable combination of the two; in such a way that, if
substituting the constraints in $f$ is difficult, the necessary
Lagrange multipliers are introduced to force them, and vice versa.

As a good example of this, the reader may want to check the derivation
of the Hartree equations in sec.~\ref{sec:QC_Hartree} (or the
Hartree-Fock ones in sec.~\ref{sec:QC_Hartree-Fock}). There, we start
by proposing a particular form for the total wavefunction $\Phi$ in
terms of the one-electron orbitals $\phi_i$ (see
eq.~(\ref{eq:chQC_constraint_total_H})) and we write the functional
$\mathcal{F}$ (which is the expected value of the energy) in terms of
that special $\Phi$ (see
eq.~(\ref{eq:chQC_constrained_functional_H})). In a second step, we
impose the constraints that the one-particle orbitals be normalized
($\langle\phi_{i}|\phi_{i}\rangle=1,i=1,\ldots,N$) and force them by
means of $N$ Lagrange multipliers $\lambda_i$. Despite the different
treatments, both conditions are constraints standing on the same
footing. The only difference is not conceptual, but operative: for the
first condition, it would be difficult to write it as a constraint;
while, for the second one, it would be difficult to define adapted
coordinates in the subspace of normalized orbitals. So, in both cases,
the easiest way for dealing with them is chosen.


\begin{thebibliography}{100}

\bibitem{Sko2005PNAS}
Skolnick, J.
\newblock {\em Proc. Natl. Acad. Sci. USA},{ \bf 102}, 2265--2266 (2005).

\bibitem{Sno2005ARBBS}
Snow, C.~D., Sorin, E.~J., Rhee, Y.~M., and Pande, V.~S.
\newblock {\em Annu. Rev. Biophys. Biomol. Struct.},{ \bf 34}, 43--69 (2005).

\bibitem{Sch2005SCI}
Schueler-Furman, O., Wang, C., Bradley, P., Misura, K., and Baker, D.
\newblock {\em Science},{ \bf 310}, 638--642 (2005).

\bibitem{Gin2005NAR}
Ginalski, K., Grishin, N.~V., Godzik, A., and Rychlewski, L.
\newblock {\em Nucleic Acids Research},{ \bf 33}, 1874--1891 (2005).

\bibitem{Bon2001ARBBS}
Bonneau, R. and Baker, D.
\newblock {\em Annu. Rev. Biophys. Biomol. Struct.},{ \bf 30}, 173--189 (2001).

\bibitem{Hao1999COSB}
Hao, M.-H. and Scheraga, H.~A.
\newblock {\em Curr. Opin. Struct. Biol.},{ \bf 9}, 184--188 (1999).

\bibitem{Ech2007COP}
Echenique, P.
\newblock {\em Contemp. Phys.},{ \bf 48}, 81--108 (2007).

\bibitem{Mor2006JPCB}
Morozov, A.~V., Tsemekhman, K., and Baker, D.
\newblock {\em J. Phys. Chem. B},{ \bf 110}, 4503--4505 (2006).

\bibitem{Jen2005ARCC}
Jensen, F.
\newblock {\em Ann. Rep. Comp. Chem.},{ \bf 1}, 1--17 (2005).

\bibitem{Mac2004JCC}
MacKerell~Jr., A.~R., Feig, M., and Brooks~III, C.~L.
\newblock {\em J. Comp. Chem.},{ \bf 25}, 1400--1415 (2004).

\bibitem{Mor2004PNAS}
Morozov, A.~V., Kortemme, T., Tsemekhman, K., and Baker, D.
\newblock {\em Proc. Natl. Acad. Sci. USA},{ \bf 101}, 6946--6951 (2004).

\bibitem{Bor2003JPCB}
Bordner, A.~J., Cavasotto, C.~N., and Abagyan, R.~A.
\newblock {\em J. Phys. Chem. B},{ \bf 107}, 9601--9609 (2003).

\bibitem{Fri1998COSB}
Friesner, R.~A. and Beachy, M.~D.
\newblock {\em Curr. Opin. Struct. Biol.},{ \bf 8}, 257--262 (1998).

\bibitem{Bea1997JACS}
Beachy, M., Chasman, D., Murphy, R., Halgren, T., and Friesner, R.
\newblock {\em J. Am. Chem. Soc.},{ \bf 119}, 5908--5920 (1997).

\bibitem{Bar2000PAC}
Barden, C.~J. and Schaffer~III, H.~F.
\newblock {\em Pure Appl. Chem.},{ \bf 72}, 1405--1423 (2000).

\bibitem{Sim1991JPC}
Simons, J.
\newblock {\em J. Chem. Phys.},{ \bf 95}, 1017--1029 (1991).

\bibitem{Lev1999BOOK}
Levine, I.~N.
\newblock {\em Quantum Chemistry},
\newblock 5th edn, Prentice Hall, Upper Saddle River (1999).

\bibitem{Jen1998BOOK}
Jensen, F.
\newblock {\em Introduction to Computational Chemistry},
\newblock John Wiley \& Sons, Chichester (1998).

\bibitem{Sza1996BOOK}
Szabo, A. and Ostlund, N.~S.
\newblock {\em Modern Quantum Chemistry: Introduced to Advanced Electronic
  Structure Theory},
\newblock Dover Publications, New York (1996).

\bibitem{Tay1995TR}
Taylor, B.~N.
\newblock Guide for the {U}se of the {I}nternational {S}ystem of {U}nits
  ({SI}),
\newblock NIST Special Publication 811, National Institute of Standards and
  Technology (1995).

\bibitem{Har1927PCPS}
Hartree, D.~R.
\newblock {\em Proc. Camb. Philos. Soc.},{ \bf 24}, 89 (1927).

\bibitem{Shu1959NAT}
Shull, H. and Hall, G.~G.
\newblock {\em Nature},{ \bf 184}, 1559 (1959).

\bibitem{Bor1954BOOK}
Born, M. and Huang, K.
\newblock {\em Dynamical Theory of Crystal Lattices}, appendices VII
  and VIII,
\newblock Oxford University Press, London (1954).

\bibitem{Bor1927APL}
Born, M. and Oppenheimer, J.~R.
\newblock {\em Ann. Phys. Leipzig},{ \bf 84}, 457--484 (1927).

\bibitem{Mar2000BOOK}
Marder, M.~P.
\newblock {\em Condensed Matter Physics},
\newblock Wiley-Interscience, New York (2000).

\bibitem{Shi2006BOOK}
Shida, T.
\newblock {\em The Chemical Bond: A Fundamental Quantum-Mechanical Picture},
\newblock Springer Series in Chemical Physics, Springer-Verlag, Berlin (2006).

\bibitem{Cra2002BOOK}
Cramer, C.~J.
\newblock {\em Essentials of Computational Chemistry: Theories and Models},
\newblock 2nd edn, John Wiley \& Sons, Chichester (2002).

\bibitem{Par1989BOOK}
Parr, R.~G. and Yang, W.
\newblock {\em Density-Functional Theory of Atoms and Molecules}, vol.~16 of
  {\em International series of monographs on chemistry},
\newblock Oxford University Press, New York (1989).

\bibitem{Sut2005PCCP}
Sutcliffe, B.~T. and Woolley, R.~G.
\newblock {\em Phys. Chem. Chem. Phys.},{ \bf 7}, 3664--3676 (2005).

\bibitem{Sut1997AQC}
Sutcliffe, B.~T.
\newblock {\em Adv. Quantum Chem.},{ \bf 28}, 65--80 (1997).

\bibitem{Sut1993JCSFT}
Sutcliffe, B.~T.
\newblock {\em J. Chem. Soc. Faraday Trans.},{ \bf 89}, 2321--2335 (1993).

\bibitem{Hun1975IJQC}
Hunter, G.
\newblock {\em Intl. J. Quant. Chem.},{ \bf 9}, 237--242 (1975).

\bibitem{Yse2003TR}
Yserentant, H.
\newblock On the electronic {S}chr{\"o}dinger equation.
\newblock Technical report, Universit{\"a}t T{\"u}bingen (2003).
\newblock {Available at:} {www.math.tu-berlin.de/$\sim$yserenta/}.

\bibitem{Sim2000JMP}
Simon, B.
\newblock {\em J. Math. Phys.},{ \bf 41}, 3523--3555 (2000).

\bibitem{Hun2000JMP}
Hunziger, W. and Sigal, I.~M.
\newblock {\em J. Math. Phys.},{ \bf 41}, 3448--3510 (2000).

\bibitem{Rus1992LNP}
Ruskai, M.~B. and Solovej, J.~P.
\newblock {\em Lect. Notes Phys.},{ \bf 403}, 153--174 (1992).

\bibitem{Hun1966HPA}
Hunziker, W.
\newblock {\em Helv. Phys. Acta},{ \bf 39}, 451--462 (1966).

\bibitem{VWi1964MFSDVS}
Van~Winter, C.
\newblock {\em Mat. Fys. Skr. Dan. Vid. Selsk},{ \bf 2}, 1--60 (1964).

\bibitem{Fri2003ARMA}
Friesecke, G.
\newblock {\em Arch. Rational Mech. Anal.},{ \bf 169}, 35--71 (2003).

\bibitem{Zhi1960TMMO}
Zhislin, G.~M.
\newblock {\em Trudy Moskovskogo matematiceskogo obscestva},{ \bf 9}, 81--120
  (1960).
\newblock (in Russian).

\bibitem{Hra2005BOOK}
Hratchian, H.~P. and Schlegel, H.~B.
\newblock in {\em {Theory and Applications of Computational Chemistry: The
  First Forty Years}}, C. Dykstra, G. Frenking, K. Kim and G. Scuseria (Eds),
  chapt.~10, Elsevier (2005).

\bibitem{Bro1983JCC}
Brooks, B.~R., Bruccoleri, R.~E., Olafson, B.~D., States, D.~J., Swaminathan,
  S., and Karplus, M.
\newblock {\em J. Comp. Chem.},{ \bf 4}, 187--217 (1983).

\bibitem{Mac1998BOOK}
MacKerell~Jr., A.~D., Brooks, B., Brooks~III, C.~L., Nilsson, L., Roux, B.,
  Won, Y., and Karplus, M.
\newblock in {\em The Encyclopedia of Computational Chemistry}, 
  P.~v.~R. Schleyer, P.~R. Schreiner, N.~L. Allinger, T. Clark, J. Gasteiger,
  P. Kollman and H.~F. Schaefer III (Eds), pp. 217--277,
  John Wiley \& Sons, Chichester (1998).

\bibitem{Pea1995CPC}
Pearlman, D.~A., Case, D.~A., Caldwell, J.~W., Ross, W.~R., Cheatham~III,
  T.~E., DeBolt, S., Ferguson, D., Seibel, G., and Kollman, P.
\newblock {\em Comp. Phys. Commun.},{ \bf 91}, 1--41 (1995).

\bibitem{Pon2003APC}
Ponder, J.~W. and Case, D.~A.
\newblock {\em Adv. Prot. Chem.},{ \bf 66}, 27--85 (2003).

\bibitem{Che2001BP}
Cheatham~III, T.~E. and Young, M.~A.
\newblock {\em Biopolymers},{ \bf 56}, 232--256 (2001).

\bibitem{Jor1988JACS}
Jorgensen, W.~L. and Tirado-Rives, J.
\newblock {\em J. Am. Chem. Soc.},{ \bf 110}, 1657--1666 (1988).

\bibitem{Coh1977BOOK}
Cohen-Tannoudji, C., Diu, B. and Lalo{\"e}, F.
\newblock {\em Quantum Mechanics},
\newblock Hermann and John Wiley \& Sons, Paris (1977).

\bibitem{Dir1929PRSL}
Dirac, P. A.~M.
\newblock {\em Proc. Roy. Soc. London},{ \bf 123}, 714 (1929).

\bibitem{See1977JCP}
Seeger, R. and Pople, J.~A.
\newblock {\em J. Chem. Phys.},{ \bf 66}, 3045--3050 (1977).

\bibitem{Mar1992EPL}
Marinari, E. and Parisi, G.
\newblock {\em Europhys. Lett.},{ \bf 19}, 451--458 (1992).

\bibitem{Cer1985JOTA}
Cerny, V.
\newblock {\em J. Optimiz. Theory App.},{ \bf 45}, 41--51 (1985).

\bibitem{Kir1983SCI}
Kirkpatrick, S., Gelatt, C.~D., and Vecchi, M.~P.
\newblock {\em Science},{ \bf 220}, 671--680 (1983).

\bibitem{Sla1930PR}
Slater, J.~C.
\newblock {\em Phys. Rev.},{ \bf 35}, 210--211 (1930).

\bibitem{Can2003BOOK}
Canc{\`{e}}s, E., DeFranceschi, M., Kutzelnigg, W., Le~Bris, C., and Maday, Y.
\newblock in {\em Handbook of numerical analysis. {V}olume {X}: {S}pecial
  volume: {C}omputational chemistry}, P.~Ciarlet and
  C.~Le~Bris (Eds), pp. 3--270, Elsevier (2003).

\bibitem{Lio1987CMP}
Lions, P.~L.
\newblock {\em Commun. Math. Phys.},{ \bf 109}, 33--97 (1987).

\bibitem{Lie1974JCP}
Lieb, E.~H. and Simon, B.
\newblock {\em J. Chem. Phys.},{ \bf 61}, 735--736 (1974).

\bibitem{Lie1977CMP}
Lieb, E.~H. and Simon, B.
\newblock {\em Commun. Math. Phys.},{ \bf 53}, 185--194 (1977).

\bibitem{Foc1930ZP}
Fock, V.
\newblock {\em Z. Phys.},{ \bf 61}, 126 (1930).

\bibitem{Koo1934PHY}
Koopmans, T.
\newblock {\em Physica},{ \bf 1}, 104 (1934).

\bibitem{Sch1991BOOK}
Schlegel, H.~B. and McDouall, J. J.~W.
\newblock in {\em {Computational Advances in Organic Chemistry: Molecular
  Structure and Reactivity}}, C.~{\"O}gretir and I.~G.
  Csizmadia (Eds), pp. 167--185, Kluwer Academic, The Netherlands (1991).

\bibitem{Onu2004COSB}
Onuchic, J.~N. and Wolynes, P.~G.
\newblock {\em Curr. Opin. Struct. Biol.},{ \bf 14}, 70--75 (2004).

\bibitem{Plo2002QRB}
Plotkin, S.~S. and Onuchic, J.
\newblock {\em Quart. Rev. Biophys.},{ \bf 35}, 111--167 (2002).

\bibitem{Dil1999PSC}
Dill, K.~A.
\newblock {\em Prot. Sci.},{ \bf 8}, 1166--1180 (1999).

\bibitem{Dob1998ACIE}
Dobson, C.~M., {\v{S}}ali, A., and Karplus, M.
\newblock {\em Angew. Chem. Int. Ed.},{ \bf 37}, 868--893 (1998).

\bibitem{Bry1995PRO}
Bryngelson, J.~D., Onuchic, J.~N., Socci, N.~D., and Wolynes, P.~G.
\newblock {\em Proteins},{ \bf 21}, 167--195 (1995).

\bibitem{Bry1987PNAS}
Bryngelson, J.~D. and Wolynes, P.~G.
\newblock {\em Proc. Natl. Acad. Sci. USA},{ \bf 84}, 7524--7528 (1987).

\bibitem{Bal1999RCC}
Bally, T. and Borden, W. T. 
\newblock {\em Rev. Comp. Chem.},{ \bf 13}, 1--97 (1999).

\bibitem{Pop1954JCP}
Pople, J.~A. and Nesbet, R. K.
\newblock {\em J. Chem. Phys.},{ \bf 22}, 571--572 (1954).

\bibitem{Var2002JPCA}
Vargas, R., Garza, J., Hay, B.~P., and Dixon, D.~A.
\newblock {\em J. Phys. Chem. A},{ \bf 106}, 3213--3218 (2002).

\bibitem{Lan2005PSFB}
L{\'{a}}ng, A., Csizmadia, I.~G., and Perczel, A.
\newblock {\em PROTEINS: Struct. Funct. Bioinf.},{ \bf 58}, 571--588 (2005).

\bibitem{Per2003JCC}
Perczel, A., Farkas, {\"O}., J{\'a}kli, I., Topol, I.~A., and Csizmadia, I.~G.
\newblock {\em J. Comp. Chem.},{ \bf 24}, 1026--1042 (2003).

\bibitem{Yu2001JMS}
Yu, C.-H., Norman, M.~A., Sch{\"{a}}fer, L., Ramek, M., Peeters, A., and van
  Alsenoy, C.
\newblock {\em J. Mol. Struct.},{ \bf 567--568}, 361--374 (2001).

\bibitem{Els2001CP}
Elstner, M., Jalkanen, K.~J., Knapp-Mohammady, M., and Suhai, S.
\newblock {\em Chem. Phys.},{ \bf 263}, 203--219 (2001).

\bibitem{Bal2000JMS}
Baldoni, H.~A., Zamarbide, G., Enriz, R.~D., Jauregui, E.~A., Farkas, {\"{O}}.,
  Perczel, A., Salpietro, S.~J., and Csizmadia, I.~G.
\newblock {\em J. Mol. Struct.},{ \bf 500}, 97--111 (2000).

\bibitem{Rod1998JMS}
Rodr{\'{\i}}guez, A.~M., Baldoni, H.~A., Suvire, F., Nieto~V{\'{a}}zquez, R.,
  Zamarbide, G., Enriz, R.~D., Farkas, {\"{O}}., Perczel, A., McAllister,
  M.~A., Torday, L.~L., Papp, J.~G., and Csizmadia, I.~G.
\newblock {\em J. Mol. Struct.},{ \bf 455}, 275--301 (1998).

\bibitem{Csa1999PBMB}
Cs{\'{a}}sz{\'{a}}r, A.~G. and Perczel, A.
\newblock {\em Prog. Biophys. Mol. Biol.},{ \bf 71}, 243--309 (1999).

\bibitem{Fre1992JACS}
Frey, R.~F., Coffin, J., Newton, S.~Q., Ramek, M., Cheng, V. K.~W., Momany,
  F.~A., and Sch{\"{a}}fer, L.
\newblock {\em J. Am. Chem. Soc.},{ \bf 114}, 5369--5377 (1992).

\bibitem{Roo1951RMP}
Roothaan, C. C.~J.
\newblock {\em Rev. Mod. Phys.},{ \bf 23}, 69--89 (1951).

\bibitem{HalPRSLA1951}
Hall, G.~G.
\newblock {\em Proc. Roy. Soc. London Ser. A},{ \bf 205}, 541--552 (1951).

\bibitem{Coo2005BOOK}
Cook, D. B.
\newblock {\em Handbook of Computational Quantum Chemistry},
\newblock Dover Publications, Mineola, New York (2005).

\bibitem{Car1977BOOK}
Carbo, R. and Riera, J. M.
\newblock {\em A General {SCF} Theory},
\newblock Lecture Notes in Chemistry, Springer-Verlag, New York (1977).

\bibitem{Hur1976BOOK}
Hurley, A. C.
\newblock {\em Introduction to the Electron Theory of Small Molecules},
\newblock Academic Press, New York (1976).

\bibitem{McW1992BOOK}
McWeeny, R.
\newblock {\em Methods of Molecular Quantum Mechanics},
\newblock Academic Press (1992).

\bibitem{Roo1960RMP}
Roothaan, C. C.~J.
\newblock {\em Rev. Mod. Phys.},{ \bf 32}, 179--185 (1960).

\bibitem{Bin1974MP}
Binkley, J. S., Pople, J. A. and Dobosh, P. A.
\newblock {\em Mol. Phys.},{ \bf 28}, 1423--1429 (1974).

\bibitem{Pet1992SOFT}
Peterson M. and Pourier R.,
\newblock {\em MONSTERGAUSS-92},
\newblock Department of Chemistry,
University of Toronto and Memorial University of Newfoundland, St.
John's, Newfoundland, Canada.

\bibitem{Pla2006JCP}
Plakhutin, B. N., Gorelik, E. V. and Breslavskaya, N. N.
\newblock {\em J. Chem. Phys.},{ \bf 125}, 204110 (2006).

\bibitem{Hir1974JCP}
Hirao, K.
\newblock {\em J. Chem. Phys.},{ \bf 60}, 3125--3133 (1974).

\bibitem{Gue1974MP}
Guest, M. F. and Saunders, V. R.
\newblock {\em Mol. Phys.},{ \bf 28}, 819--828 (1974).

\bibitem{Kob1997AQC}
Kobus, J.
\newblock {\em Adv. Quantum Chem.},{ \bf 28}, 1--14 (1997).

\bibitem{Jen2005TCA}
Jensen, F.
\newblock {\em Theo. Chem. Acc.},{ \bf 113}, 267--273 (2005).

\bibitem{Hea1988JPC}
Head-Gordon, M. and Pople, J. A.
\newblock {\em J. Phys. Chem.},{ \bf 92}, 3063--3069 (1988).

\bibitem{Pop1999RMP}
Pople, J.~A.
\newblock {\em Rev. Mod. Phys.},{ \bf 71}, 1267--1274 (1999).

\bibitem{Bra1971IJQC}
Brailsford, D. F. and Hall, G. G.
\newblock {\em Intl. J. Quant. Chem.},{ \bf 5}, 657--668 (1971).

\bibitem{Gar2003BOOK}
Garc{\'{\i}}a de~la Vega, J.~M. and Miguel, B.
\newblock in {\em {Introduction to Advanced Topics of Computational
  Chemistry}}, L.~A. Montero, L.~A. D{\'{\i}}az and
  R.~Bader (Eds), chapt.~3, pp. 41--80, Editorial de la Universidad de
  la Habana (2003).

\bibitem{Hel1995BOOK}
Helgaker, T. and Taylor, P.~R.
\newblock in {\em {Modern Electronic Structure Theory. Part II}},
  D.~R. Yarkony (Ed), pp. 725--856, World Scientific, Singapore (1995).

\bibitem{Abr1964BOOK}
Abramowitz, M. and Stegun, I.~A.
\newblock {\em Handbook of Mathematical Functions with Formulas, Graphs, and
  Mathematical Tables},
\newblock 9th edn, Dover, New York (1964).

\bibitem{Sla1930PRb}
Slater, J.~C.
\newblock {\em Phys. Rev.},{ \bf 36}, 57--54 (1930).

\bibitem{Zen1930PR}
Zener, C.
\newblock {\em Phys. Rev.},{ \bf 36}, 51--56 (1930).

\bibitem{Mat2002IJQC}
Mathar, R.~J.
\newblock {\em Intl. J. Quant. Chem.},{ \bf 90}, 227--243 (2002).

\bibitem{Kat1957CPAM}
Kato, T.
\newblock {\em Commun. Pure Appl. Math.},{ \bf 10}, 151--177 (1957).

\bibitem{Boy1950PRSLA}
Boys, S.~F.
\newblock {\em Proc. Roy. Soc. London Ser. A},{ \bf 200}, 541--554 (1950).

\bibitem{Sch1995IJQC}
Schlegel, H.~B. and Frisch, M.~J.
\newblock {\em Intl. J. Quant. Chem.},{ \bf 54}, 83--87 (1995).

\bibitem{Ech2006JCCb}
Echenique, P., Calvo, I., and Alonso, J.~L.
\newblock {\em J. Comp. Chem.},{ \bf 27}, 1748--1755 (2006).

\bibitem{Ech2006JCCa}
Echenique, P. and Alonso, J.~L.
\newblock {\em J. Comp. Chem.},{ \bf 27}, 1076--1087 (2006).

\bibitem{Heh1969JCP}
Hehre, W.~J., Stewart, R.~F., and Pople, J.~A.
\newblock {\em J. Chem. Phys.},{ \bf 51}, 2657--2664 (1969).

\bibitem{Dit1971JCP}
Ditchfield, R., Hehre, W.~J., and Pople, J.~A.
\newblock {\em J. Chem. Phys.},{ \bf 54}, 724--728 (1971).

\bibitem{Heh1972JCP}
Hehre, W.~J., Ditchfield, R., and Pople, J.~A.
\newblock {\em J. Chem. Phys.},{ \bf 56}, 2257--2261 (1972).

\bibitem{Har1973TCHA}
Hariharan, P.~C. and Pople, J.~A.
\newblock {\em Theor. Chim. Acta},{ \bf 28}, 213--222 (1973).

\bibitem{Fri1984JCP}
Frisch, M.~J., Pople, J.~A., and Binkley, J.~S.
\newblock {\em J. Chem. Phys.},{ \bf 80}, 3265--3269 (1984).

\bibitem{Kri1980JCP}
Krishnan, R., Binkley, J.~S., Seeger, R., and Pople, J.~A.
\newblock {\em J. Chem. Phys.},{ \bf 72}, 650--654 (1980).

\bibitem{Bin1980JACS}
Binkley, J.~S., Pople, J.~A., and Hehre, W.~J.
\newblock {\em J. Am. Chem. Soc.},{ \bf 102}, 939--947 (1980).

\bibitem{Spi1982JCC}
Spitznagel, G.~W., Clark, T., Chandrasekhar, J., and Schleyer, P. v.~R.
\newblock {\em J. Comp. Chem.},{ \bf 3}, 363--371 (1982).

\bibitem{Cla1983JCC}
Clark, T., Chandrasekhar, J., Spitznagel, G.~W., and Schleyer, P. v.~R.
\newblock {\em J. Comp. Chem.},{ \bf 4}, 294--301 (1983).

\bibitem{Dun1989JCP}
Dunning Jr., T. H.
\newblock {\em J. Chem. Phys.},{ \bf 90}, 1007--1023 (1989).

\bibitem{Woo1983BOOK}
Woodgate, G.~K.
\newblock {\em Elementary Atomic Structure},
\newblock 2nd edn, Oxford University Press, USA (1983).

\bibitem{Sha2006PCCP}
Shao, Y., Molnar L. F., Jung, Y., Kussmann, J., Ochsenfeld, C., Brown, S. T.,
Gilbert, A. T. B., Slipchenko, L. V., Levchenko, S. V., O'Neill, D. P.,
DiStasio, R. A., Lochan, R. C., Wang, T., Beran, G. J. O., Besley, N. A.,
Herbert, J. M., Lin, C. Y., Van Voorhis, T., Chien, S. H., Sodt, A,
Steele, R. P., Rassolov, V. A., Maslen, P. E., Korambath, P. P.,
Adamson, R. D., Austin, B., Baker, J., Byrd, E. F. C., Dachsel, H.,
Doerksen, R. J., Dreuw, A, Dunietz, B. D., Dutoi, A. D., Furlani, T. R.,
Gwaltney, S. R., Heyden, A., Hirata, S., Hsu, C. P., Kedziora, G.,
Khalliulin, R. Z., Klunzinger, P., Lee, A. M., Lee, M. S., Liang, W.,
Lotan, I., Nair, N., Peters, B., Proynov, E. I., Pieniazek, P. A.,
Rhee, Y. M., Ritchie, J., Rosta, E., Sherrill, C. D., Simmonett, A. C.,
Subotnik, J. E., Woodcock, H. L., Zhang, W., Bell, A. T.,
Chakraborty, A. K., Chipman, D. M., Keil, F. J., Warshel, A.,
Hehre, W. J., Schaefer, H. F., Kong, J., Krylov, A. I.,
Gill, P. M. W., Head-Gordon, M.
\newblock {\em Phys. Chem. Chem. Phys.},{ \bf 8}, 3172--3191 (2006).

\bibitem{Cha1996BOOK}
Challacombe, M., Schwegler, R. and Alml\"of, J.
\newblock in {\em {Computational Chemistry: Review of Current Trends}},
  J. Leczszynski (Ed), pp. 53--107, World Scientific (1996).

\bibitem{Och2007RCC}
Ochsenfeld, C., Kussmann, J. and Lambrecht, D. S.
\newblock in {\em {Reviews in Computational Chemistry}},
  K. B. Lipkowitz, T. R. Cundari (Eds), Vol. 23, pp. 1--82,
 John Wiley \& Sons (2007).

\bibitem{Gil1991IJQC}
Gill, P. M. W., Johnson, B. G. and Pople, G. A.
\newblock {\em Intl. J. Quant. Chem.},{ \bf 40}, 745--652 (1991).

\bibitem{Gil1994AQC}
Gill, P. M. W.
\newblock {\em Adv. Quant. Chem.},{ \bf 25}, 141--173 (1994).

\bibitem{Pop1978JCOP}
Pople, J. A. and Hehre, W. J.
\newblock {\em J. Comp. Phys.},{ \bf 27}, 161--168 (1978).

\bibitem{McM1978JCOP}
McMurchie, L. E. and Davidson, E. R.
\newblock {\em J. Comp. Phys.},{ \bf 26}, 218--231 (1978).

\bibitem{Sch1989JCP}
Schlegel, H. B.
\newblock {\em J. Chem. Phys.},{ \bf 90}, 5630--5634 (1989).

\bibitem{Oba1988JCP}
Obara, S. and Saika, A.
\newblock {\em J. Chem. Phys.},{ \bf 89}, 1540--1559 (1988).

\bibitem{Hea1988JCP}
Head-Gordon, M. and Pople, J. A.
\newblock {\em J. Chem. Phys.},{ \bf 89}, 5777--5786 (1988).

\bibitem{Gil1989IJQC}
Gill, P. M. W., Head-Gordon, M. and Pople, G. A.
\newblock {\em Intl. J. Quant. Chem., Symp.},{ \bf 23}, 269 (1989).

\bibitem{Gil1990JCP}
Gill, P. M. W., Head-Gordon, M. and Pople, G. A.
\newblock {\em J. Chem. Phys.},{ \bf 94}, 5564 (1990).

\bibitem{Gil1991IJQCb}
Gill, P. M. W. and Pople, G. A.
\newblock {\em Intl. J. Quant. Chem., Symp.},{ \bf 40}, 753 (1991).

\bibitem{Ada1997JCP}
Adams, T. R., Adamson, R. S. and Gill, P. M. W.
\newblock {\em J. Chem. Phys.},{ \bf 107}, 124--131 (1997).

\bibitem{Goe1999RMP}
Goedecker, S.
\newblock {\em Rev. Mod. Phys.},{ \bf 71}, 1085--1123 (1999).

\bibitem{Koh1996PRL}
Kohn, W.
\newblock {\em Phys. Rev. Lett.},{ \bf 76}, 3168--3171 (1996).

\bibitem{Dyc1973TCA}
Dyczmons, V.
\newblock {\em Theoret. Chim. Acta},{ \bf 28}, 307--310 (1973).

\bibitem{Ahl1974TCA}
Ahlrichs, R.
\newblock {\em Theoret. Chim. Acta},{ \bf 33}, 157--167 (1974).

\bibitem{Alm1982JCC}
Alml\"of, J., Faegri Jr., K. and Korsell, K.
\newblock {\em J. Comp. Chem.},{ \bf 3}, 385--399 (1982).

\bibitem{Has1989JCC}
H\"aser, M. and Ahlrichs, R.
\newblock {\em J. Comp. Chem.},{ \bf 10}, 104--111 (1989).

\bibitem{Lam2005JCP}
Lambrecht, D. S. and Ochsenfeld, C.
\newblock {\em J. Chem. Phys.},{ \bf 123}, 184101 (2005).

\bibitem{Gil1994CPL}
Gill, P. M. W., Johnson, B. G. and Pople, G. A.
\newblock {\em Chem. Phys. Lett.},{ \bf 217}, 65--68 (1994).

\bibitem{Jia1991IJQC}
Jiancheng, X. and Shouping, J.
\newblock {\em Intl. J. Quant. Chem.},{ \bf 39}, 123--130 (1991).

\bibitem{Str1995JCP}
Strout, D. L. and Scuseria, G. E.
\newblock {\em J. Chem. Phys.},{ \bf 102}, 8448--8452 (1995).

\bibitem{Pan1991IJQC}
Panas, I., Alml\"of, J. and Feyereisen, M. W.
\newblock {\em Intl. J. Quant. Chem.},{ \bf 40}, 797--807 (1991).

\bibitem{Ter1994CPL}
Termath, V. and Handy, N. C.
\newblock {\em Chem. Phys. Lett.},{ \bf 230}, 17--24 (1994).

\bibitem{Bur1996JCP}
Burant, J. C., Scuseria, G. E. and Frisch, M. J.
\newblock {\em J. Chem. Phys.},{ \bf 105}, 8969--8972 (1996).

\bibitem{Och1998JCP}
Ochsenfeld, C., White, C. A. and Head-Gordon, M.
\newblock {\em J. Chem. Phys.},{ \bf 109}, 1663--1669 (1998).

\bibitem{Sch1997JCP}
Schwegler, E., Challacombe, M. and Head-Gordon, M.
\newblock {\em J. Chem. Phys.},{ \bf 106}, 9708--9717 (1997).

\bibitem{Och2000CPL}
Ochsenfeld, C.
\newblock {\em Chem. Phys. Lett.},{ \bf 327}, 216 (2000).

\bibitem{Whi1994CPL}
White, C. A., Johnson, B. G., Gill, P. M. W. and Head-Gordon, M.
\newblock {\em Chem. Phys. Lett.},{ \bf 230}, 8--16 (1994).

\bibitem{Gre1987JCOP}
Greengard, L. and Rokhlin, C.
\newblock {\em J. Comp. Phys.},{ \bf 73}, 325--348 (1987).

\bibitem{Bar1986NAT}
Barnes, J. and Hut P.
\newblock {\em Nature},{ \bf 104}, 446--449 (1986).

\bibitem{Kut1995CPL}
Kutteh, R., Apr\`a, E. and Nichols, J.
\newblock {\em Chem. Phys. Lett.},{ \bf 238}, 173--179 (1995).

\bibitem{Cha1995JCP}
Challacombe, M., Schwegler, E. and Alml\"of, J.
\newblock {\em J. Chem. Phys.},{ \bf 104}, 4685 (1995).

\bibitem{Str1996SCI}
Strain, M. C., Scuseria, G. E. and Frisch, M. J.
\newblock {\em Science},{ \bf 271}, 51--53 (1996).

\bibitem{Pet1994JCP}
Petersen, H. G., Soelvason, D. and Perram, J. W.
\newblock {\em J. Chem. Phys.},{ \bf 101}, 8870--8876 (1994).

\bibitem{Whi1996JCP}
White, C. A. and Head-Gordon, M.
\newblock {\em J. Chem. Phys.},{ \bf 104}, 2620--2629 (1996).

\bibitem{Izm2006JCP}
Izmaylov, A. F., Scuseria, G. E. and Frisch, M. J.
\newblock {\em J. Chem. Phys.},{ \bf 125}, 104103 (2006).

\bibitem{Fus2005JCP}
F\"usti-Moln\'ar, L. and Kong, J.
\newblock {\em J. Chem. Phys.},{ \bf 122}, 074108 (2005).

\bibitem{Fus2002JCP}
F\"usti-Moln\'ar, L. and Pulay, P.
\newblock {\em J. Chem. Phys.},{ \bf 117}, 7827 (2002).

\bibitem{Scu1999JPCA}
Scuseria, G. E.
\newblock {\em J. Phys. Chem. A},{ \bf 103}, 4782--4790 (1999).

\bibitem{Bow2002JPCM}
Bowler, D. R., Miyazaki, T. and Gillan, M. J.
\newblock {\em J. Phys.: Condens. Matter},{ \bf 14}, 2781--2798 (2002).

\bibitem{Ort1957BOOK}
Ortega, J.
\newblock in {\em {Mathematical Methods for Digital Computers}},
  A. Ralston and H. S. Wilf (Eds), Vol. 2, p. 94, Wiley, New York (1957).

\bibitem{Mas1998JPCA}
Maslen, P. E., Ochsenfeld, C., White, C. A., Lee, M. S. and Head-Gordon, M.
\newblock {\em J. Phys. Chem. A},{ \bf 102}, 2215--2222 (1998).

\bibitem{Sha2003JCP}
Shao, Y., Saravanan, C., Head-Gordon, M. and White, C. A.
\newblock {\em J. Chem. Phys.},{ \bf 118}, 6144--6151 (2003).

\bibitem{Mau1993PRB}
Mauri, F., Galli, G. and Car, R.
\newblock {\em Phys. Rev. B},{ \bf 47}, 9973--9976 (1993).

\bibitem{Ste1996IJQC}
Stewart, J. J. P.
\newblock {\em Intl. J. Quant. Chem.},{ \bf 58}, 133--146 (1996).

\bibitem{Ste1982JCC}
Stewart, J. J. P., Cs\'asz\'ar, P. and Pulay, P.
\newblock {\em J. Comp. Chem.},{ \bf 3}, 227--228 (1982).

\bibitem{Och1997CPL}
Ochsenfeld, C. and Head-Gordon, M.
\newblock {\em Chem. Phys. Lett.},{ \bf 270}, 399--405 (1997).

\bibitem{Sal2007JCP}
Salek, P., H{\o}st, S., Th{\o}gersen, L., J{\o}rgensen, P.,
Manninen, P., Olsen, J., Jans\'{\i}k, B., Reine, S.,
Pawlowski, F., Tellgren, E., Helgaker, T. and Coriani, S.
\newblock {\em J. Chem. Phys.},{ \bf 126}, 114110 (2007).

\bibitem{Mil1997JCP}
Millam, J. M. and Scuseria, G. E.
\newblock {\em J. Chem. Phys.},{ \bf 106}, 5569--5577 (1997).

\bibitem{Ord1993PRB}
Ordej\'on, P., Drabold, D. A., Grumbach, M. P. and Martin, R. M.
\newblock {\em Phys. Rev. B},{ \bf 48}, 14646--14649 (1993).

\bibitem{Lia2003JCP}
Liang, W., Saravanan, C., Shao, Y., Baer, R., Bell, A. T. and Head-Gordon, M.
\newblock {\em J. Chem. Phys.},{ \bf 119}, 4117--4125 (2003).

\bibitem{Cul1985BOOK}
Cullum, J. K. and Willoughby, R. A.
\newblock {\em {Lanczos Algorithms for Large Symmetric Eigenvalue
 Computations. Vol 2}},
 Birkh\"auser, Boston (1985).

\bibitem{Eri1980MC}
Ericsson, T. and Ruhe, A.
\newblock {\em Math. Comput.},{ \bf 35}, 1251 (1980).

\bibitem{Hel2000CPL}
Helgaker, T., Larsen, H., Olsen, J. and J{\o}rgensen, P.
\newblock {\em Chem. Phys. Lett.},{ \bf 327}, 397--403 (2000).

\bibitem{Tho2004JCP}
Th{\o}rgensen, L., Olsen, J., Yeager, D., J{\o}rgensen, P., Salek, P. and
Helgaker, T.
\newblock {\em J. Chem. Phys.},{ \bf 121}, 16--27 (2004).

\bibitem{Mar2000TR}
Marx, D. and Hutter, J.
\newblock in {\em {Modern Methods and Algorithms of Quantum
  Chemistry}},
  J. Grotendorst (Ed), Vol. 3, pp. 329--477,
  John von Neumann Institute for Computing, J\"ulich (2000).

\bibitem{Pul1987ACP}
Pulay, P.
\newblock {\em Adv. Chem. Phys.},{ \bf 69}, 241 (1987).

\bibitem{Sch1987ACP}
Schlegel, H. B.
\newblock {\em Adv. Chem. Phys.},{ \bf 67}, 249 (1987).

\bibitem{Sha2001JCP}
Shao, Y., White, C. A. and Head-Gordon, M.
\newblock {\em J. Chem. Phys.},{ \bf 114}, 6572--6577 (2001).

\bibitem{Bur1996CPL}
Burant, J. C., Strain, M. C., Scuseria, G. E. and Frisch, M. J.
\newblock {\em Chem. Phys. Lett.},{ \bf 248}, 42--49 (1996).

\end{thebibliography}
\end{document}